%% file: White_Paper_ngEHT_Fundamental_Physics_SWG.tex
\documentclass[smallcondensed]{svjour3}     

\usepackage[T1]{fontenc}
\usepackage[utf8]{inputenc}

\smartqed  
\usepackage{natbib}
\usepackage{comment}
\usepackage{amssymb,amsmath}
\usepackage{graphicx}
\usepackage{booktabs}
\usepackage{ulem}
\usepackage{color}
\usepackage{physics}
\usepackage[colorlinks=true,citecolor=blue,urlcolor=blue,breaklinks]{hyperref}

\usepackage{changepage}
\newenvironment{cwid}{\begin{adjustwidth}{1.5em}{}}{\end{adjustwidth}}

\usepackage{makecell}
\usepackage{xspace}
\usepackage{bbm}
\usepackage[dvipsnames]{xcolor}   
\usepackage{lscape}
\usepackage{longtable}
\RequirePackage{fix-cm}

\def\m87{{M87$^{\ast}\xspace$}}
\def\sgra{{Sgr\,A$^{\ast}\xspace$}}

\defcitealias{ngEHT_refarray}{ngEHT~Array}
\defcitealias{ngEHT_KSG}{ngEHT~Science}

\journalname{Living Reviews in Relativity}

\newcommand\be{\begin{equation}}
\newcommand\ee{\end{equation}}
\def\be{\begin{equation}}
\def\ee{\end{equation}}
\def\bea{\begin{eqnarray}}
\def\eea{\end{eqnarray}}
\newcommand{\beq}{\begin{eqnarray}}
\newcommand{\eeq}{\end{eqnarray}} 
\newcommand{\ba}{\begin{align}}
\newcommand{\ea}{\end{align}}

\begin{document}

\title{Fundamental Physics Opportunities with the Next-Generation Event Horizon Telescope}

\author{Dimitry~Ayzenberg, Lindy Blackburn, Richard~Brito, Silke~Britzen, Avery~E.~Broderick, Ra\'ul~Carballo-Rubio, Vitor~Cardoso, Andrew~Chael, Koushik Chatterjee, Yifan~Chen, Pedro~V.~P.~Cunha, Hooman~Davoudiasl, Peter~B.~Denton, Sheperd~S.~Doeleman, Astrid~Eichhorn, Marshall~Eubanks, Yun~Fang, Arianna~Foschi, Christian~M.~Fromm, Peter~Galison, Sushant~G.~Ghosh, Roman~Gold, Leonid~I.~Gurvits, Shahar~Hadar, Aaron~Held, Janice~Houston, Yichao~Hu, Michael~D.~Johnson, Prashant Kocherlakota, Priyamvada~Natarajan, H\'ector~Olivares, Daniel~Palumbo, Dominic W.~Pesce, Surjeet~Rajendran, Rittick~Roy, Saurabh, Lijing~Shao, Shammi~Tahura, Aditya~Tamar, Paul~Tiede, Fr\'ed\'eric~H.~Vincent, Luca~Visinelli, Zhiren~Wang, Maciek~Wielgus, Xiao~Xue, Kadri~Yakut, Huan~Yang, Ziri~Younsi}

%
%
%
\authorrunning{The ngEHT Fundamental Physics SWG}
\titlerunning{Fundamental Physics Opportunities with the ngEHT}
\maketitle
%
%
%
\begin{abstract}
The Event Horizon Telescope (EHT) Collaboration recently published the first images of the supermassive black holes in the cores of the Messier 87 and Milky Way galaxies.
These observations have provided a new means to study supermassive black holes and probe physical processes occurring in the strong-field regime.
We review the prospects of future observations and theoretical studies of supermassive black hole systems with the next-generation Event Horizon Telescope (ngEHT), which will greatly enhance the capabilities of the existing EHT array.
These enhancements will open up several previously inaccessible avenues of investigation, thereby providing important new insights into the properties of supermassive black holes and their environments.
This review describes the current state of knowledge for five key science cases, summarising the unique challenges and opportunities for fundamental physics investigations that the ngEHT will enable.

%
\end{abstract}

\setcounter{tocdepth}{3}
\newpage
\tableofcontents


\section{Introduction} \label{sec:intro}
Recent very-long-baseline interferometry (VLBI) observations of supermassive black holes (SMBHs) have opened a new path to observe and study strong field gravity.
Black holes (BHs) lie at the edge of our understanding of the fundamental laws of physics.
The mechanisms governing their genesis and evolution are poorly understood, but there is substantial evidence for the pivotal role they play in star formation, galactic evolution, cosmic energy exchange and transport, accretion and outflows, and the generation of ultra-high-energy (UHE) emissions.
From a fundamental physics perspective, BHs hold a tremendous potential for advancing scientific knowledge.
They are considered central to possible energy extraction mechanisms {\it from vacuum} of their surrounding deep gravitational potentials, wherein gravitational lensing can be so strong that the trajectories of light rays are ``closed''.
The Einstein field equations (EFEs) break down in BH interiors, raising the prospect that the geometry close to the event horizon may carry observable imprints which will prove crucial in the development of a more comprehensive description of the gravitational interaction.
Such developments will require conclusive experimental evidence for the existence of astrophysical BH event horizons and for a detailed mapping of their geometry.
The last decade has been vital for the field.
In 2015 the first direct detection of gravitational waves (GWs) opened up the remarkable new tool of GW astronomy to study compact objects.
GWs have been successfully used to probe BHs and neutron stars in the highly dynamical regime.

\begin{figure}[htpb]
	\centering
	\includegraphics[width=1.0\textwidth]{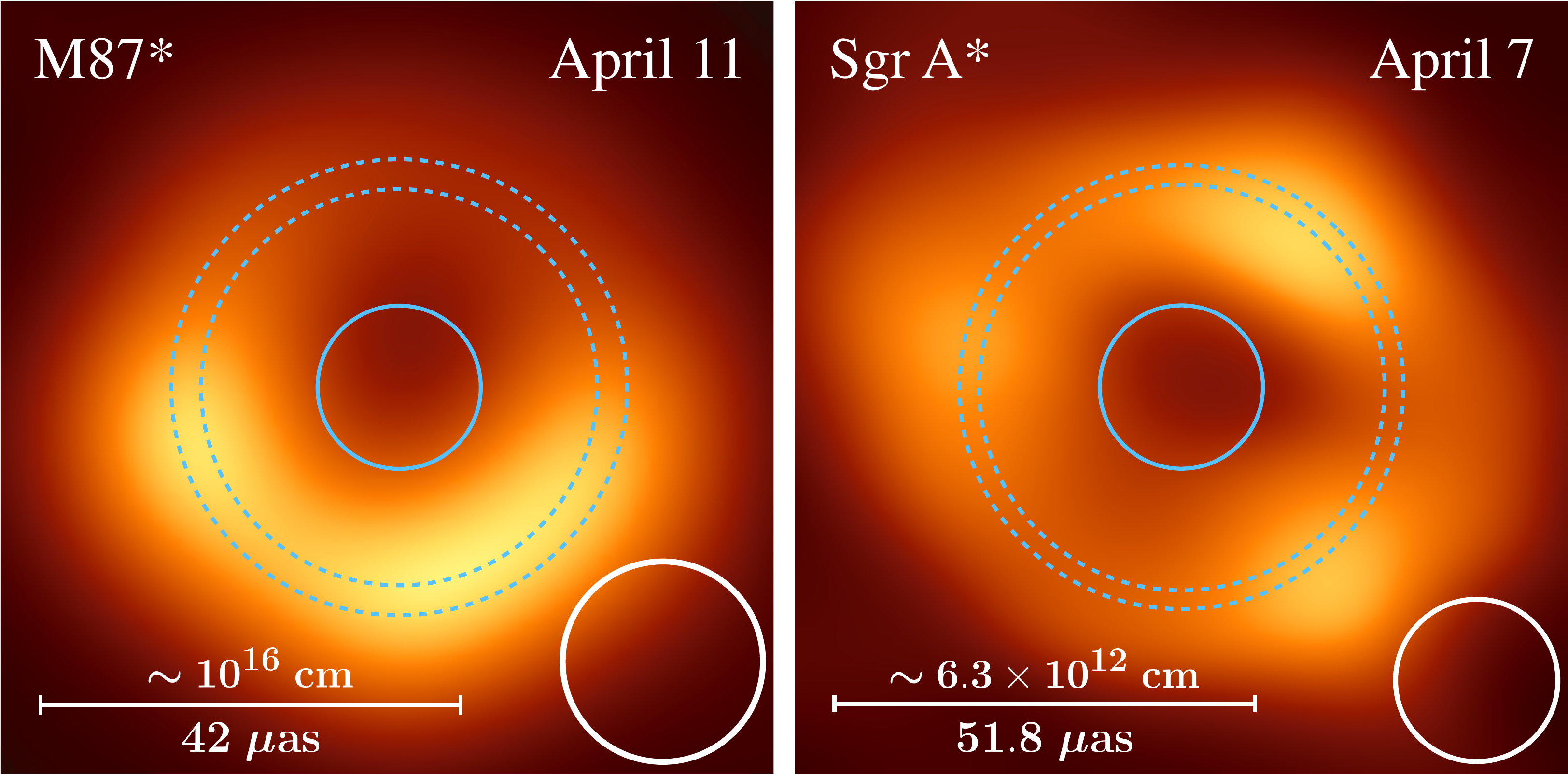}
	\caption{EHTC images of M87* (left) and \sgra~(right).
The central solid blue circles show the largest possible diameter (in GR) of each BH's event horizon, i.e., the Schwarzschild value of $4~r_{\rm g}$, where $r_{\rm g}$ is the BH gravitational radius (see Sec.~\ref{sec:Light_ring}).
The size of the event horizon in the observed image would appear slightly larger than the solid blue circles, due to gravitational lensing.
Note that the event horizons fit within the central dark regions of both images (the central brightness depression).
Pairs of dashed blue circles delineate the estimated diameter range of the bright ring from image domain analysis of M87* ($42~\pm~3 ~\mu$as) and \sgra~($51.8~\pm~2.3 ~\mu$as).
These ranges are consistent with the prediction of the Schwarzschild BH shadow diameter ($2\sqrt{27}~r_{\rm g}$).
The white circles in the lower right of both panels show the $20~\mu$as FWHM circular Gaussian beam (EHT 2017).
See \cite{EventHorizonTelescope:2019dse} and \cite{EventHorizonTelescope:2022xnr} for further information.
Figure reproduced from \cite{Younsi_review_in_prep}.
}
\label{fig:M87_SgrA_SbyS}
\end{figure}

The achievement of hitherto unprecedented resolving power in traditional observational astronomy using optical, infrared, and radio VLBI over the past few decades has led to significant progress in the study of BH systems.
Of particular importance in the context of this review are the multidecadal observations of the orbital motions of S-stars  in the Galactic Centre (GC) by the Keck telescope \citep{do2019} and the VLTI \citep{GRAVITY:2021xju}.
These observations constrained the central mass of the GC to be $\sim 4 \times 10^{6} M_{\odot}$ (see Table~\ref{tab:masses_sgrA}), providing compelling evidence for the existence of a supermassive compact object, presumably a SMBH.
Parallel advances in mm wavelength VLBI during these years provided a means to spatially resolve the immediate environment of the compact object in the GC at frequencies where the surrounding hot plasma becomes optically thin.
This led to the discovery of horizon-scale features of the GC SMBH Sagittarius A*, hereafter \sgra, \citep{Doeleman_2008_sgra}, and similar scale structures around the M87 SMBH, hereafter M87* \citep{Doeleman_2012_m87}.
Subsequent growth of mm-VLBI arrays led to the Event Horizon Telescope (EHT) Collaboration (EHTC) producing the first ever images of these two SMBHs \citep{EventHorizonTelescope:2019dse,EventHorizonTelescope:2022xnr}.
These results present a bright ring feature in both SMBH images, demonstrating the persistence of this feature across a scale of more than three orders of magnitude in mass (see Figure \ref{fig:M87_SgrA_SbyS}), as anticipated from the scale-invariance of general relativity (GR).
Both images also demonstrate manifestly similar image morphologies, with pronounced central brightness depressions and ring diameters consistent with the predictions of GR.
These recent breakthroughs are complementary to GW observations, probing spacetime geometries which can be understood as ``static'' (i.e., effectively stationary) by detecting radiation produced by matter in the vicinity of the BH's purported event horizon.

In this review we assess the potential of a next-generation Event Horizon Telescope (ngEHT) to extract information on foundational issues related to BHs and near-horizon physics, presenting an overview of the most promising future prospects for studies of BHs and strong field gravity in the next decade.
The ngEHT program is expected to transform our understanding of SMBH sources via substantial improvements in angular resolution, image dynamic range, multi-wavelength capabilities, and long-term monitoring, as well as rendering several new SMBH systems accessible to event horizon-scale study.
The structure of the persistent ring feature in EHT images of \sgra and M87* is currently not sufficiently resolved to unambiguously confirm the presence of a ``photon ring''\footnote{In the absence of attenuating material media, gravitational field theory predicts a formally infinite hierarchy of successively thinner and fainter photon rings: see Sec.~\ref{sec:Light_ring}.}. 
The ngEHT will use measurements of the emission near SMBHs to probe their spacetime geometry, measuring the BH mass and spin, as well as enabling tests of the no-hair theorem of astrophysical BHs.
%
\subsection{ngEHT: array architecture and vision} \label{sec:intro_bg}

\begin{figure}
	\centering
	\includegraphics[width=1.0\textwidth]{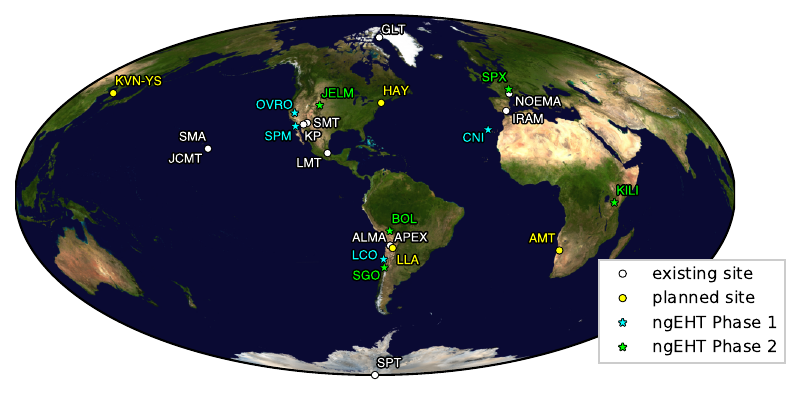}
	\caption{Vision for the ngEHT array. Current EHT sites are shown in white, candidate ngEHT Phase 1 sites are blue, and candidate ngEHT Phase~2 sites are green. In addition, yellow markers show four additional sites that are planned to come online over the next five years: the 37~m Haystack Telescope \citep[HAY;][]{Kauffmann_2023}, the 15~m Africa Millimetre Telescope \citep[AMT;][]{Backes_2016}, the Large Latin American Millimeter Array \citep[LLA;][]{Romero_2020}, and the Yonsei Radio Observatory of the Korea VLBI Network \citep[KVN-YS;][]{Asada_2017}. For additional details on the ngEHT array, see \citetalias{ngEHT_refarray}. Figure reproduced from \citetalias{ngEHT_KSG}.
}
\label{fig:globe}
\end{figure}

\begin{figure}
	\centering
	\includegraphics[width=1.0\textwidth]{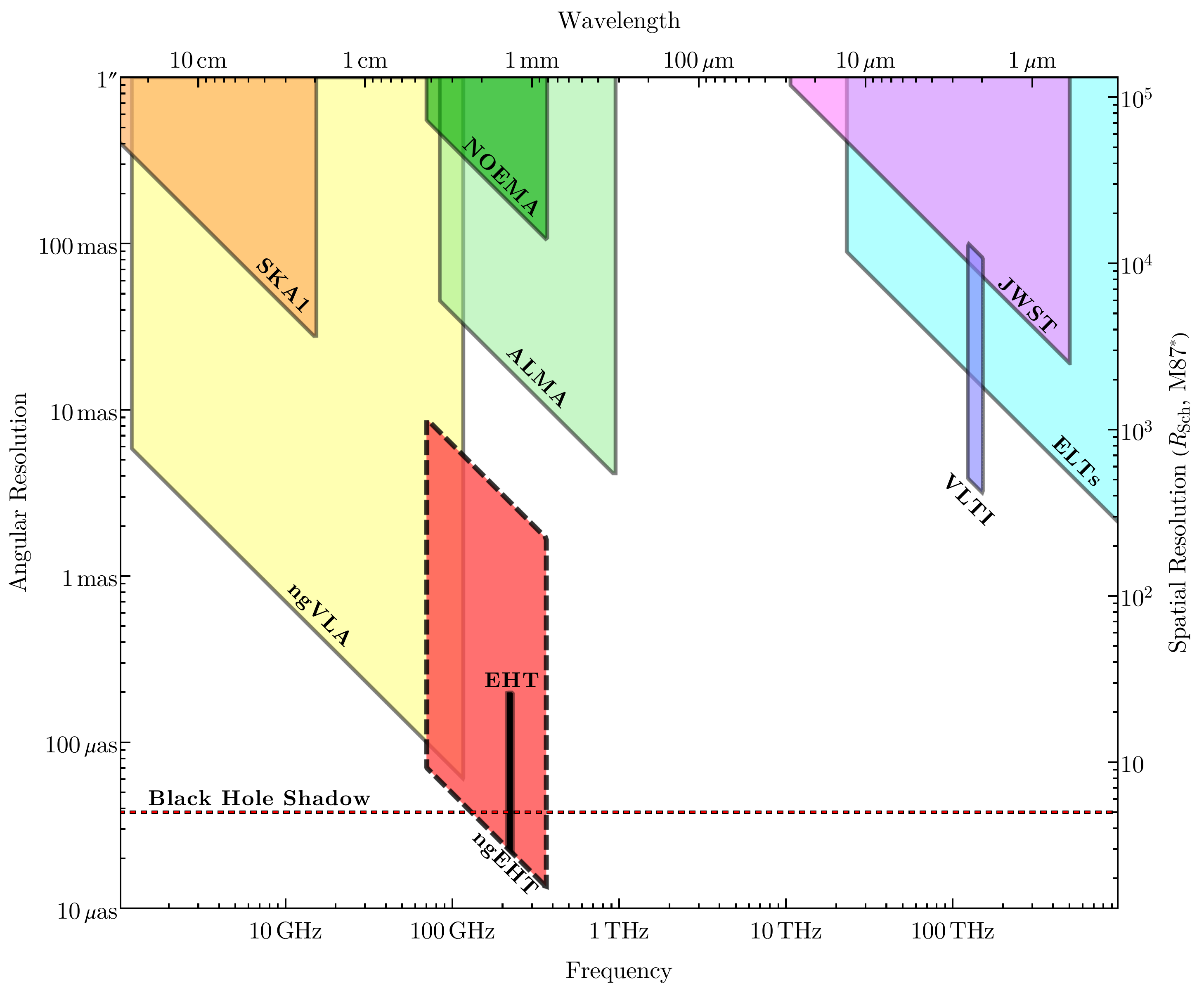}
	\caption{Observing frequencies and imaging angular resolutions for current and next generation facilities. The EHT and ngEHT have significantly finer imaging resolution than any other telescope. The ngEHT will significantly expand the frequency coverage of the EHT and will provide access to larger angular scales. Figure reproduced from \citetalias{ngEHT_KSG}.
}
\label{fig:resolution}
\end{figure}

The current EHT results were all achieved using an observing campaign in 2017 that included 8 telescopes at 6 geographic sites. In an interferometric array such as the EHT, each baseline joining a pair of telescopes samples a single Fourier component of the sky image. EHT observations of \sgra~had 15 intersite baselines, while EHT observations of \m87 had 10 intersite baselines (the South Pole Telescope cannot observe \m87). The EHT observations recorded a single 4\,GHz band, centered on 228\,GHz. Since the initial EHT observing campaign, the array has been expanded to include 3 additional sites, has doubled the recorded bandwidth, and and has recently added the ability to observe at 345\,GHz \citep{Crew_2023}.

The ngEHT is a project that is designing an array that would substantially enhance the observational capabilities of the EHT \citep{Doeleman2019Studying}. 
Over two phases of deployment, the ngEHT will add up to ${\sim}10$ additional sites worldwide by ${\sim}2030$ (see \autoref{fig:globe}). The ngEHT will also include three simultaneous observing bands at 86, 230, and 345\,GHz. Apart from providing spectral information, this configuration will allow substantially improved phase coherence at high frequencies because the dominant sources of ngEHT phase errors are non-dispersive \citep[see, e.g.,][]{Issaoun_2022_86GHz,Rioja_2023}. 
Together, these improvements will augment the angular resolution of current EHT images by approximately 50\% and will increase the angular scales that the array can image by an order of magnitude (see \autoref{fig:resolution}). They will also increase the dynamic range of BH images by 1--2 orders of magnitude and will ultimately support dense observations year-round, significantly improving the current temporal coverage of EHT observations (${\sim}1\,{\rm week}$ per year). 
For a summary of the ngEHT array, see \citet[][hereafter \citetalias{ngEHT_refarray}]{ngEHT_refarray}. For a summary of the complete set of ngEHT science goals, see \citet[][hereafter \citetalias{ngEHT_KSG}]{ngEHT_KSG}.

\subsection{Outline and notation} \label{sec:intro_outline}
This review article is organised as follows. 
In Sec.~\ref{sec:Light_ring} we review the status and prospects of ngEHT studies of the photon ring to provide evidence of strong field gravitational lensing by SMBHs and yield novel tests of GR.
Next, in Sec.~\ref{sec:Mass_spin} we present an overview of ngEHT observables for measuring SMBH mass and spin, followed by an overview of other studies which can complement those the ngEHT will perform.
In Sec.~\ref{sec:Ultralight_fields} we discuss the capability of the ngEHT to search for ultralight bosonic fields below the eV scale, together with the implications of these studies for the structure of the central compact object and its surrounding plasma.
Section \ref{sec:Tests_GR_Kerr} presents an overview of tests of GR and the Kerr hypothesis with the ngEHT. Different physical scenarios in beyond-GR theories are classified, with their implications for testable violations of the Kerr hypothesis discussed, together with a summary of key science cases for testing features of compact objects in and beyond GR.
Finally, in Sec.~\ref{sec:introbinaries} we discuss the prospects of studies of SMBH binaries with the ngEHT. Candidate sources are identified and the challenges and prospects of probing SMBH binaries with the ngEHT are explored.

In this review we venture to provide a framework for the exploration and development of these exciting new prospects.
Studies such as those discussed in this review will consolidate our understanding of the different ways ngEHT-driven VLBI observations can probe the properties of SMBHs and enable new tests of fundamental physics.

The acronyms used in this review are summarised in Table \ref{table:acronyms}.
We adopt the geometrical unit convention in which $G=c=1$, and unless otherwise stated assume the metric signature to be $[-,+,+,+]$.
When specifying vectors and tensors, Greek indices (e.g., $\mu$, $\nu$) span $(0,\, 1,\, 2,\, 3)$ and Latin indices (e.g., $i$, $j$) span $(1,\, 2,\, 3)$, where $0$ denotes the temporal component and $(1,\, 2,\, 3)$ denote spatial components.

\subsubsection{List of acronyms used in this review}
%
\begin{longtable}[h]{ll}
\label{table:acronyms}
 AGN & Active Galactic Nucleus \\
 Athena & Advanced Telescope for High ENergy Astrophysics \\
 BAL & Broad absorption line \\
 BH & Black hole \\
 BBH & Binary black hole \\
 CFT & Conformal field theory \\
 CMB & Cosmic microwave background \\
 CCTP & Conjugate closure trace product \\
 CTA & Cherenkov Telescope Array \\
 DM & Dark matter \\
 DAGN & Dual AGN \\
 ECO & Exotic compact object \\
 EdGB & Einstein dilaton Gauss Bonnet \\
 EdM & Einstein dilaton Maxwell \\
 EFEs & Einstein field equations \\ 
 EFT & Effective field theory \\
 EHT & Event Horizon Telescope \\
 EHTC & Event Horizon Telescope Collaboration \\
 EMRI & Extreme mass-ratio inspiral \\ 
 EVPA & Electric vector position angle \\
 FFE & Force-Free Electrodynamics \\
 FWHM & Full width half maximum \\
 GC & Galactic Center \\
 GR & General relativity \\ 
 GRMHD & General-relativistic magnetohydrodynamics \\
 GRRT & General-relativistic radiative transfer \\
 GW & Gravitational wave \\
 GWB & Gravitational wave background \\
 HD & Hellings \& Downs \\
 IMBH & Intermediate-mass black hole \\
 IR & Infrared \\
 ISCO & Innermost stable circular orbit \\
 JWST & James Webb Space Telescope \\
 LIGO & Laser Interferometer Gravitational-Wave Observatory \\
 LISA & Laser Interferometer Space Antenna \\
 LLAGN & Low luminosity AGN \\
 LT & Lense-Thirring \\
 M87 & Messier 87 \\
 MAD & Magnetically arrested disk \\
 MCMC & Markov Chain Monte Carlo \\
 MRI & Magnetorotational instability \\
 ngEHT & Next-Generation Event Horizon Telescope \\
 NGC & New General Catalogue \\
 ngVLA & Next-Generation Very Large Array \\
 NHEK & Near horizon extremal Kerr \\
 NIR & Near Infrared \\
 PN & Post-Newtonian \\
 PTA & Pulsar Timing Array \\
 QPO & Quasi-periodic oscillation \\
 RIAF & Radiatively inefficient accretion flow \\
 SANE & Standard and normal evolution \\
 SCO & Stellar compact object \\
 SDSS & Sloan Digital Sky Survey \\
 SEP & Strong equivalence principle \\ 
 SFPR & Source Frequency Phase Referencing \\
 Sgr A* & Sagittarius A* \\
 SKA & Square kilometer array \\
 SMBH & Supermassive black hole \\
 SMBHB & Supermassive black hole binary\\
 SNR & Signal-to-noise ratio \\ 
 UHE & Very high energy \\ 
 VHE & Very high energy \\
 VLA & Very Large Array \\
 VLBI & Very-Long-Baseline Interferometry \\
 VLBA & Very Long Baseline Array \\
 VLTI & Very Large Telescope Interferometer \\
 SWG & Science working group \\
\end{longtable}

\subsection{Author contributions}
The ngEHT Fundamental Physics Science Working Group (SWG) was coordinated by {\bf Vitor Cardoso and Ziri Younsi}.
The topic ``Studies of the photon Ring'' was led by {\bf Shahar Hadar and Daniel Palumbo}.
The topic ``Measuring black hole mass and spin'' was led by {\bf Dimitry Ayzenberg, Lijing Shao and Huan Yang}.
The topic ``Searching for ultralight fields with the ngEHT'' was led by {\bf Richard Brito and Yifan Chen}.
The topic ``Tests of GR and the Kerr hypothesis with the ngEHT'' was led by {\bf Astrid Eichhorn and Aaron Held}.
The topic ``Exploring binary black holes with the ngEHT'' was led by {\bf Silke Britzen and Roman Gold}.
All authors contributed to the writing of the document.

\newpage
\input{Light_ring.tex}


\newpage
\input{Mass_spin.tex}

\newpage
\input{Ultralight_fields}


\newpage
\input{Tests_GR_Kerr.tex}


\newpage
\input{Binaries.tex}


\newpage

\phantomsection
\addcontentsline{toc}{section}{\protect\numberline{}Conclusions}
\section*{Conclusions} \label{sec:Conclusions}
The past four years have been a short yet exciting time for BH physicists, first with the EHT measurement of the M87* SMBH shadow in 2019, and shortly thereafter with the measurement of Sgr A*'s shadow in 2022.
These pioneering measurements heralded the start of a new era in BH research and have enabled, for the first time, direct imaging of matter in the vicinity of event horizons.
In the years ahead, improvements will be made in the instrument specifications of telescopes within the interferometric array, which will in turn open several new avenues for exploring physical phenomena around BHs.
This review provides an overview of the major fundamental physics themes and current status of BH imaging observations which will guide the development of a future ngEHT array and enable significantly more sensitive studies of SMBHs.

In the context of these future improvements we have summarised the current status of several key science topics underpinning BH imaging studies: studies of the photon ring, measurement of BH mass and spin, searches for ultralight fields, tests of GR and Kerr, and potential studies of binary SMBHs.
In the years to come, advances will be made in both the sensitivity of measurements, and in the data analysis techniques which are applied to these data, which will further advance our understanding of BHs and their environments.
This review presents an overview of the exciting scientific potential of future BH imaging studies, and is intended to be useful as a reference for researchers interested in utilising the ngEHT as a distinct new tool for probing BHs and studying fundamental physics.

\phantomsection
\addcontentsline{toc}{section}{\protect\numberline{}Acknowledgments}
\section*{Acknowledgments}
%
We are grateful to Kfir Blum, Sam Gralla, Paolo Pani, Odele Straub and Nico Yunes for useful discussions and important feedback.
We express our thanks to the EHT Collaboration's Publication Committee and the three anonymous internal referees.
%

\newpage

\phantomsection
\addcontentsline{toc}{section}{\protect\numberline{}References}
\bibliographystyle{spbasic-FS.bst}
\bibliography{References}

\newpage

\phantomsection
\addcontentsline{toc}{section}{\protect\numberline{}Author affiliations}
\section*{Author affiliations}
{\it D.~Ayzenberg}
\begin{cwid}Theoretical Astrophysics, Eberhard-Karls Universit\"at T\"ubingen, D-72076, T\"ubingen, Germany. \\
ORCID: 0000-0003-0238-6181
\end{cwid}
\vspace*{0.5em}{\it Lindy~Blackburn}
\begin{cwid}Center for Astrophysics $\vert$ Harvard \& Smithsonian, 60 Garden Street, Cambridge, MA 02138, USA,\\
Black Hole Initiative at Harvard University, 20 Garden Street, Cambridge, MA 02138, USA. \\
ORCID: 0000-0002-9030-642X
\end{cwid}
\vspace*{0.5em}{\it R.~Brito}
\begin{cwid}CENTRA, Departamento de F\'{\i}sica, Instituto Superior T\'ecnico -- IST, Universidade de Lisboa -- UL, Avenida Rovisco Pais 1, 1049-001 Lisboa, Portugal. \\
ORCID: 0000-0002-7807-3053
\end{cwid}
\vspace*{0.5em}{\it S.~Britzen}
\begin{cwid}Max-Planck-Institut f\"ur Radioastronomie, Auf dem H\"ugel 69, D-53121 Bonn, Germany. \\
ORCID: 0000-0001-9240-6734
\end{cwid}
\vspace*{0.5em}{\it A.~Broderick}
\begin{cwid}Perimeter Institute for Theoretical Physics, 31 Caroline Street North, Waterloo, ON, N2L 2Y5, Canada,\\
Department of Physics and Astronomy, University of Waterloo, 200 University Avenue West, Waterloo, ON, N2L 3G1, Canada,\\
Waterloo Centre for Astrophysics, University of Waterloo, Waterloo, ON N2L 3G1 Canada. \\
ORCID: 0000-0002-3351-760X
\end{cwid}
\vspace*{0.5em}{\it R.~Carballo-Rubio} 
\begin{cwid}CP3-Origins, University of Southern Denmark, Campusvej 55, DK-5230 Odense M, Denmark,\\
Florida Space Institute, University of Central Florida, 12354 Research Parkway, Partnership 1, 32826 Orlando, FL, USA. \\
ORCID: 0000-0001-6389-6105
\end{cwid}
\vspace*{0.5em}{\it V.~Cardoso}
\begin{cwid}Niels Bohr International Academy, Niels Bohr Institute, Blegdamsvej 17, 2100 Copenhagen, Denmark,\\
CENTRA, Departamento de F\'{\i}sica, Instituto Superior T\'ecnico -- IST, Universidade de Lisboa -- UL, Avenida Rovisco Pais 1, 1049-001 Lisboa, Portugal. \\
ORCID: 0000-0003-0553-0433
\end{cwid}
\vspace*{0.5em}{\it A.~Chael}
\begin{cwid}Princeton Gravity Initiative, Jadwin Hall, Princeton University, Princeton NJ 08544, USA.\\
ORCID: 0000-0003-2966-6220
\end{cwid}
\vspace*{0.5em}{\it K.~Chatterjee}
\begin{cwid}Department of Physics, University of Maryland, College Park, MD 20742, USA,\\
Institute for Research in Electronics and Applied Physics, University of Maryland, College Park, MD 20742, USA,\\
Black Hole Initiative at Harvard University, 20 Garden Street, Cambridge, MA 02138, USA,\\
Harvard-Smithsonian Center for Astrophysics, 60 Garden Street, Cambridge, MA 02138, USA. \\
ORCID: 0000-0002-2825-3590
\end{cwid}
\vspace*{0.5em}{\it Y.~Chen}
\begin{cwid}Niels Bohr International Academy, Niels Bohr Institute, Blegdamsvej 17, 2100 Copenhagen, Denmark.\\
ORCID: 0000-0002-2507-8272
\end{cwid}
\vspace*{0.5em}{\it P.~V.~P.~Cunha}
\begin{cwid}Departamento de Matem\'{a}tica da Universidade de Aveiro and Centre for Research and Development in Mathematics and Applications (CIDMA), Campus de Santiago, 3810-183 Aveiro, Portugal. \\
ORCID: 0000-0001-8375-6943
\end{cwid}
\vspace*{0.5em}{\it H.~Davoudiasl}
\begin{cwid}High Energy Theory Group, Physics Department,
Brookhaven National Laboratory, Upton, NY 11973, USA.\\
ORCID: 0000-0003-3484-911X
\end{cwid}
\vspace*{0.5em}{\it P.~B.~Denton}
\begin{cwid}High Energy Theory Group, Physics Department,
Brookhaven National Laboratory, Upton, NY 11973, USA. \\
ORCID: 0000-0002-5209-872X
\end{cwid}
\vspace*{0.5em}{\it S.~S.~Doeleman}
\begin{cwid}Center for Astrophysics $\vert$ Harvard \& Smithsonian, 60 Garden Street, Cambridge, MA 02138, USA,\\
Black Hole Initiative at Harvard University, 20 Garden Street, Cambridge, MA 02138, USA.\\
ORCID: 0000-0002-9031-0904
\end{cwid}
\vspace*{0.5em}{\it A.~Eichhorn}
\begin{cwid}CP3-Origins, University of Southern Denmark, Campusvej 55, DK-5230 Odense M, Denmark.\\
ORCID: 0000-0003-4458-1495
\end{cwid}
\vspace*{0.5em}{\it M.~Eubanks}
\begin{cwid}Asteroid Initiatives, LLC, USA.\\
ORCID: 0000-0001-9543-0414
\end{cwid}
\vspace*{0.5em}{\it Y.~Fang}
\begin{cwid}Perimeter Institute for Theoretical Physics, Waterloo, ON N2L2Y5, Canada,\\
Kavli Institute for Astronomy and Astrophysics, Peking University, Beijing 100871, China. \\
ORCID: 0000-0003-0065-8622
\end{cwid}
\vspace*{0.5em}{\it A.~Foschi}
\begin{cwid}CENTRA, Departamento de F\'{\i}sica, Instituto Superior T\'ecnico -- IST, Universidade de Lisboa -- UL, Avenida Rovisco Pais 1, 1049-001 Lisboa, Portugal. \\
ORCID: 0000-0002-4636-637X
\end{cwid}
\vspace*{0.5em}{\it C.~M.~Fromm}
\begin{cwid}Institut für Theoretische Physik und Astrophysik, Universität Würzburg, Emil-Fischer-Strasse 31, 97074 Würzburg, Germany,\\
Institut für Theoretische Physik, Goethe Universität Frankfurt, Max-von-Laue-Str.1, 60438 Frankfurt am Main, Germany,\\
Max-Planck-Institut für Radioastronomie, Auf dem Hügel 69, D-53121 Bonn, Germany.\\
ORCID: 0000-0002-1827-1656
\end{cwid}
\vspace*{0.5em}{\it P.~Galison}
\begin{cwid}Black Hole Initiative at Harvard University, 20 Garden Street, Cambridge, MA 02138, USA,\\
Department of History of Science, Harvard University, Cambridge, MA 02138, USA,\\
Department of Physics, Harvard University, Cambridge, MA 02138, USA.\\
ORCID: 0000-0002-6429-3872
\end{cwid}
\vspace*{0.5em}{\it S.~G.~Ghosh}
\begin{cwid}Centre for Theoretical Physics, Jamia Millia Islamia, New Delhi 110025, India,\\
Astrophysics and Cosmology Research Unit, School of Mathematics, Statistics and Computer Science, University of KwaZulu-Natal, Private Bag 54001, Durban 4000, South Africa.\\
ORCID: 0000-0002-0835-3690
\end{cwid}
\vspace*{0.5em}{\it R.~Gold}
\begin{cwid}CP3-Origins, University of Southern Denmark, Campusvej 55, DK-5230 Odense M, Denmark. \\
ORCID: 0000-0003-2492-1966
\end{cwid}
\vspace*{0.5em}{\it L.~I.~Gurvits}
\begin{cwid}Joint Institute for VLBI ERIC (JIVE), Oude Hoogeveensedijk 4, NL-7991~PD Dwingeloo, the Netherlands,\\
Faculty of Aerospace Engineering, Delft University of Technology, Kluyverweg 1, NL-2629~HS Delft, the Netherlands \\
ORCID: 0000-0002-0694-2459
\end{cwid}
\vspace*{0.5em}{\it S.~Hadar}
\begin{cwid}Department of Mathematics and Physics, University of Haifa at Oranim, Kiryat Tivon 3600600, Israel,\\
Haifa Research Center for Theoretical Physics and Astrophysics, University of Haifa, Haifa 3498838, Israel. \\
ORCID: 0000-0002-6960-0704
\end{cwid}
\vspace*{0.5em}{\it A.~Held}
\begin{cwid}Theoretisch-Physikalisches Institut, Friedrich-Schiller-Universit\"at Jena, Max-Wien-Platz 1, 07743 Jena, Germany,\\
The Princeton Gravity Initiative, Jadwin Hall, Princeton University, Princeton, New Jersey 08544, USA. \\
ORCID: 0000-0003-2701-9361
\end{cwid}
\vspace*{0.5em}{\it J.~Houston}
\begin{cwid}Center for Astrophysics $\vert$ Harvard \& Smithsonian, 60 Garden Street, Cambridge, MA 02138, USA.\end{cwid}
\vspace*{0.5em}{\it Y.~Hu}
\begin{cwid}Mullard Space Science Laboratory, University College London, Holmbury St.~Mary, Dorking, Surrey, RH5 6NT, United Kingdom. \\
ORCID: 0000-0001-9252-0246
\end{cwid}
\vspace*{0.5em}{\it M.~D.~Johnson}
\begin{cwid}Center for Astrophysics $\vert$ Harvard \& Smithsonian, 60 Garden Street, Cambridge, MA 02138, USA,\\
Black Hole Initiative at Harvard University, 20 Garden Street, Cambridge, MA 02138, USA.\\
ORCID: 0000-0002-4120-3029
\end{cwid}
\vspace*{0.5em}{\it P.~Kocherlakota}
\begin{cwid}Black Hole Initiative at Harvard University, 20 Garden St., Cambridge, MA 02138, USA,\\
Center for Astrophysics, Harvard \& Smithsonian, 60 Garden St., Cambridge, MA 02138, USA. \\
ORCID: 0000-0001-7386-7439
\end{cwid}
\vspace*{0.5em}{\it P.~Natarajan}
\begin{cwid}Department of Astronomy, Yale University, 52 Hillhouse Avenue, New Haven, CT 06511, USA,\\
Department of Physics, Yale University, P.O. Box 208121, New Haven, CT 06520, USA,\\
Black Hole Initiative at Harvard University, 20 Garden Street, Cambridge, MA 02138, USA. \\
ORCID: 0000-0002-5554-8896
\end{cwid}
\vspace*{0.5em}{\it H.~Olivares}
\begin{cwid}Department of Astrophysics/IMAPP, Radboud University Nijmegen, P.O. Box 9010, NL-6500 GL Nijmegen, The Netherlands. \\
ORCID: 0000-0001-6833-7580
\end{cwid}
\vspace*{0.5em}{\it D.~Palumbo}
\begin{cwid}Center for Astrophysics $\vert$ Harvard \& Smithsonian, 60 Garden Street, Cambridge, MA 02138, USA,\\
Black Hole Initiative at Harvard University, 20 Garden Street, Cambridge, MA 02138, USA.\\
ORCID: 0000-0002-7179-3816
\end{cwid}
\vspace*{0.5em}{\it D.~W.~Pesce}
\begin{cwid}Center for Astrophysics $\vert$ Harvard \& Smithsonian, 60 Garden Street, Cambridge, MA 02138, USA,\\
Black Hole Initiative at Harvard University, 20 Garden Street, Cambridge, MA 02138, USA.\\
ORCID: 0000-0002-5278-9221
\end{cwid}
\vspace*{0.5em}{\it S.~Rajendran}
\begin{cwid}Department of Physics and Astronomy, Johns Hopkins University, Baltimore, MD 21218, USA. \\
ORCID: 0000-0001-9915-3573
\end{cwid}
\vspace*{0.5em}{\it R.~Roy}
\begin{cwid}Anton Pannekoek Institute for Astronomy, University of Amsterdam, Science Park 904, 1098 XH, Amsterdam, The Netherlands. \\
ORCID: 0000-0003-3714-5310
\end{cwid}
\vspace*{0.5em}{\it Saurabh}
\begin{cwid}P. D. Patel Institute of Applied Sciences, Charusat University, Anand, 388421 Gujarat, India. \\
ORCID: 0000-0001-7156-4848
\end{cwid}
\vspace*{0.5em}{\it L.~Shao}
\begin{cwid}Kavli Institute for Astronomy and Astrophysics, Peking University, Beijing 100871, China,\\
National Astronomical Observatories, Chinese Academy of Sciences, Beijing 100012, China.  \\
ORCID: 0000-0002-1334-8853
\end{cwid}
\vspace*{0.5em}{\it S.~Tahura}
\begin{cwid}University of Guelph, Guelph, Ontario N1G 2W1, Canada,\\
Perimeter Institute for Theoretical Physics, Ontario, N2L 2Y5, Canada. \\
ORCID: 0000-0001-5678-5028
\end{cwid}
\vspace*{0.5em}{\it A.~Tamar}
\begin{cwid}124 Bank Enclave, First Floor, Laxmi Nagar, Delhi-110092, India. \\
ORCID: 0000-0001-8763-4169
\end{cwid}
\vspace*{0.5em}{\it P.~Tiede}
\begin{cwid}Center for Astrophysics $\vert$ Harvard \& Smithsonian, 60 Garden Street, Cambridge, MA 02138, USA,\\
Black Hole Initiative at Harvard University, 20 Garden Street, Cambridge, MA 02138, USA. \\
ORCID: 0000-0003-3826-5648
\end{cwid}
\vspace*{0.5em}{\it F.~H.~Vincent}
\begin{cwid} LESIA, Observatoire de Paris, CNRS, Universit\'e Pierre et Marie Curie, Universit\'e
Paris Diderot, 5 place Jules Janssen, 92190 Meudon, France.\\
ORCID: 0000-0002-3855-0708
\end{cwid}
\vspace*{0.5em}{\it L.~Visinelli}
\begin{cwid}Tsung-Dao Lee Institute (TDLI), 520 Shengrong Road, 201210 Shanghai, P.~R.~China,\\
School of Physics and Astronomy, Shanghai Jiao Tong University, 800 Dongchuan Road, 200240 Shanghai, P.\ R.\ China. \\
ORCID: 0000-0001-7958-8940
\end{cwid}
\vspace*{0.5em}{\it Z.~Wang}
\begin{cwid}Department of Physics and Astronomy, University of Waterloo, 200 University Avenue West, Waterloo, ON, N2L 3G1, Canada,\\
Waterloo Centre for Astrophysics, University of Waterloo, Waterloo, ON N2L 3G1, Canada. \\
ORCID: 0009-0004-9417-2213
\end{cwid}
\vspace*{0.5em}{\it M.~Wielgus}
\begin{cwid}Department of Physics and Astronomy, University of 2 Max-Planck-Institut f\"ur Radioastronomie, Auf dem H\"ugel 69, 53121 Bonn, Germany,\\
Research Centre for Computational Physics and Data Processing, Institute of Physics, Silesian University in Opava, Bezru\v{c}ovon\'am.~13, CZ-746 01 Opava, Czech Republic.\\
ORCID: 0000-0002-8635-4242
\end{cwid}
\vspace*{0.5em}{\it X.~Xue}
\begin{cwid}Institute of Theoretical Physics, Universit\"{a}t  Hamburg, 22761, Hamburg, Germany,\\
Deutsches Elektronen-Synchrotron DESY, Notkestr. 85, 22607, Hamburg, Germany.\\
ORCID: 0000-0002-0740-1283
\end{cwid}
\vspace*{0.5em}{\it K.~Yakut}
\begin{cwid}Department of Astronomy and Space Sciences, Faculty of Science, University of Ege, 35100, \.Izmir, Turkey,\\
Ege Gravitational Astrophysics Research Group (eGRAVITY), University of Ege, 35100, \.Izmir, Turkey. \\
ORCID: 0000-0003-2380-9008
\end{cwid}
\vspace*{0.5em}{\it H.~Yang}
\begin{cwid}University of Guelph, Department of Physics, Guelph, ON N1G2W1, Canada,\\
Perimeter Institute for Theoretical Physics, Waterloo, ON N2L2Y5, Canada. \\
ORCID: 0000-0002-9965-3030
\end{cwid}
\vspace*{0.5em}{\it Z.~Younsi}
\begin{cwid}Mullard Space Science Laboratory, University College London, Holmbury St.~Mary, Dorking, Surrey, RH5 6NT, United Kingdom. \\
ORCID: 0000-0001-9283-1191
\end{cwid}

\end{document}

%% file: Light_ring.tex
\section{Studies of the photon ring} \label{sec:Light_ring}

There are three classical tests of GR: the precession of the planet Mercury, the deflection of light by the Sun and the gravitational redshift of light.
The first observation of light deflection by the Sun was measured by Arthur Eddington and his team during the 1919 eclipse \citep{Crispino:2019yew,Will:2014kxa}.
They recorded a deflection angle of $\delta \sim 1.7$ arcseconds.
This is consistent with the weak field prediction of GR where, for a small dimensionless compactness $r_{\rm g}/r_0$, the angle is given by $\delta \approx 4\,r_{\rm g}/r_0$.
Here $r_0$ is the perihelion of the light ray's trajectory and $r_{\rm g} := GM/c^{2}$ is the gravitational radius of the deflector, wherein $M$ is the mass of the Sun, $G$ is Newton's gravitational constant and $c$ is the speed of light.

In the strong field regime, near extremely compact objects such as BHs, the very same principles of GR predict that the deflection angle should become unbounded at the so-called \textit{photon shell}: the spacetime region, close to the BH event horizon, where light rays may orbit indefinitely at fixed (Boyer-Lindquist) radius \citep[see][for a review and historical account]{Perlick:2004}.
The Schwarzschild radius of a BH is defined as $r_{\rm S}:= 2GM/c^{2}\equiv2\,r_{\rm g}$ and corresponds to the radius of the event horizon of a Schwarzschild BH.
For the Schwarzschild metric, the photon shell is located at $3\,r_{\rm g}$, i.e., at a coordinate radius of only $0.5\,r_{\rm S}$ away from the event horizon.

The light rays of the photon shell are unstable; when they are slightly perturbed, time evolution drives them away from the shell, and eventually they either reach asymptotic infinity or fall into the BH. 
Nevertheless, in this process they experience extreme lensing, with order-unity deflection angles (in radians), and carry an imprint of the spacetime geometry at the photon shell region.
The extreme gravitational lensing close to the photon shell is accompanied by an extreme ``Shapiro-like'' delay, directly responsible for the late-time appearance of collapsing spacetimes or of transient electromagnetic phenomena in the BH vicinity \citep{1965SvA.....8..868P, 1968ApJ...151..659A, Cardoso:2021sip, Ferrari:1984zz, Cardoso:2008bp}. 

The detection of extreme gravitational lensing may therefore provide new ways to observe gravitational phenomena in the strong-field regime.
The goal of this Chapter is to discuss how EHT and ngEHT observations can be used to provide evidence and quantification of the strong deflection of light by BHs.

\subsection{The photon shell and ring}\label{sec:background}
Geodesic motion in the Kerr spacetime has been studied since the pioneering work of \cite{Carter_1968} and numerous papers thereafter \citep[see, e.g.,][]{Walker_Penrose_1970, MTB, Hackmann:2010zz}. In this section we discuss the unstably bound null geodesics of Kerr, which are of central importance for the interpretation of BH images. These are null-geodesic orbits with fixed Boyer-Lindquist radial coordinate \citep{BoyerLindquist1967}.
They are pivotal for astrophysical observations of BHs since they define the universal features in the observational appearance of the BH, as will be described below.

In what follows we adopt Boyer-Lindquist coordinates $x^\mu=(t,r,\theta,\phi)$ and denote the photon 4-momentum by $p_\mu$.
We denote the BH mass by $M$ and its angular momentum by $J$.
The BH spin parameter is defined as $a:=J/M$, and the dimensionless spin parameter is defined as $a_{*}:=J/M^{2}\equiv a/M$.
Null geodesics (photons) in the Kerr spacetime are specified by two constants of motion: the energy-scaled angular momentum component parallel to the axis of symmetry $\lambda=L/E$, where $L \equiv p_{\phi}$ and $E \equiv -p_t$, and the energy-scaled Carter constant $\eta = Q/E^2$, where $Q = p_{\theta}^{2} + \left( p_{\phi}^{2}\csc^{2}\theta - a^{2} p_{t}^{2} \right) \cos^{2}\theta$.

The energy itself only determines the frequency of the photon moving along the geodesic and not its trajectory.
The existence of a quadratic Killing tensor and the resulting separability of the Hamilton-Jacobi equation for the Kerr spacetime give rise to the Carter separation constant, $Q$, which characterizes the polar motion.
Together with the null condition $p_{\mu}p^{\mu}=0$, the geodesic motion may be reduced to a problem of quadratures, i.e., expressed as a system of four coupled first-order ordinary differential equations for the geodesic path.

We now focus on a specific, special subset of the Kerr BH's null geodesics which orbit at a fixed Boyer-Linquist radius $r=\tilde{r}$. These orbits play a central role in this section and they are interchangeably referred to as spherical/bound/critical photon orbits.
As mentioned above, the radial equation is cast in the form of one-dimensional radial motion in an effective potential, $V_{r}(r)$. 
Solving $V_{r}(\tilde{r})=V'_{r}(\tilde{r})=0$ determines a one-parameter family of critical parameters $\lambda(\tilde{r})$ and $\eta(\tilde{r})$ for which spherical photon orbits exist. These bound photon orbits comprise the \textit{photon shell}, and are labeled by their radius $\tilde{r}$ \citep{Darwin1959, Bardeen1972, Luminet_1979,Teo2003, Gralla:2019xty, Johnson:2019ljv}. The fact that $V''_{r}(\tilde{r})<0$ for all these orbits shows that they are \textit{unstable}. The impact parameters $\lambda$ and $\eta$ may be thought of as coordinates on an observer's screen, and the above-described set $\{\lambda(\tilde{r}),\eta(\tilde{r})\}$ defines the critical curve, with $\tilde{r}$ a parameter along it. 

The spherical photon orbits, which constitute the photon shell, exist in the range $\tilde{r}_-<\tilde{r}<\tilde{r}_+$, where the outermost (retrograde, $+$) and innermost (prograde, $-$) equatorial circular photon orbits are located at the radii:
\begin{equation}
\tilde{r}_{\pm} = 2M\left\{ 1 + \cos\left[ \frac{2}{3}\mathrm{arccos}\left(\pm |a_*| \right) \right] \right\} \,.
\end{equation}
For $a_*=0$, these two radii coincide and there is a unique spherical photon orbit radius which defines the so-called Schwarzschild photon sphere. Note that in the Schwarzschild geometry geodesics are planar, as a result of spherical symmetry. 

Near-critical null geodesics are governed by the properties of the photon shell. Essentially, they are controlled by a triplet of critical parameters: the Lyapunov exponent $\gamma(\tilde{r})$ \citep{Johnson:2019ljv}, describing the instability rate of the orbits, and the temporal and azimuthal periods, $\tau(\tilde{r})$, $\delta(\tilde{r})$, respectively \citep{Teo2003,Gralla:2019drh,Gralla:2019ceu}. 

\begin{figure}
	\centering
	\includegraphics[width=0.99\textwidth]{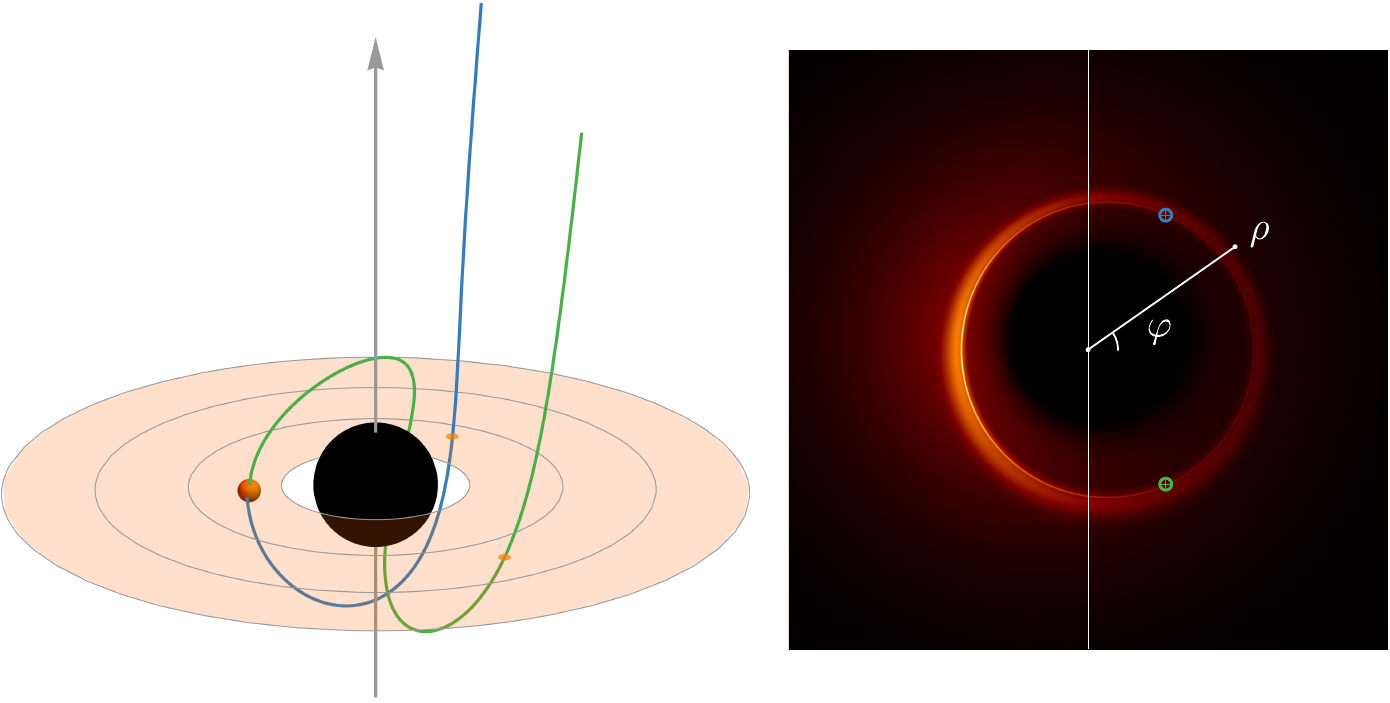}
	\caption{Left: near-critical null geodesics emanating from a flare (orange sphere) in an optically thin equatorial emission disk around a Kerr BH with $a_*=0.94$. The blue light ray has half-orbit number $n=1$, while the green ray has $n=2$. Right: image of the disk as would be seen by an infinite-resolution distant observer at an inclination of $17^\circ$. Strongly lensed light rays, which undergo multiple half-orbits, appear on the observer screen close to the ``critical curve'', displaying enhanced brightness, and compose the photon ring. Correlated images of the same spacetime event (e.g., the flare) appear at different angles and times along the ring (blue and green dots on the right image). Figure from \cite{Hadar:2020fda}.
}
\label{fig:photon shell and ring}
\end{figure}

In an optically thin setting, each light source in the vicinity of the BH will have multiple (mathematically, an infinite number of) images on an observer's screen. The different images may be indexed by the number of half-orbits executed by the photons that create them, where the half-orbit number $n$ is the number of polar turning points (i.e., in the motion in $\theta$) the photon undergoes between its emission and observation. 
The weakly-lensed direct image, $n=0$, is therefore accompanied by extremely lensed subrings $n=1,\,2,\,\ldots$, composed of photons that orbited the BH before detection. The subring images are increasingly thin, and appear exponentially close to the critical curve, both as $\sim e^{-\gamma n}$. More precisely, if $(\rho,\varphi)$ are polar coordinates on the observer screen and $\langle I^n \rangle$ is the $n^{th}$ subring's contribution to the time-averaged specific intensity, the subrings obey the asymptotic relation \citep{Johnson:2019ljv}:
\begin{align}
    \langle I^{n+1}(\tilde{\rho}+\delta \rho,\varphi) \rangle = \langle I^{n}(\tilde{\rho}+e^\gamma \delta \rho,\varphi) \rangle \,,
\end{align}
where $\tilde{\rho}(\varphi)$ parameterizes the critical curve on the screen, $\delta \rho$ is a small deviation from it, and we assume that, on average, the source is axisymmetric and reflection-symmetric with respect to the equatorial plane.
Practically, for most configurations the first indirect image, $n=1$, coarsely straddles the critical curve, while the $n\geq2$ very closely follows it.

The \textit{photon ring} \citep{Bardeen:1973tla, Luminet_1979, Johannsen:2010ru, Gralla:2019xty, Johnson:2019ljv} is the sum of the $n \geq 1$ subrings.
Due to the exponential demagnification of subsequent rings, the majority of the photon ring flux in a general viewing geometry will lie in the $n=1$ subring. 
For time-averaged images (or equivalently, axisymmetric flows), only the Lyapunov exponent, which describes the demagnification between subsequent winding numbers, is necessary to describe the relative structure of each photon ring. 
In other words, for a flow viewed over many realizations of the turbulence, the spatial structures that emerge on average are all simply related to the Lyapunov exponent. 
For the temporal and transient observables discussed in Sec.~\ref{sec:time dependent observables}, $\tau$ and $\delta$ also come into play.

The photon ring, in particular its thickness, has not yet been resolved by the EHT. The ngEHT is expected to sample the $n=1$ with its longest baselines, especially at 345 GHz. This could provide a unique probe of spacetime, given low enough optical and Faraday depths~\citep[see, e.g.][]{Himwich_2020, Mosci_2021, Ricarte_2021,Palumbo_2022}. The relation between the $n=0$ and $n=1$ sub-images is a general probe of the spacetime, and its measurement could provide some constraining power on BH parameters, as well as enable novel tests of GR \citep{Wielgus:2021peu, 2022ApJ...927....6B, Staelens_2023}. 

Below, we consider the science cases enabled by detection of the photon ring, namely: demonstrating the existence of the photon ring, static model fitting of the $n=1$ ring, and dynamical tracking of features of the space time. 
Both spatial and temporal sensitivity to photons in the $n=1$ ring offer opportunities for sensitive measurements of mass and spin, as will be further described in Sec.~\ref{sec:Mass_spin}.

\subsection{Science Cases}

\subsubsection{Demonstrating existence of the \texorpdfstring{$n=1$}{} ring}
\label{sec:existence}
\begin{figure}
    \centering
    \includegraphics[width=0.98\textwidth]{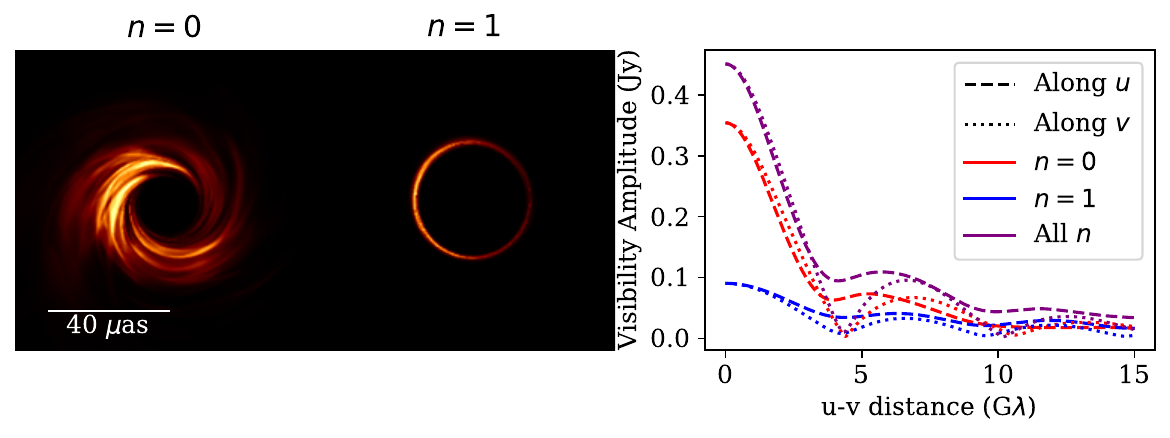}
    \caption{Left: $n=0$ and $n=1$ images from a MAD GRMHD simulation viewed with parameters appropriate for M87* at 230 GHz. Right: visibility response along the $u$ and $v$ axes of the decomposed and full image. At the baseline lengths accessible to the ngEHT, the $n=0$ and $n=1$ image have comparable correlated flux density.}
    \label{fig:GRMHD_decomposition}
\end{figure}

The presence of a photon ring arises as a qualitative consequence of extreme compactness of the central compact object and the existence of a photon shell. 
Hence, its qualitative detection alone constitutes an important confirmation of our understanding of strong gravity in a broader sense than precision-testing GR or constraining alternative spacetime metrics. 
Equipped with 345~GHz detectors, the ngEHT could approach a 13~G$\lambda$ baseline length. At this baseline length a hint of $n=1$ photon ring presence could be detected simply as a systematic excess of long baseline flux density with respect to the values measured on shorter baselines, informing us about additional power at high spatial frequencies. 
However, robust, physically-informed probes are necessary to make strong statements about photon ring existence. We outline a few potential approaches here.

As shown in \citet{Johnson:2019ljv} and reproduced here in Figure~\ref{fig:GRMHD_decomposition}, the $n=1$ ring in realistic simulations of \m87 will be detectable above the 10 mJy thermal noise level of typical EHT baselines \citep{EventHorizonTelescope:2019uob}. However, we observe that teasing out the $n=1$ structure in the general case of turbulent general-relativistic magnetohydrodynamics (GRMHD) will involve distinguishing two emission sources entering at comparable correlated flux density in VLBI measurements. Though the $n=1$ ring will not be strictly resolved, data analysis methods that permit super-resolved structure (geometric and emissivity modeling, as well as some imaging methods) may enable measurements of photon ring properties.
The main challenge for the ngEHT will be demonstrating that the data prefer the presence of a thin ring. 

Fortunately, the BH spacetime is stationary relative to the evolving accretion flow, meaning that large volumes of data taken over many realizations of the turbulent plasma should indicate a single value of BH parameters like mass and spin. 
Nonetheless, demonstrating that any detailed structure consistent with the photon ring is present is challenging: of all the improvements to the ngEHT, by far the most important in realizing this goal is the expansion to 345~GHz with associated frequency phase transfer from simultaneous observations at 86~GHz (for \m87) or 230~GHz (for \sgra). Higher frequency means longer baselines and thus sharper angular resolution.
Moreover, at higher frequencies the characteristic optical and Faraday depths in the accretion flow decrease, tending to favor larger fractions of the observed flux in the photon ring where the optical path length is longer \citep[see, e.g.,][]{EventHorizonTelescope:2021srq}.  

\begin{figure}[t]
    \centering
    \includegraphics[width=0.99\textwidth]{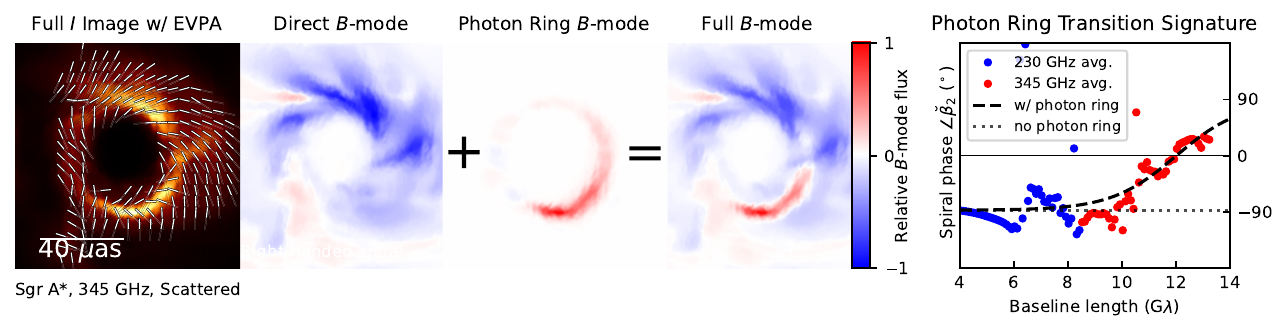}
    \caption{Polarized interferometric indication of the \sgra~photon ring. Left: a MAD GRMHD simulation of \sgra with $R_{\rm low}=1$, $R_{\rm high}=80$, and viewed at $30^\circ$, after corruption by interstellar scattering. Middle panels: divergence-free $B$-mode polarization defined relative to the image center, showing the sign flip between the direct and indirect image. Right panel: the phase of the polarimetric spiral quotient defined in \citet{Palumbo_2023} after averaging over 24 hours of the simulation movie at left, which reveals the presence of the photon ring $B$-mode reversal even without phase information constraining the image center. This detection mechanism is only possible with long-baseline 345 GHz detections which just barely reach the indirect image-dominated regime. }
    \label{fig:interferometric_prpol}
\end{figure}

One promising application of the ngEHT's novel 345 GHz coverage is a polarimetric test for the existence of the photon ring. As discussed in greater detail in Sec.~\ref{subsec:polspiral}, in the low-inclination, low-spin limit, the photon ring exhibits a simple negation of the divergence-free ``$B$-mode'' of polarization. 
Recent studies of favored models for \m87 and Sgr A* suggest that a gain-insensitive polarized interferometric observable, $\breve{\beta}_2$, can detect this reversal, with the first hints of the \sgra~photon ring available on the longest 345 GHz baselines of the ngEHT \citep{Palumbo_2023}. This observable, which extends the analysis of rotationally symmetric polarization described in \citet{PWP_2020} and used in \citet{EventHorizonTelescope:2021srq}, effectively contains the same information as the interferometric fractional polarization $\breve{m}$ expressed in a rotating basis. Figure~\ref{fig:interferometric_prpol} demonstrates that, for magnetically arrested disk (MAD) flows of modest inclination in \sgra, the longest baselines in the 345~GHz ngEHT detect the phase transition to the $n=1$ ring under varying model assumptions.

Given the reality of the comparable signal-to-noise ratio (SNR) of $n=0$ and $n=1$ interferometric visibilities at ngEHT baselines in realistic accretion flows, the ngEHT will necessarily report measurements of the $n=1$ ring that are strongly dependent on model specification. Whether the ngEHT measurement is treated as proof of existence by the astronomical community is dependent on the defensibility of the assumptions made in model fitting and in testing our methods against simulations. As an example, \citet{2020ApJ...898....9B} demonstrated a hybrid approach in which a geometrically agnostic image grid is model fit in parallel with an optional ring component; demonstration of the existence of the ring is then reliant on Bayesian information criteria, evaluating whether the data prefers the presence of the ring based on Bayesian evidence. In Sec.~\ref{sec:static_modeling} we outline several approaches for static modeling of the $n=1$ ring that are usable not only as probes of parameter values, but also as tests of detection.

\subsubsection{Static modeling of the \texorpdfstring{$n=1$}{} ring}
\label{sec:static_modeling}
As discussed in \citet{Johnson:2019ljv}, the exponential demagnification of subsequent photon rings leads to cascading baseline regimes where individual subrings dominate.
However, at the longest baselines accessible to the Earth, i.e., $\sim10$~G$\lambda$ at 230~GHz and $\sim15$~G$\lambda$ at 345~GHz, the direct $n=0$ and indirect $n=1$ image structure are of comparable flux density, as discussed earlier in Sec.~\ref{sec:existence}. Here, we mention a few approaches under development and highlight the philosophical path forward for static modeling of the $n=1$ ring.

\begin{figure}
    \centering
    \includegraphics[width=0.98\textwidth]{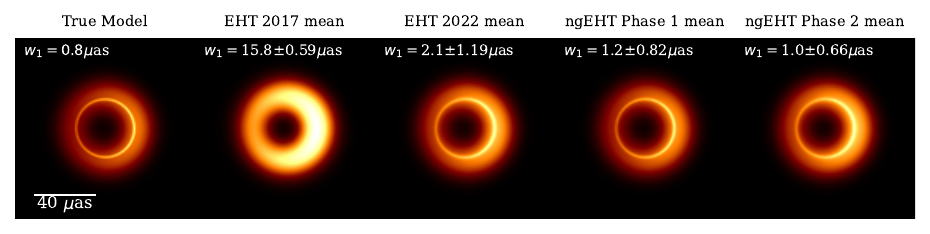}
    \caption{Geometric model fits containing two geometric ``m-rings'' with identical prior volume adapted from \citet{Tiede_2022_photon_rings}. The fits assume that the two rings, 0 and 1, have hierarchical widths. Increasing data quality and coverage eventually requires the presence of a sharp ring. Here $w_1$ specifies the mean and $1\sigma$ uncertainty on the thinner ring's thickness. Starting with the EHT 2022 array, all arrays have joint 230~GHz and 345~GHz coverage, the most crucial difference in capability of recovering sharp features.}
    \label{fig:mring_fits}
\end{figure}

Due to a tendency for imaging algorithms to favor smoothness and structure at a single (pixel) spatial scale, imaging VLBI data is ill-suited to experiments seeking to measure photon ring structures. In order to measure a sharp feature like the $n=1$ ring, methods which permit (or enforce) the sharp sub-image are necessary. Geometric model fitting is ideal for general probes of what size of features may be permitted by data.
For example, one can imagine fitting a pair of smooth rings with priors on diameter and thickness which permit one to be thick (presumably capturing the direct image) and another to be razor-thin (presumably capturing the indirect image).
A sufficiently lenient prior would yield a posterior on these parameters which indicates whether the data rejects, permits, or demands a sharp feature to be present.
An example of such a fit for a variety of array architectures and frequencies is shown in Figure~\ref{fig:mring_fits}. The underlying true model contains a thin ring and a thick ring, but only the combined data of the ngEHT 230 and 345~GHz arrays are sufficient to require recovery of the thin ring. This type of test, in which priors are deliberately uninformative, is a useful pessimistic test of whether the ngEHT will find a photon ring. Alternatively, one may enforce the presence of a thin ring through the prior volume and allow its diameter to vary widely.
This approach was shown to be very successful in finding the photon rings present in GRMHD simulations \citep{2020ApJ...898....9B}.

Emissivity modeling, however, provides the most direct approach. By taking GR as given, one may fit ngEHT  data with lensed models of the underlying accretion flow. Such approaches innately capture features of the BH accretion system that are not directly probed by imaging or modeling of the sky intensity distribution.
In particular, given a specification of the emitting material, the assumption of GR predicts direct and indirect lensed images of the flow without the typical cost of adding additional model components. Thus, any emissivity model-fitting approach will elegantly enable parallel constraints on the properties of the accretion flow and the spacetime itself.
The effects of non-GR spacetimes on photon ring properties are discussed in Sec.~\ref{sec:Tests_GR_Kerr}.

Fluid models of turbulent plasma are generally too expensive to evaluate in a forward modeling framework, so simplifications of the flow are required. The most established model for use with VLBI data is the radiatively inefficient accretion flow (RIAF), a static axisymmetric three-dimensional model of the emission that has been shown to successfully capture quiescent structures of GRMHD \citep{Broderick:2009, Broderick_2011_RIAF, Broderick_2014_RIAF}. Meanwhile, \citet{Tiede:2020jgo} demonstrated promising temporal sensitivity to mass and spin with a model consisting of infalling hotspots. Most recently, \citet{palumbo_bayesian_2022} used the equatorial toy models in \citet{Gelles_2021} and \citet{Narayan2021_polarized_image} to produce a simple, axisymmetric forward model for the polarized image of the accretion flow while avoiding radiative transfer. 

At first glance, emissivity modeling obviates the typical calibration to GRMHD simulations performed in previous EHT analysis by directly measuring spacetime parameters while marginalizing over potential emitting structures \citep{EventHorizonTelescope:2019pgp, EventHorizonTelescope:2019ggy}. 
Though the possibility for (potentially artificially) wider varieties of emission structures is useful for creating reliable measurements of BH parameters and would address common criticisms of EHT measurements such as those in \citet{Gralla:2019xty}, model misspecification provides a more fundamental limitation on the success of these approaches. The task remains to find emissivity model specifications that perform well on realistic GRMHD simulations, and to build trust in the results from these approaches, which typically produce (occasionally erroneously) tighter posteriors on system parameters than other methods.

\subsubsection{Harnessing time dependence for photon ring detection}
\label{sec:time dependent observables}
A complementary method to measure the photon ring relies on image variability. Light rays emitted by any source near the BH travel to the telescope along multiple curved paths, arriving at different times and image positions. Therefore, the variability of an optically-thin source must induce intensity correlations between different image positions and times in the (time-dependent) image. The indirect, $n\geq1$ ``light echoes'' are part of the photon ring and the temporal and angular separations between them are largely universal, i.e. depend on the spacetime geometry in the photon shell. Consequently, a successful measurement of light echoes will constitute a detection of the photon ring. Moreover, determining their quantitative details will allow measurements of mass and spin (see Secs.~\ref{sec:background} and~\ref{sec:Mass_spin}) and, eventually, strong-field tests of GR, as we detail below. 

The turbulent environment in the vicinity of a SMBH is expected to produce significant emissivity fluctuations in large regions of the parameter space.
This has been observationally demonstrated using archival \m87 data by \citet{Wielgus_2020b}. Horizon-scale variability was also recently confirmed with sub-mm VLBI with the EHT as reported in  \citet{Wielgus_2022_lightcurves} and \citet{SgrA_PaperIV}. As explained above, the BH spacetime convolves these source fluctuations in an intricate yet universal manner determined by its lensing properties. This leads to spatio-temporal correlations of intensity fluctuations across the image, and especially across the photon ring, where the light echoes appear. 

An observable which efficiently distills these correlations from the data is the two-point correlation function of intensity fluctuations on the photon ring \citep{Hadar:2020fda}:
\begin{align} \label{eq:photon ring correlator}
    \mathcal{C}(T,\varphi,\varphi') = \langle \Delta I(t,\varphi) \Delta I(t+T,\varphi') \rangle \,,
\end{align}
where $\Delta I(t,\varphi)$ is the intensity fluctuation (integrated over the width of the ring) at time $t$ and angle $\varphi$ in the image of the ring. 

\begin{figure}
	\centering
	\includegraphics[width=0.7\textwidth]{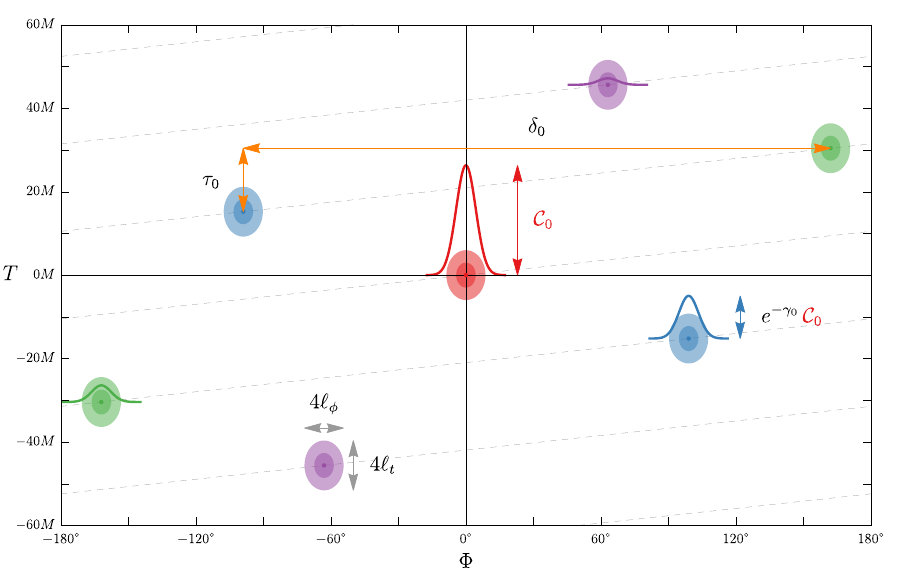}
	\caption{Universal, self-similar structure in the autocorrelation function $\mathcal{C}(T,\Phi=\varphi-\varphi')$, Equation~\eqref{eq:photon ring correlator}, for polar observations of an equatorial disk surrounding a Kerr BH with $a_{*}=0.94$. The colored peaks arise from pairs of correlated photons with the same half-orbit numbers $n=n'$ (red), and different half-orbit numbers $\left|n-n'\right|=1$ (blue), $\left|n-n'\right|=2$ (green), and $\left|n-n'\right|=3$ (purple). The (identical) shapes of the correlation peaks are determined by source properties, while their locations and relative magnitudes are determined by the spacetime geometry. Here $\gamma_0$, $\delta_0$, and $\tau_0$ are photon ring critical exponents \citep{Johnson:2019ljv,Gralla:2019drh}. Figure from \cite{Hadar:2020fda}.}
	\label{fig:autocorrelation}
\end{figure}

Importantly, the structure of this correlation function displays universal features. For optically thin emission, it will display a series of peaks in the space spanned by the angles and time separation $\{T,\varphi,\varphi' \}$. The peaks indicate a high degree of correlation, arising from the fact that different image fluctuations arise from the same source fluctuation in the BH's vicinity. The locations and relative magnitudes of the correlation peaks depend only on the BH's spacetime geometry and not on the emission details and we therefore refer to them as universal. The shape of the peaks does depend strongly on the geometric and statistical properties of the accretion flow. An example of the expectation for $\mathcal{C}$ in the case of polar observation is shown in Figure~\ref{fig:autocorrelation}.

Preliminary order-of-magnitude estimates in \citet{Hadar:2020fda} indicate an apparent possibility of measuring
the correlation function $\mathcal{C}$ using an Earth-based array by upgrading certain aspects of the EHT, in particular the overall observation time.
The main advantage of this observable is that it is in principle sensitive to the photon ring even when the latter's width is unresolved.
Resolving the diameter (which was already achieved by the EHT) and essentially attaining adequate temporal resolution should be sufficient for its measurement.
The ngEHT is expected to provide an orders-of-magnitude improvement over the EHT in the relevant SNR thanks to the addition of dedicated stations, which will greatly enhance the overall observation time. The time cadence of image reconstruction, therefore, must be comparable or smaller than the BH's natural geometric timescale. For M87* observations ($r_{\rm g}/c \approx 9$ hours) this is not an issue, but for \sgra~($r_{\rm g}/c \approx 21$ seconds) this requirement poses a challenge.

An observability estimate follows from the general principle: $\mathrm{SNR}\sim\sqrt{N_\mathrm{eff}}$, where $N_\mathrm{eff}$ is the effective number of samples. 
Here $N_\mathrm{eff}$ depends on the source temporal and angular correlation lengths, and on the magnitude of fluctuations. These are uncontrolled parameters which can be estimated for M87* and \sgra. Furthermore, $N_\mathrm{eff}$ also depends on improvable parameters like temporal and angular resolution, and, importantly, on the observation time (linearly) and the number of stations (roughly quadratically). 
For M87*, \citet{Hadar:2020fda} estimated that $\sim\mathrm{months}$ of monitoring with ngEHT may give first estimates of $\mathcal{C}$. 
For \sgra, estimates crucially depend on the expected temporal resolution, as discussed above. Note that the estimate described here is but a preliminary step and assessing in full the observability of $\mathcal{C}$ demands significant further work.
See Sec.~\ref{sec: open questions} for a discussion of possible future improvements of the method.

A related approach to the utilization of time-dependent emission and its light echoes was described in \citet{Wong2021}, wherein the characteristic signatures of localized emission events, such as orbiting hotspots, were considered.
Such events were shown to lead to ``BH glimmer'', created by light echoes of the direct image. The higher-order images appear around the photon ring at multiple angles and times, the values of which are determined by the lensing properties of the Kerr geometry. The glimmer pattern across positions and times on the ring carries geometric information on the mass, spin and Kerr nature of the lensing object. 

Another interesting proposal concerning autocorrelations was put forward in \citet{Chesler:2020gtw}. This work proposed the coherent (i.e., phase-dependent) two-point autocorrelation function as a potential observable. In principle, electric fields contain information that intensities (which are phase-independent) do not convey. 
However, for $\sim$mm observations \citet{Chesler:2020gtw} argued that this observable was out of practical reach since it is suppressed by the ratio of the observing wavelength to the BH length-scale. It remains to be seen whether coherent correlators could be relevant for other types of observations, or other observables.

The optical signatures of orbiting hotspots around BHs were first considered in \citet{Broderick:2005my}. Their observability with the EHT was the focus of \citet{Tiede:2020jgo}, where the effects of shearing of the spot were incorporated within a semi-analytical model. In addition, it was argued that the observation of multiple such hotspots would allow to ``tomographically'' map the spacetime in the vicinity of the BH. Theoretical predictions for the the astrometric signatures and flux variability of horizon-scale flares were studied in \citet{saida2017} and \citet{GRAVITY2020_flares}. These were then applied to \sgra's three flares observed by GRAVITY in 2018, showing their consistency with a hotspot closely orbiting the BH. The time-dependent signatures of infalling gas clouds in VLBI, and their dependence on BH parameters, were studied in \citet{Moriyama2015} and \citet{Moriyama2019}. 

In \citet{Narayan2021_polarized_image}, \citet{Gelles:2021}, and \citet{Vos2022}, the polarimetric signatures of orbiting hotspots were investigated, studying their dependence on the BH's parameters, the magnetic field structure, and the hotspot parameters. Subsequently, these models were applied in \citet{Wielgus2022_hotspot} to data recorded by ALMA immediately after the Sgr A* X-ray flare, on 11 April 2017. The observed variability was interpreted as arising from an equatorial hotspot, orbiting clockwise in a vertical magnetic field.

In fact, an important signature of strong lensing already lies at the level of the total luminosity. The latter's late-time fall-off after a transient accretion process is predicted to provide an imprint of the photon ring \citep{1965SvA.....8..868P, 1968ApJ...151..659A, Cardoso:2021sip}. As emitting matter falls towards the BH, e.g., a star or inhomogeneity in the accreting material, the late-time dependence of the luminosity is {\it not} that due to redshift close to the horizon, but is actually governed by the Lyapunov exponent, $\gamma$, due to the extreme lensing of photons. In particular, the luminosity of bodies being accreted onto non-spinning BHs decreases as ${\cal L}\sim e^{-t/(3\sqrt{3}M)}$ as the object approaches the horizon.
Detection of this time dependence would be a strong, complementary indication of extreme gravitational lensing.

\subsection{Open Questions} \label{sec: open questions}

The work outlined and referenced in this section has demonstrated that ngEHT measurements of the photon ring have the capacity to constrain the mass-to-distance ratio and spin of M87*.
However, much work remains to formalize what will constitute a reliable detection of the photon ring, as well as the specific pathways to connect measurements of the $n=1$ ring to constraints on non-GR spacetimes.

For instance, estimates of the $n=1$ subring size using the EHT 2017 data of \m87 were published by \citet{Broderick:2022tfu}. The measurement used the hybrid modeling approach discussed above, which combines a low-resolution image raster with a sharp ring \citep{2020ApJ...898....9B}. The authors demonstrated that this method successfully measures the $n=1$ photon ring properties in a set of five GRMHD images. However, because any photon ring detection with the EHT or ngEHT will require some degree of superresolution, these measurements are strongly dependent upon the underlying methodology and assumptions.

The \citet{Broderick:2022tfu} results do not constitute a detection of the photon ring for a number of reasons. First, tests of hybrid imaging in \citet{Tiede_2022_photon_rings} find that the hybrid imaging methodology readily produces false positives: hybrid imaging strongly prefers a photon ring even when applied to synthetic data from images with no photon ring, even if the fitted ring is constrained to be thin. In addition, the fitted ring parameters are substantially biased by the direct ($n=0$) emission. Moreover, distinct tests in \citet{Tiede_2022_photon_rings}, \citet{palumbo_bayesian_2022} and \citet{Lockhart_2022} each show that the EHT 2017 data of \m87 do not constrain the presence or absence of the photon ring. In short, the use of hybrid imaging for photon ring detection and measurement requires additional study and development to be reliable, and its applicability to both the EHT and ngEHT is an active area of research.

Regarding the constraints on non-Kerr metrics that may be possible with ngEHT photon ring measurements, \citet{Wielgus:2020uqz} and \citet{Kocherlakota2023} have worked out observable differences in ring size in face-on viewing geometries for a number of GR and non-GR BH alternatives.  
For example, in parameterised tests of GR, the spacetime is modified by the presence of extra deformation parameters other than the spin and mass defined  as the ``hairs''. Constraining the values of these parameters can act as a test of GR in the strong field limit. These deformation parameters can distort the characteristic shape and size of the photon ring, allowing for some stringent constraints given a strong detection of the ring. The same can be extrapolated to other compact objects and solutions of modified gravity theories where these photon rings tend to have distinct features which can further help in ruling out some of these models. 
In addition, it remains to be seen whether and how ngEHT observations could provide robust, universal ways to infer spacetime symmetries.

There remains much work to be done before static or dynamical modeling of the accretion flow emissivity distribution will be able to closely approximate all structures observed in GRMHD simulations. In particular, the RIAF, hotspot, and equatorial models mentioned in this text do not typically include outflows (self-consistent or otherwise), while simulation efforts in \citet{EventHorizonTelescope:2021srq} show significant emission along the jet funnel as opposed to the disk in models consistent with the EHT data on \m87. Capturing the full space of possible emission geometries is the most natural way to produce well-motivated uncertainties on BH parameters from sub-mm VLBI. \citet{Levis_tomography} has provided a useful first step towards inference of arbitrary emission regions in a fixed Schwarzschild spacetime without detailed radiative transfer.
Future, more general approaches will be crucial for understanding the level of confidence of ngEHT measurements and their sensitivity to GRMHD calibration.

Finally, it will be important to look into several open issues regarding time-domain signatures of the photon ring. As already alluded to, translation of the autocorrelation observable directly into (semi-)raw visibility amplitudes, bypassing the need for image reconstruction with each step of a ``movie'', could be very useful for analyzing rapidly varying sources such as \sgra. It will be important to obtain a good heuristic grip on the observables described in Sec.~\ref{sec:time dependent observables} away from small inclinations, for varying source models, and at all possible spins, as well as to discern the effects of deviations from the universal regime, i.e., the contribution of the correlation between the $n=0$ and $n=1$ subrings.
Ultimately, an end-to-end study of realistic synthetic data generated from ``slow light'' GRMHD movies of both \sgra~and \m87 sampled with realistic cadence is required, imaging each snapshot with the best imaging algorithms available.
This effort is already underway.
An outstanding challenge will be to flesh out additional interesting observables which may benefit from source variability and to define their associated requirements for the ngEHT.

%% file: Mass_spin.tex
\section{Measuring black hole mass and spin\label{sec:Mass_spin}}
\subsection{Introduction}
According to the uniqueness results, a stationary BH in vacuum is fully characterized by its mass, spin, and electric charge~\citep{Robinson:1975bv,Chrusciel:2012jk,Cardoso:2016ryw}. Astrophysical BHs are expected to be electrically neutral due to quantum discharge effects, electron-positron pair production, and charge neutralization by astrophysical plasmas~\citep{Gibbons:1975kk,1969ApJ...157..869G,1975ApJ...196...51R,Blandford:1977ds}. Therefore, mass and spin are the only fundamental quantities that determine the BH geometry within GR~\citep{Zajacek:2018ycb}. Measuring the masses and spins of BHs will help constrain their formation channels and growth mechanisms \citep{Volonteri:2010wz,Volonteri:2021sfo}, map out the population demographics \citep{shankar2004supermassive}, and examine BH feedback models \citep{terrazas2020relationship}. The spin of BHs is also relevant for probing ultra-light fields, as discussed in Sec.~\ref{sec:Ultralight_fields}.

Observables obtained by the ngEHT have been proposed to probe the strong-gravity regime of the BH spacetime, which in turn constrain the BH mass and spin, assuming the Kerr metric. Some of them (i.e., BH/light ring imaging) have already been employed for the measurement of mass and spin for M87* and Sgr A*. Others may require specific signatures of accretion flows (e.g., hotspots), additional information (e.g., polarization) and/or temporal measurements (e.g., autocorrelation), which should become accessible through the ngEHT. In particular, they can provide an unprecedented opportunity to probe properties of near-extremal BHs, which are deeply connected to the holography principle.

The mass and spin measurement of Sgr A* will likely be conducted earlier and more accurately by other means, e.g., through a combination of S-star orbits, possible clouds, pulsars, or lurking stellar-mass BHs. The ngEHT measurements can thus most likely be verified by comparing to these alternative methods \citep{Will:2007pp, Psaltis:2015uza}. Here, we discuss how the combined information can be used to break possible degeneracy of parameters in any individual approach, in order to achieve better measurement precision.

\subsection{Overview of ngEHT observables for measuring mass and spin}
%
The EHT collaboration has announced the observation results for two supermassive BHs so far: M87*~\citep{EventHorizonTelescope:2019dse} and Sgr~A*~\citep{EventHorizonTelescope:2022xnr}. By measuring the emission region diameter and combining with calibrations from GRMHD simulations, the mass of M87* is determined to be  $(6.5\pm 0.7)\times 10^9 M_\odot$, corresponding to an angular diameter of the shadow $ 42 \pm 3 \,\mu$as. The angular diameter of Sgr A* is measured to be $51.8 \pm 2.3 \,\mu$as and the corresponding mass is constrained within $4.0^{+1.1}_{-0.6}\times 10^6 M_\odot$, consistent with constraints from the S-stars. The most precise measurements of the mass are reported by the VLTI as $\left(4.297\pm0.013\right)\times 10^{6}~M_{\odot}$ \citep{GRAVITY:2021xju}, and by Keck as $\left(3.951\pm0.047\right)\times 10^{6}~M_{\odot}$ \citep{Do:2019txf}.
These differ because of the difference in $R_0$ between the two results: scaling the Keck measured mass to the VLTI distance yields $4.299\pm0.063 \times 10^{6}~M_{\odot}$ \citep{GRAVITY:2021xju}. The current EHT-based mass value, due to its 20--80 times larger error, is not sufficiently precise to distinguish between these two aforementioned values.

The spin magnitude of M87* is poorly constrained but the image morphology is consistent with  the shadow of a spinning (instead of non-spinning) Kerr BH. If the spin axis and M87's large-scale jet are aligned, then the BH spin vector is pointed away from the Earth, with recent studies showing this can still be the case even for misaligned accretion flows \citep{Chatterjee2020}.
On the other hand, the Sgr A* images disfavor scenarios where the BH is viewed at high inclination, as well as non-spinning BHs and those with retrograde accretion disks.

\begin{table}[htbp!]
  \caption{Mass measurements for the SMBH at the center of the Milky Way.}
    \centering{
    \begin{tabular}{c|c|c|c}
    \hline
    \hline
            & GRAVITY/VLTI$^{a}$  & Keck$^{b}$ & EHT$^{c}$\\
    \hline
    Sgr A*   &$(4.297\pm 0.012)\times 10^6 M_{\odot}$ &$(3.975\pm 0.058)\times 10^6 M_{\odot}$ &$(4.0^{+0.1}_{-0.6})\times 10^6 M_{\odot}$ \\
    \hline
    \hline
    \end{tabular}}
    \label{tab:masses_sgrA}
{\noindent \small $^{a}$ \citet{GRAVITY:2021xju}, $^{b}$ \citet{Do:2019txf}, $^{c}$ \citet{EventHorizonTelescope:2022xnr}.}
\end{table}
%

\subsubsection{Light ring imaging}
As discussed in Sec.~\ref{sec:Light_ring}, within the EHT image there exists a theoretically infinite sequence of lensed images of the emission region. The locations of these photon rings asymptote to the boundary of the shadow, the size and shape of which encodes the mass and spin of the BH \citep{Medeiros2020, Johannsen:2010ru, Takahashi2004, Falcke:1999pj, Luminet_1979, Hilbert1917}. 
Actual observables are likely, by virtue of being luminous, the low-order images, i.e., primary, secondary, and tertiary images, corresponding to the $n=0,\,1,\, \mathrm{and}~2$ photon rings in \citet{2022ApJ...927....6B} and \citet{Johnson:2019ljv}.
The relative locations of the lensed images at different orders depend on mass and spin, enabling a measurement of both. 
Because this remains true even for polar observers, observing the secondary presents a unique pathway to measuring spins in M87* and Sgr A*.

The expected width of the $n=1$ photon ring is $\lesssim1\,{\rm \mu as}$, and thus well below the diffraction limit of Earth-bound mm-VLBI.  Nevertheless, combined modeling and imaging techniques might provide an ability to both extract the highly uncertain primary image and constrain properties of an additional narrow ring-like image feature \citep{2020ApJ...898....9B}.
This procedure leverages high SNR to separate the $n=1$ photon ring and diffuse primary emission on Earth-sized baselines, and assumes that one does not encounter a noise floor of systematic uncertainties.  

While it is possible to extract the ring diameter with EHT coverage \citep{2020ApJ...898....9B}, measuring the width and total flux in the $n=1$ photon ring will only be possible with the additional stations and sensitivity of the ngEHT.
The first EHT analyses of M87* already super-resolve the source, achieving a typical resolution of $\sim10\,{\rm \mu as}$. 
Given the strong priors that accompany a specified ring model, the degree of super-resolution in the determination of the ring width, $w$, is approximately: 
\begin{equation}
\begin{aligned}
    \frac{\sigma_w}{b}
    &\sim
    \frac{1}{2\pi u w N \sqrt{n_s} \, {\rm SNR}}\\
    &\sim
    1\% \times \left(\frac{\rm SNR}{7}\right)^{-1} 
    \left(\frac{u}{10\,{\rm G\lambda}}\right)^{-1}
    \left(\frac{w}{1\,{\rm \mu as}}\right)^{-1}
    \left(\frac{N}{20}\right)^{-1}
    \left(\frac{n_s}{10}\right)^{-1/2}\,,
\end{aligned}
\end{equation}
where $N$ is the number of stations, $n_s$ is the number of independent scans, $b=1/u\sim20\,{\rm\mu as}$ is the nominal beam, $u$ is the maximum baseline length in $\lambda$, and SNR is the thermal signal-to-noise ratio.  Thus, $N \times{\rm SNR} \gtrsim 140$ is needed to eventually achieve the sub-$1\,{\rm \mu as}$ precision needed to resolve the $n=1$ photon ring width in a single observation.

\begin{figure}
    \centering
    \includegraphics[width=0.99\textwidth]{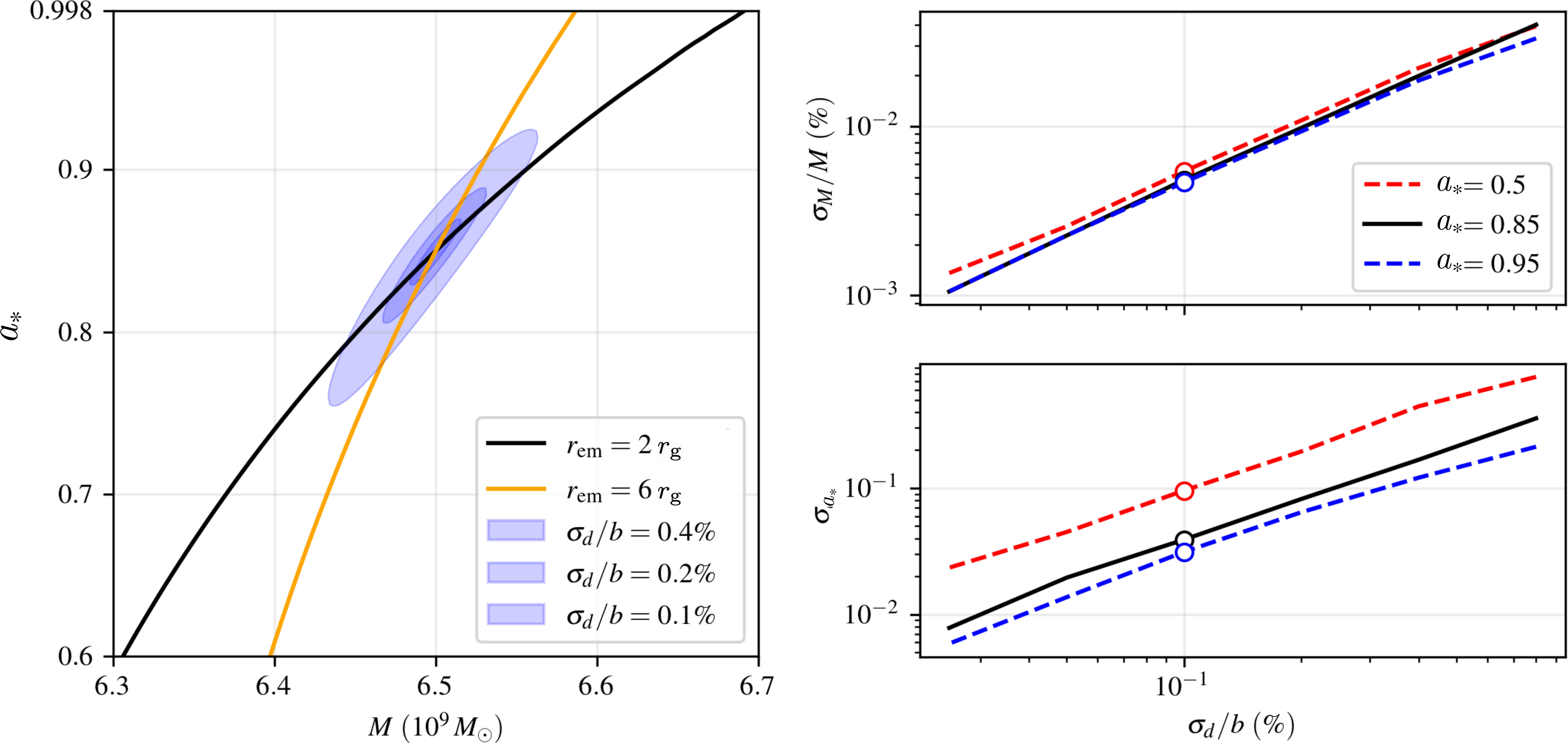}
    \caption{Left: predicted joint constraints on mass and spin from measurements of the primary and secondary images (i.e., the $n=0$ and $n=1$ photon rings) of the emission about a BH with mass $M = 6.5\times10^9~M_{\odot}$ and spin $a_{*} = 0.85$, appropriate for M87*. The two lines show the degenerate constraint when the emission is dominated by that at $2~r_{\rm g}$ (black) and $6~r_{\rm g}$ (orange).  The combined $1\sigma$ regions are indicated in blue for diameter measurements of various precision, ranging from $\sigma_d/b=0.1\%$ to $0.4\%$. Right: estimates of the precision of mass (top) and spin (bottom) for different intrinsic BH spins, as a function of diameter measurement precision. The open points show the fiducial value in Equation~\ref{eq:diameter_precision}. 
     }
    \label{fig:photon_ring_spin_est}
\end{figure}

A single measurement of the diameter of the $n=1$ photon ring alone would provide a mass measurement that has a bounded systematic uncertainty. For equatorial emission seen by a polar observer, the diameter of the $n=1$ photon ring ranges from $4.30~M/D$ to $6.17~M/D$ as the radius of the peak emission moves from the horizon to infinity~\citep{2022ApJ...927....6B}, where $D$ is the source distance. Thus, the conclusive detection of a photon ring necessarily eliminates the current dominant systematic uncertainty for mass estimates of M87*.

The differing behavior of the primary and secondary image dependence on the emission location provides a means to probe spin.  A single simultaneous measurement of the primary emission location and $n=1$ photon ring results in a degenerate constraint on the mass and spin of M87*.  Similar to the ring width, it should be possible to constrain the diameter of the photon ring considerably better than the nominal beam:
\begin{equation}
    \frac{\sigma_d}{b} \sim \frac{1}{\pi N \sqrt{n_s} {\rm SNR}}
    \sim
    0.1\% \times \left(\frac{\rm SNR}{7}\right)^{-1} 
    \left(\frac{N}{20}\right)^{-1}
    \left(\frac{n_s}{10}\right)^{-1/2},
    \label{eq:diameter_precision}
\end{equation}
where the significant improvement arises because the ring diameter, unlike the ring width, is well resolved by Earth-sized baselines.
Because the emission region in M87* is highly variable \citep{EventHorizonTelescope:2019dse,EventHorizonTelescope:2019uob,EventHorizonTelescope:2019jan,EventHorizonTelescope:2019ths,EventHorizonTelescope:2019pgp,EventHorizonTelescope:2019ggy,EventHorizonTelescope:2021bee,EventHorizonTelescope:2021srq}, two such measurements of the primary emission and the $n=1$ photon ring diameter made at widely separated times, and thus for different characteristic emission radii, produce estimates for both.
An illustrative example for M87* is shown in Figure~\ref{fig:photon_ring_spin_est}, in which the radius of the peak emission moves from $2~r_{\rm g}$ to $6~r_{\rm g}$ between observations.
Measuring the radii of the direct emission and the $n=1$ photon ring in either epoch results in degenerate measurements of mass and spin shown by the two lines.
Combining the two epochs produces a joint measurement of mass and spin, where the precision depends on the difference in the emission region location, the true BH spin, and the degree of super-resolution in the measurement of the $n=1$ photon ring diameter.

\subsubsection{Hotspot tracking} \label{hotspot}

M87* and Sgr~A* both exhibit localized variability in the emission region.  In M87* this variability appears as ejections within the jet, which are instrumental to measuring the jet velocity on milliarcsecond scales \citep{Jeter2020,Ly2007,Walker2016,Hada2016}. In Sgr~A* it appears as broad-spectrum flaring, extending from the millimeter to the X-ray \citep{Genzel2003, Gillesen2006, Dodds-Eden:2009wba, Witzel2012, Neilsen2013, Ponti:2017grl, Fazio2018}, and at least a subset of these have been associated with orbiting features \citep{2018A&A...618L..10G}.
Simultaneous X-ray and infrared (IR) observations of Sgr A* variability hint at the multiwavelength emission properties of orbiting hotspots \citep{Boyce2019} and their close links to particle acceleration mechanisms \citep[e.g.,][]{Ball2019}.
The dynamical nature of the ``hotspots'' enable high-precision measurements of BH mass and spin \citep{Broderick:2005my, Broderick:2005jj, Doeleman:2009, 2018A&A...618L..10G, Tiede:2020jgo}.

Flare reconstructions are fundamentally strong lensing experiments, relating the direct emission and higher-order lensed images from a dominant, compact emission region.  The chief systematic uncertainty is the astrophysics of the flare emission itself, including its unknown orbit, temporal evolution, density, temperature, and rate of shear.  For these reasons, relating flare orbital periods to those expected for circular geodesics (i.e., Keplerian motion), is not a direct measure of spacetime properties.  However, the relationship between the primary and secondary images is independent of the orbital dynamics, requiring only motion to selectively relate different regions of the image plane \citep{Broderick:2005my,Broderick:2005jj}. Reconstructing a single flare observed over a handful of orbits would yield a sub-1\% accuracy spin measurement, while simultaneously recovering the astrophysical hotspot parameters over a wide range of flare models, BH parameters, and in the presence of an obscuring accretion flow and intervening scattering in the Galactic disk \citep{Tiede:2020jgo}. Yet, flares could cool on shorter time scales than the orbital time scale, and shearing may render it impossible to observe significantly more than one full orbit.

Each flare observed by ngEHT would produce an independent measurement of the gas rotation, which can provide constraints of BH spin \citep{Conroy2023}. The observation of flares occurring at multiple orbital radii probes the spacetime at different locations, therefore providing an immediate test of the Kerr metric, which demands all such spin measurements be consistent with each other \citep{Tiede:2020jgo}. The observation of flares at different frequencies provide a means to test the achromatic nature of lensing in GR. Thus, observations of multiple flaring epochs enables a tomographic mapping of the BH spacetime.

The most significant practical limitation is the need to observe multiple, high-brightness flares, dominated by orbiting features.  For both M87* and Sgr A*, this is most readily accomplished with a monitoring campaign that triggers target of opportunity observations.
While the full $(u,v)$-coverage of the ngEHT is preferable, modeling with even a 10 station subset, e.g., only the proposed new ngEHT sites, would suffice for dynamical flare modeling to produce high-precision spin estimates.

\subsubsection{Photon ring autocorrelations}

As discussed in Sec.~\ref{sec:Light_ring}, the two-point correlation of intensity fluctuations on the photon ring encodes information about the background spacetime, which can be used to measure the BH mass and spin. Here we provide more details about the underlying principle and the estimation of the SNR.

The correlation function $\mathcal{C}$ is expected to be described by localized peaks with separations in time and azimuthal angle around the ring. For example, if two light rays are emitted from the same point source and perform half-orbits of the BH $k$ and $k'$ times, respectively, before reaching the observer, they will contribute to a peak in the correlation function.
These peaks can be labeled by $m=k-k'$ and should share an identical profile based on the source statistics. The peak width is set by the correlation length of fluctuations in the source, while the locations and relative heights of the peaks are dependent on the BH parameters. 
Specifically, each successive peak is suppressed by $e^{-\gamma}$ and is translated by $(\tau,\delta)$, where $\gamma$, $\tau$, and $\delta$ are the critical exponents that describe geodesics near the critical curve. 
The critical exponents have been computed analytically for the Kerr spacetime~\citep{Johnson:2019ljv, Gralla:2019drh}.
For a geodesic approaching the critical radius, the ratio of distances from the critical radius of successive half-orbits $k$ is:
\begin{equation}
    \frac{\delta r_{k+1}}{\delta r_{k}}\approx e^{-\gamma}\,.
\end{equation}
While these near-critical geodesics stay near the critical radius, they continue to move in the other directions.
The elapsed time $\Delta t$ and the swept azimuthal angle $\Delta \phi$ for each half-orbit, which approach a constant value for large half-orbit number $k$, are given by:
\begin{equation}
    \Delta t \approx \tau + \delta t_{k}\,, \qquad \Delta\phi \approx \delta + \delta\phi_{k}\,,
\end{equation}
and $\delta t_{k},\delta\phi_{k}\sim e^{-k\gamma}\rightarrow{}0$ as $k\rightarrow{}\infty$. 
Observing the correlation structure would provide measurements of the critical parameters $(\gamma,\tau,\delta)$ and, as these are dependent on the BH spacetime, would in turn allow for estimates of the BH mass and spin.
The SNR for the correlation of $N$ independently sampled pairs of  intensity functions for individual images can be estimated as~\citep{Hadar:2020fda}:
\begin{equation}\label{eq:snr}
    \text{SNR}\sim\frac{l_{\phi}\theta_{\text{ph}}}{\theta_{\text{obs}}}\text{SNR}_{\infty}\,,
\end{equation}
where $\text{SNR}_{\infty}$ is the SNR for an idealized case of infinite resolution and the $m^{\text{th}}$ correlation peak
\begin{equation}
    \text{SNR}_{\infty}\sim e^{-|m|\gamma}\sqrt{\frac{2\pi t_{\text{obs}}}{l_{\phi}l_{t}}}\,.
\end{equation}
Here $t_{\text{obs}}$ is the observing time, $l_{t}$ and $l_{\phi}$ are the correlation lengths in the $t$ and $\phi$ directions, respectively, $\theta_{\text{ph}}$ is the angular radius of the photon ring, and $\theta_{\text{obs}}$ is the finite angular resolution of the observation.

Based on this SNR estimate and as mentioned in Sec.~\ref{sec:Light_ring}, detecting the $m=1$ peak of M87* with the EHT would require observations every few days over a span of many months or even several years. Due to Sgr A*'s significantly shorter gravitational timescale, the limiting factor becomes the ability to form an image with very short ($\sim$ minutes) observations. An ngEHT-like array with the capability to create movies of Sgr A* would be sufficient and the SNR for the $m=1$ peak of  Sgr A* would be about two orders of magnitude larger than the SNR for M87* for the same observing duration, primarily because of the shorter coherence time for Sgr A* \citep{Hadar:2020fda}. Reaching a high enough SNR to detect the $m=2$ and higher-order correlation peaks, however, would require a significantly improved array and many years of observations.

Whilst observations that can detect the $m=1$ correlation peak may be possible with the proposed ngEHT array, more work needs to be done to pin down the various technical requirements. In addition, it is not yet clear how the SNR of the peaks translates into uncertainties in the mass and spin measurements.  
More work is also required to determine if correlations in the astrophysical structure of the disk could contaminate the correlation function.

\subsubsection{Polarization spirals in direct and indirect images}
\label{subsec:polspiral}
\begin{figure}[ht!]
    \centering
    \includegraphics[width=0.98\textwidth]{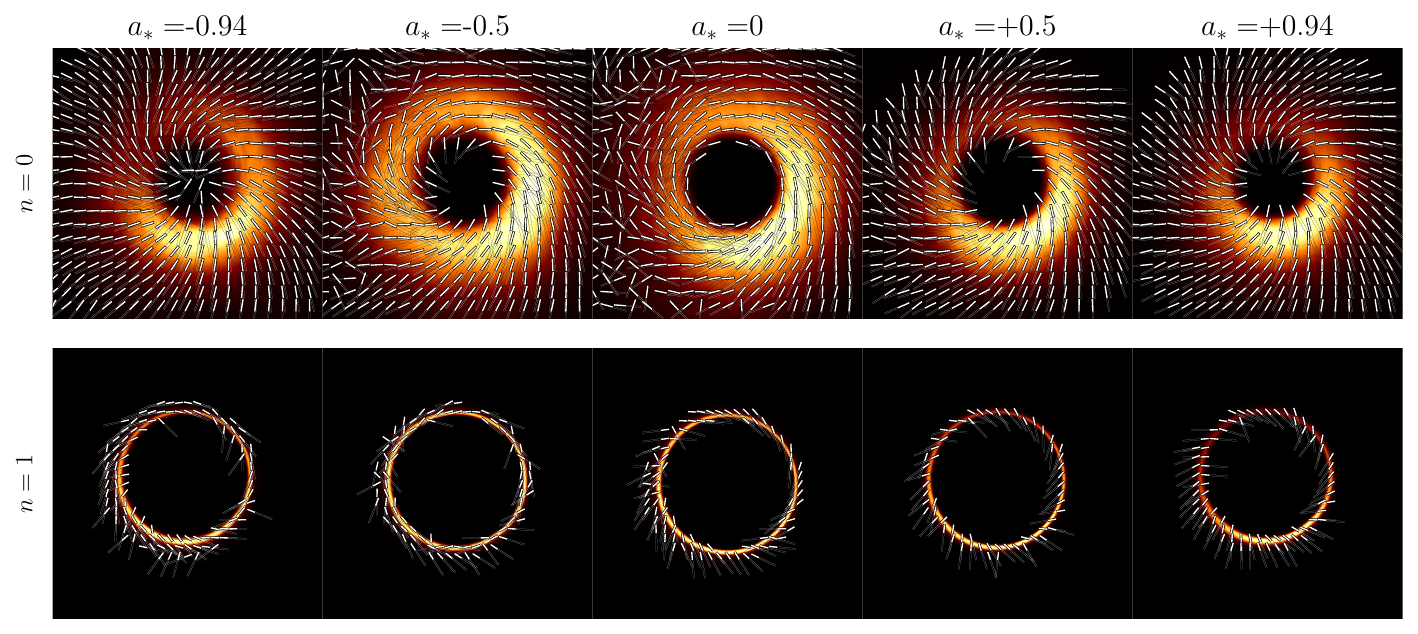}
    \caption{Time-averaged images of the direct ($n=0$) image and first lensed ($n=1$) image from MAD GRMHD simulations at various spins, rotated so that the approaching jet is oriented $288^\circ$ East of North \citep[adapted from][]{Palumbo_2022}. Ticks show the EVPA. Polarization spirals about the ring become more radial at higher spin magnitudes, reverse direction over radius in retrograde flows ($a_{*}<0$), and approximately reflect through the origin across sub-image index $n$. These images use the $R_{\rm high}$ electron heating prescription, each having $R_{\rm high}=80$, a reasonable value for both M87* and Sgr A* \citep{Mosci_2016, EventHorizonTelescope:2019pgp,Collaboration2019_V}. The simulations themselves were generated with \texttt{iharm3d} \citep{Gammie_HARM_2003} and ray traced with \texttt{ipole} \citep{IPOLE_2018}.
    See \citet{Wong_2022} for additional details on the ray tracing.}
    \label{fig:MAD_spin_tavgs}
\end{figure}

Although the astrophysical details of the accreting plasma often confound measurements of spacetime properties, occasionally the emission illuminates BH properties. Figure~\ref{fig:MAD_spin_tavgs} shows time-averaged images of MAD simulations decomposed into direct and indirect images for a variety of spins.
In these images a number of features with a direct connection to spin are seen.
Most apparent is the electric vector position angle (EVPA) spiral, which becomes more radial at higher spin magnitudes, as first identified in \citet{PWP_2020}. 
\citet{Emami_2023} studied the spiral trend in detail, finding that this effect arises from frame dragging causing the plasma velocity and magnetic field to become more toroidal in structure at higher spins, causing the polarization (perpendicular to the magnetic field) to appear more radial in structure.
We see also that the sub-image polarization spiral has opposite handedness compared to the direct image spiral.
This feature arises from the complex conjugation of the Walker-Penrose constant \citep{Walker_Penrose_1970} across sub-images, derived in \citet{Himwich_2020}. 
This conjugation has a simple behavior in the face-on, zero-spin limit, and leads to an approximate reflection of the polarization through the origin, as observed in GRMHD in \citet{Palumbo_2022} and is related to a depolarization near the photon ring observed in \citet{Alejandra_2021}. 
The detailed relationship between the direct and indirect image polarization is a direct probe of spin. Lastly, we see that in retrograde models, the spiral changes direction from the large scale accretion rotation direction at large radii to the interior, strongly frame-dragged region, as investigated in \citet{Ricarte_2022_framedragging}.

Taken together, there are several pathways through which images with high resolution and dynamic range can be used to elucidate spin, given prior assumptions and understanding of the underlying accretion flow properties. 
These features are more difficult to observe in models undergoing standard and normal evolution (SANE), but both EHT results and results from the GRAVITY collaboration support the conclusion that M87* and Sgr~A* are in the MAD state \citep{Gravity_2020_MAD}. 
These features are also more difficult to observe in individual snapshots, but the ngEHT is proposed to observe M87* with a roughly weekly cadence, enabling analysis of average, quiescent structure, while the same can be done with Sgr~A* in the course of a few days. Polarized emission from the accretion flow will be a crucial tool in probing the BH spacetimes of M87* and Sgr~A*, and should complement more general approaches which attempt to circumvent astrophysical details. Cases with a single-baseline for polarimetry were studied by \citet{Palumbo:2023auc}.

\subsubsection{Probes of extremal BH/signatures of NHEK}
As rotating BHs approach the extremal limit, a non-degenerate near-horizon region emerges with an enhanced conformal symmetry, often referred as near horizon extremal Kerr (NHEK). Thanks to the presence of NHEK, there are a set of zero-damping quasinormal modes with slow decay rates (zero in the extremal limit) \citep{Yang:2013uba}, and the field perturbations display self-similar behavior \citep{Gralla:2018xzo}. The NHEK perturbations are important as their prescription determines how an infalling particle never overspins a near-extremal BH to produce a naked singularity, as required by Weak Cosmic Censorship \citep{Sorce2017}. The additional conformal symmetry also allows the construction of the Kerr/CFT conjecture \citep{Guica:2008mu} which relates a $2+1$ conformal field to the NHEK region. It is therefore of fundamental theoretical interest to probe and test the signatures of the NHEK, with ngEHT and/or other observations.

Astrophysical BHs may reach dimensionless spins of up to $a_{*}\sim0.998$, assuming Shakura-Sunyaev thin-disk accretion \citep{Shakura:1972te}.
For these rapidly spinning BHs, the Green's function exhibits a transient power-law growth in the near-horizon region and a transient power-law decay at far distances \citep{Yang:2013uba}, as compared with the exponential signature of the Green's function for generic spins. This power-law behavior of the Green's function, as a manifestation of collectively excited zero-damping modes, is a direct signature of NHEK and its enhanced symmetry. One possible way to test this signature is through the measurement of photon ring auto-correlations. For near extremal Kerr BHs, the correlation function follows a power law (instead of exponential) relation between different peaks separated by angle and time, which may be resolved through performing measurements with sufficient accuracy. 
Roughly speaking, in order to distinguish between a power-law decay from an exponential decay in the correlation function, at least the correlations of the $m=0, m=1$ pair and the $m=0, m=2$ pair are needed, although it is challenging to detect the $m=2$ peaks since this requires observation over many years (see discussion in previous subsection). 
With only the $m=0, m=1$ pair, it is possible to perform a consistency test between the measured correlation and the inferred correlation if the Green's function follows a $1/t$ power-law decay. 
Such correlation may also be interpreted as $e^{-\gamma(M,a_*)}$ for appropriate $(M,a_*)$, but the degeneracy may be broken by including constraints of $(M,a_*)$ from other observables. 
In this setting, the SNR of the correlation in Equation~\ref{eq:snr} should be modified as:
\begin{align}\label{eq:snrnhek}
     \text{SNR}\sim\frac{l_{\phi}\theta_{\text{ph}}}{\theta_{\text{obs}}} \left (\frac{C}{|m|\tau}\right ){ \sqrt{\frac{2\pi t_{\text{obs}}}{l_{\phi}l_{t}}}} \,,
\end{align}
where $C$ is a numerical factor to be determined by the Green's function.
The higher the SNR, the better the statistical confidence that can be claimed for the consistency test.
It is also noteworthy that this power-law signature is comprised of a subclass of photon orbits close to the equatorial plane of the BH. A further observer-inclination-angle-dependent modification should be incorporated in order to further refine Equation~\eqref{eq:snrnhek}.

Another possible avenue is to search for related signatures in the BH image and/or the transient images of hot spots \citep{Gralla:2017ufe}, since part of the co-rotating light rings reside on the horizon (with deviation much smaller than the gravitational radius).
Recent studies have investigated MHD accretion flows onto BHs with $a_{*}=0.998$, in both the Kerr spacetime and in other theories of gravity \citep{Younsi:2021dxe,Chatterjee2023b,Chatterjee2023a}.
Defining $\epsilon\equiv (1-a_*)^{1/3}$, the flux of the hotspot generally scales as $\epsilon /\log \epsilon $ \citep{Gralla:2017ufe}, i.e., diminishing flux in the extremal limit. The redshift factor of the emission varies with orbital phase, with peak blueshift factor being $\sqrt{3}$ and redshift factor being $1/\sqrt{3}$. The redshift/blueshift factor of Iron K$\alpha$ lines may be used for such a test.
It is, however, unclear what the most relevant observables for ngEHT measurements in this context will be.
In addition, the number of high-spin candidates for ngEHT measurement is currently highly uncertain.

\subsection{Overview of complementary measurements of mass and spin}
Accurate measurements of mass and spin are vital to understand the nature of BHs and explore their discovery potential. In order to draw firm conclusions, it is important to cross-check such measurements with other independent observations. In many cases, the mass and spin estimates may exhibit degeneracy to certain degrees.
To break such degeneracy it is useful to have complementary measurements which can be combined appropriately.
We briefly discuss below alternative measurements of BH mass and spin from S-stars, pulsars, GWs, and quasi-periodic oscillations (QPOs).

\subsubsection{Probing Sgr A* with S-stars}

The Sgr~A* SMBH in the Galactic Center is surrounded by a dense cluster of young stars, commonly referred to as S-stars. Some of them have been discovered with small
periastron distances and high eccentricities \citep{Schodel:2002py, Meyer2012Sci, GRAVITY:2021xju} and are powerful probes of the properties of Sgr A*. 
By monitoring the orbit of the star S2, the GRAVITY instrument \citep[][a near-infrared interferometer mounted at the ESO VLTI]{2017A&A...602A..94G} has detected the leading order relativistic effects, i.e., the relativistic redshift~\citep{GRAVITY:2018ofz,Do2019RelativisticRO} and Schwarzschild precession~\citep{GRAVITY:2020gka, GRAVITY:2021xju}, constraining the mass of the central object to a very good precision (see Table~\ref{tab:masses_sgrA})~\citep{GRAVITY:2021xju}. 
The typical precision of mass measurement with S-stars is in the range $0.01\% - 0.1\%$, since the tightness of the mass measurement depends on the relative ratio of the astrometric accuracy to the semimajor axis and the ratio of the redshift accuracy to the orbital velocity \citep{Weinberg2005W, Zhang_2015}, and does not depend directly on the semimajor axis or eccentricity.

The spin and quadrupole moment of Sgr~A* can be constrained by detecting spin-induced effects and quadrupole-induced precession in the motion of S-stars \citep{Will:2007pp, Psaltis:2015uza,Zhang_2015, Waisberg2018}. 
The spin of Sgr~A* induces Lense-Thirring precession on the S-star orbits and rotates the orbital plane around the spin axis. 
This precession should be observable both in astrometry and radial velocity space \citep{Will:2007pp, Zhang_2015, Waisberg2018}. 
To constrain the spin parameter, S-stars need to be found within milliparsec-scale distance of Sgr A*, where the key parameter is the pericenter distance $d_{\rm peri}$. 
Detecting quadrupole-induced effects will be even more challenging.
Explicit simulations for the GRAVITY+ project \citep{2022Msngr.189...17A} show that the simple time-averaged estimates by \citet{Merritt:2009ex} were too pessimistic, as tracking the orbits will reveal the moment of the deviation, which for spin or quadrupole effects will coincide with a pericenter passage.
The combination of astrometry from GRAVITY+ and spectroscopy from ELT-MICADO should be able to deliver a spin measurement within a few years of operation, assuming a star with suitable $d_{\rm peri}$ can be tracked.

The S-stars found to-date do not pass close enough to Sgr A*, thus the measurements of spin and quadrupole parameters have not yet been achieved \citep{Iorio:2020uet, 2022A&A...657A..82G}. 
The upgraded GRAVITY+ instrument, with higher sensitivity, may possibly find closer S-stars with sufficient brightness, and will continue to monitor S2 over a longer period of time, thereby yielding estimates of the spin and quadrupole moment of the central object~\citep{Psaltis:2015uza,Zhang_2015, Yu2016ProspectsFC, Waisberg2018,2022A&A...657A..82G}.
Two or more S-stars in closer orbits are often required, so that the combination of position and redshift data provide complementary information to measure the spin of Sgr A*~\citep{Will:2007pp, Zhang_2015}.
Figure~\ref{fig:MassSpin} shows the constraints of mass and spin parameters of Sgr~A* with two stars assumed to have orbital periapsis distances of $800~r_{\rm g}$ and $1000~r_{\rm g}$ and eccentricities of 0.9 and 0.8, respectively, whilst assuming an astrometric precision of $10~\mu$as \citep{Psaltis:2015uza}.
These studies did not account for the improvement possible with ELT spectroscopy, which will bring into reach stars of spectral types that show rich spectra in the near-IR, potentially achieving radial velocity uncertainties as low as $\sim$ 0.1~km/s \citep{evans2015science,Simon:2019kmm}, two orders of magnitude better than what is currently possible with S2.

\subsubsection{Observation of Sgr A* EMRIs with LISA}

\begin{figure}
\begin{center} 
\includegraphics[width=0.98\textwidth]{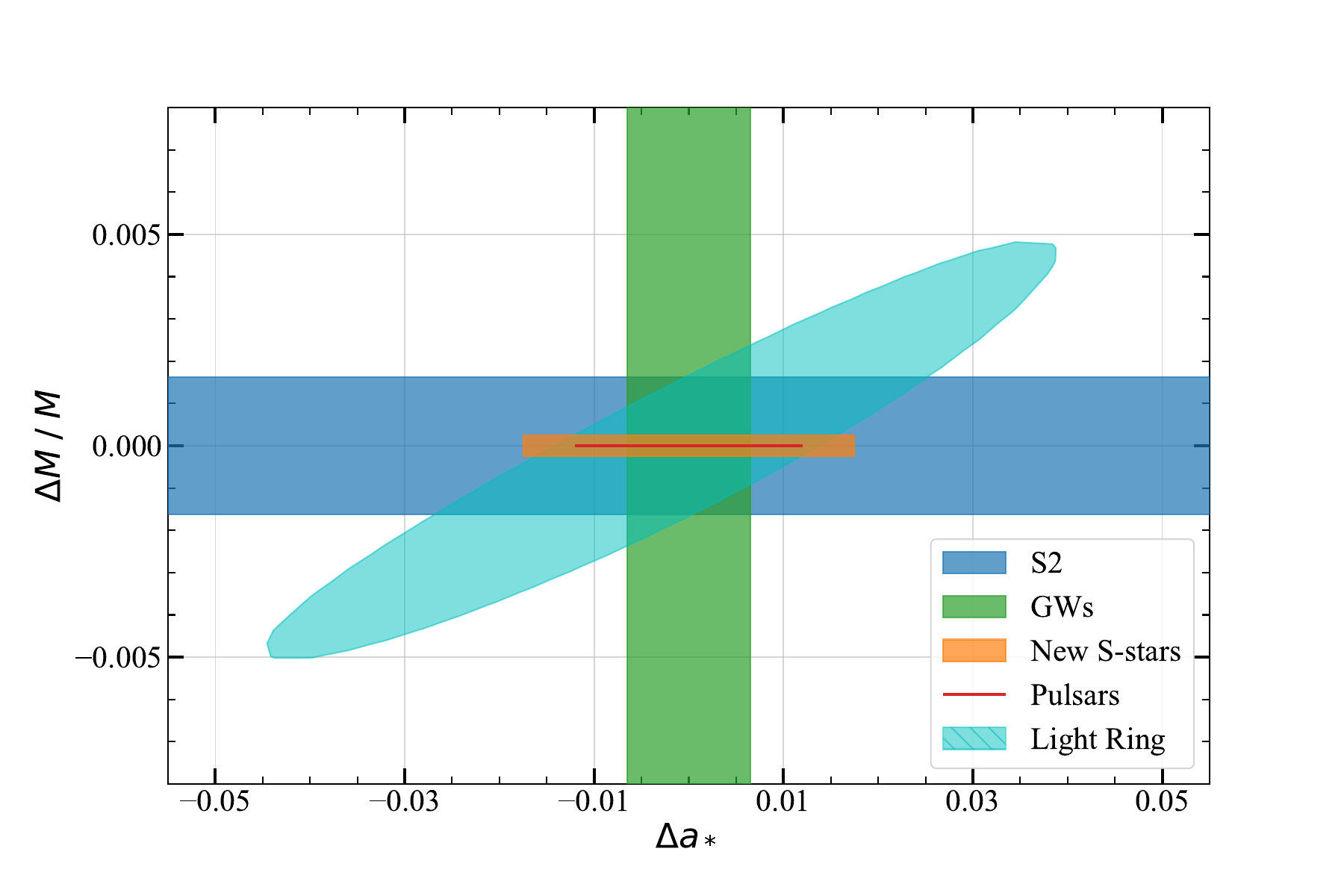}
\caption{Fractional error in the mass of Sgr A* ($\Delta M/M$ with $\Delta M$ denoting 1-$\sigma$ error in M) vs. 1-$\sigma$ error in the dimensionless spin of  Sgr A* ($\Delta a_*$) achieved from various observations. 
The dimensionless spin parameter is defined as $a_{*}=|\vec{S}|/M^2$~\citep{Kramer:2004hd}.
Observation of the star S2 with GRAVITY places a constraint on $M$ with $\Delta M=1.3\times10^4\,M_{\odot}$~\citep{GRAVITY:2021xju}, shown by the horizontal range in blue. 
Implementing such bounds on mass, GW observations with LISA can achieve $\Delta a_*=0.013$~\citep{Tahura:2022ffs} (vertical range in green). 
The orange bar shows the precision of mass and spin measurements the future GRAVITY instrument can achieve ($\Delta M /M \simeq 0.05\%$ and $\Delta a_*=0.035$), given that S-stars are found closer to Sgr A*~\citep{Psaltis:2015uza, Zhang_2015}. Projected pulsar timing observations can obtain $\Delta a_{*}=0.024$ (horizontal range of the red line), while $\Delta M/M$ is of the order $10^{-6}$~\citep{Psaltis:2015uza}, which is much smaller than the scale of the above figure. 
The $\delta_d/b=0.2\%$ case in Figure~\ref{fig:photon_ring_spin_est} is reproduced here for comparison, as labeled by ``Light Ring''. Future pulsar timing and GRAVITY experiments have the potential to provide the most precise spin measurements.
}
\label{fig:MassSpin}
\end{center}
\end{figure}

Extreme mass ratio inspirals (EMRIs) where stellar-mass compact objects orbit SMBHs are an important class of GW source for milliHertz GW detectors \citep{LISA:2017pwj,Pan:2021ksp}.
Stellar-mass BHs and compact stars orbiting around Sgr~A* are potential compact binary sources (referred to here as Sgr~A* EMRIs) that can be explored with the future Laser Interferometer Space Antenna (LISA) GW observatory~\citep{Naoz:2019sjx,Gourgoulhon:2019iyu}. 
Such observations will provide a promising direction for measuring the spin of Sgr~A*. The prospect of such measurements has been studied recently in \citet{Gourgoulhon:2019iyu} and \citet{Tahura:2022ffs} in the case of circularized binaries with orbital separations $\leq10^2~r_{\rm g}$ (here $r_{\rm g}$ denotes the gravitational radius of Sgr~A*; see also \citet{Yang:2022ror} for an analysis with brown dwarfs), which are well motivated by several formation scenarios~\citep{Emami:2019uty,Emami:2019mzi,Pan:2021lyw,Pan:2021xhv}. 
As the location and distance of Sgr~A* from the Solar System are known, the waveform template bank for such systems requires fewer parameters, leading to a threshold SNR of detection ($\sim 10$) which is much smaller than a typical threshold ($\sim 20$) for EMRIs detectable by LISA~\citep{MockLISADataChallengeTaskForce:2009wir}. 
Fisher analyses with Monte-Carlo samplings of the direction and magnitude of the spin of Sgr~A* suggest that the spin can be measured within $\sim 2\%$ uncertainty. 
Such precision may thus be better than that achievable via future S-star and pulsar observations, and a comparison among them is presented in Figure~\ref{fig:MassSpin}. 
Furthermore, one can estimate the direction of the spin of Sgr~A* with similar accuracy from GW observations with LISA~\citep{Tahura:2022ffs}.

There are various issues worth noting in this scenario of spin measurement.
First of all, it relies crucially on the abundance of stellar-mass BHs near Sgr A* and whether they are massive enough to produce signals above the threshold SNR. 
In addition, distributed dark matter may create an orbital precession degenerate to that of the spin-induced one (see also Sec.~\ref{sec:Ultralight_fields}). 
However, according to analysis in \citet{Heissel:2021pcw}, these effects will be easily distinguishable via observations of the star S2.  
Finally, eccentric Sgr A* EMRIs can be generated in the mass segregation scenario ~\citep{Linial:2022big,Emami:2019mzi,Emami:2019uty,Alexander:2008tq,binney2011galactic}, for which the higher harmonics of orbital frequency also contribute to the gravitational waveform. Since the higher harmonics of the frequency should be closer to the most sensitive band of LISA, the event SNRs for such EMRIs are expected to be higher than the circular ones considered in \citet{Tahura:2022ffs}, resulting in more precise parameter estimation. 
However, detailed studies in this regard are yet to be performed.

\subsubsection{Pulsars around the Sgr A* black hole}
Timing of pulsars near Sgr~A* provides an alternative way to
probe the spacetime geometry.
In particular, such observations are in
principle capable of measuring the three lowest-order
moments of the BH, namely its mass ($M$), spin vector ($\vec{S}$), and quadrupole moment ($Q$) with high precision~\citep{Kramer:2004hd, Bower:2018mta, 2019BAAS...51c.438B,DellaMonica:2023ydm}. 
Similar to measurements of BH shadows, observations of pulsars can be used to test the Cosmic Censorship Conjecture, the uniqueness properties of BHs, and modified gravity~\citep{Liu:2011ae, Psaltis:2015uza, Dong:2022zvh}.

Pulsar timing is essentially a ranging experiment, which measures the projected
pulsar orbit along the line of sight. Although until now no such pulsars
have been discovered in the vicinity of Sgr~A*, targeted searches are ongoing.
Population studies and the expected sensitivity and range of new telescopes, e.g., the Square Kilometre Array~\citep[SKA;][]{2012ApJ...753..108W, Shao:2014wja, Goddi:2016qax, Weltman:2018zrl} demonstrate that one might discover such a pulsar, although others have been cautious \citep{2014ApJ...783L...7D}. 
If a pulsar is
discovered in a close orbit around Sgr A*, the orbital dynamics are determined by the BH spacetime, and/or in
the sequence of the post-Newtonian expansion, the mass, spin, and quadrupole moments that show up in the equation of motion at different post-Newtonian orders. 
Different values of these parameters predict different evolution behaviours of the orbit, which in turn leave traces in the pulsar timing data. 
Therefore, with a dedicated experiment, the multipole moments of Sgr A*'s spacetime geometry can be extracted with precision.

Figure~\ref{fig:MassSpin} shows an example from simulations reported by \citet{Psaltis:2015uza}. 
An eccentric orbit with $e=0.8$ and an orbital period $P_b = 0.5~{\rm yr}$ is assumed. 
In order to account for the external perturbations from matter around Sgr A*, only data around three passages of
periapsis are used.
Within such a scenario the mass and spin can already
be constrained with sub-percent-level accuracy. 
The degenerate directions of these constraints are very different from other measurements with the S-stars and BH
shadows. 
Therefore, a combination of these with pulsar data would reduce the uncertainty in BH parameter measurements, although the precision of the pulsar measurement trumps all other experiments.
However, one must emphasize that the different independent methods will serve as important cross-validation of each other, and finding any discrepancy between methods would have important consequences.

\subsubsection{Quasi-periodic oscillations}

QPOs may probe the strong field regime of BHs as their timescales correspond roughly to those of matter orbiting in the innermost regions of the accretion disc, assuming that the cause of emission variations is due to motions in the gravitational field and not due to quasi-periodic heating mechanisms \citep[e.g.,][]{Kato:2010,Dolence:2012,Shcherbakov:2013,Dokuchaev:2014,Miyoshi:2011,Brink:2015}.

For sources suitable for dynamical imaging by the ngEHT, namely Sgr~A*, QPOs have been reported in the radio band on timescales of the order of tens of minutes, usually falling in the range of time periods $T \sim 17-57$ mins \citep{Miyoshi:2011,Shcherbakov:2013,Dokuchaev:2014} and occur roughly in integer ratios \citep{Miyoshi:2011,Brink:2015}. However, even for well studied sources such as Sgr~A*, the inference of BH spin from QPO measurements has not produced consistent results.
\citet{Shcherbakov:2013} assumed a model spin to be $a_*=0.9375$ which predicts the QPO with $T \sim 35$ mins and is different from the results of \citet{Dolence:2012}, who assumed the same BH spin. 
\citet{Dokuchaev:2014} attempted to explain the QPOs as light curves of orbiting hot spots on nearly circular orbits. 
By analysing the observed QPOs of periods 11.5 mins and 19 mins, the spin was inferred to be $a_*\approx 0.65$, which is consistent with the (broad) range inferred from mm-VLBI \citep{Broderick:2009}. 
Dokuchaev's model associated the $T \sim 11.5$ mins QPO with the period of rotation of the BH horizon and the $T \sim 19$ mins to the latitudinal oscillation period of hot spots \citep{Dokuchaev:2014} moving on nearly circular orbits. However, there are certain limitations on sensitivity of orbiting hotspots to spin measurements \citep{Matsumoto:2020,Gelles:2021}. Nevertheless, due to multi-wavelength observations of QPOs \citep{Dolence:2012}, inferring spin from QPOs modelled in terms of hotspot motion seems a viable avenue to explore. 
 The recent EHT results of Sgr A* \citep{EventHorizonTelescope:2022xnr} showed that observations were consistent with models that had spins of $a_*=0.5$ and $a_*=0.94$ and these are rather close to the QPO-based spin inference discussed above.

The importance of hot spots for probing the spacetime and its role in determining spin has already discussed in Sec.~\ref{hotspot}. 
Herein, we note that the QPO frequencies based on modelling of hot spots discussed in \citet{Dokuchaev:2014} are independent of the astrophysical model, hence it is possible that BH spin measurements can be obtained which are agnostic to at least some astrophysical uncertainties.

In a similar vein, if we further align with the hot spot model for studying QPOs, as has been done across the multi-wavelength spectrum \citep{Schnittman:2004,Schnittman:2005,Zamaninasab:2008,Johanssen:2011,Dolence:2012}, the quantity \textit{in radio astronomy} which is sensitive to modelling variability in these hotspots is the closure phases \citep{Doeleman:2009}, which can show periodicity over several cycles. 
In particular, the periodicity of orbital hotspots manifests in closure phases at both, 230~GHz and 345~GHz \citep{Doeleman:2009}. 
As was also noted in \citet{Doeleman:2009}, detection of periodicity is enhanced by the addition of more telescopes in the western hemisphere and, comparatively, increase in bandwidth is of lesser importance.
On the other hand, increased bandwidth is desirable for studying polarisation of the source \citep{Doeleman:2009a,Doeleman:2009}. 
Thus, the possibility of measuring spins from QPOs can directly inform instrumentation requirements for the ngEHT and combining the observations at 230~GHz and 345~GHz can potentially provide stronger constraints on BH spins of astrophysical sources such as Sgr~A*.

\subsubsection{Near-IR flares and orbiting hot spots}

 Near-IR observations with adaptive optics resolution at the 8m--10m-class telescopes revealed quasi-periodic light curves of Sgr~A* \citep{Genzel:2003as,Hamaus:2008yw}, although the number of cycles observed was too low to firmly conclude whether a periodic process was truly causing the variations \citep{Witzel:2012up}. If true, the oscillation period obviously sets a lower bound on the spin.

The situation has now become much more compelling with the discovery of astrometric loops during NIR flares observed with GRAVITY \citep{GRAVITY:2020gka}. The data directly show that hot spots revolve clockwise around Sgr~A* in a near-Keplerian motion at radii of around $8~r_{\rm g}$, with a close-to face-on geometry. 
Strong support for this picture comes from simultaneous polarimetry of the near-IR flares. These flares show loops in the Q-U plane, with the same revolution time. This can be explained by a poloidal magnetic field geometry and the low inclination angle of the Sgr~A* system.

Whether the flares (and other quasi-periodic features) can actually be used to probe the spacetime around Sgr~A* is not yet clear. Simulations by \cite{Ressler:2018yhi} show that the geometry of the inner accretion disc, along which the hot spot motions, are governed by the influx of angular momentum from the incoming material, i.e., stellar winds of massive stars in the case of Sgr~A*.
The orientation of the mean flares' angular momentum vector is consistent with that of the disk of massive young stars moving clockwise at radii between $1$--$10$ arcseconds \citep{Paumard:2006im,Lu:2008iz,vonFellenberg:2022lyo}.

%% file: Ultralight_fields.tex
\section{Searching for ultralight fields with the ngEHT}\label{sec:Ultralight_fields}
Shortly after Peccei and Quinn proposed a resolution of the strong CP
problem~\citep{Peccei:1977hh,Peccei:1977ur},{\it i.e.}, the puzzling
smallness of the CP violating parameter in QCD, it was realized that
it would lead to the appearance of a light pseudo-scalar, the
``axion''~\citep{Weinberg:1977ma,Wilczek:1977pj}. Laboratory and
astrophysical data constrain the axion to be an ``ultralight boson''
with a mass below the eV scale~\citep{ParticleDataGroup:2022pth}. 
Cosmological constraints imply a lower bound on the typical QCD axion
mass~\citep{Preskill:1982cy,Abbott:1982af,Dine:1982ah}, close to which
it could be a viable dark matter candidate. 
Similar ultralight bosons have since been proposed in a plethora of beyond Standard
Model theories~\citep{Svrcek:2006yi,Abel:2008ai,Arvanitaki:2009fg,Goodsell:2009xc,Marsh:2015xka,Freitas:2021cfi}.
Such particles, like the QCD axion, can be compelling dark matter
candidates, but are extremely hard to detect or exclude with usual particle
detectors. Their low mass makes them a special type of dark matter
candidate with de Broglie wavelengths that can be as large as a
galaxy~\citep{Hu:2000ke,Robles:2012uy,Schive:2014dra,Hui:2016ltb,Ferreira:2020fam}.
This feature can lead to interesting unique properties when compared to
other dark matter candidates. For example, ultralight bosons can form
solitonic structures where the balance between gravitation and ``quantum'' pressure leads to a flat core profile in the inner region of galaxies. This mechanism, proposed to address small-scale puzzles in the observations of galaxies~\citep{Hu:2000ke,Robles:2012uy,Schive:2014dra,Hui:2016ltb,Broadhurst:2019fsl,DeMartino:2018zkx,Pozo:2020fft}, also provides a lower limit on the mass of dark matter~\citep{Bar:2018acw}, comparable to limits from cosmology~\citep[see, e.g.,][]{Kobayashi:2017jcf}.
Depending on the masses and couplings of the bosons, such self-gravitating structures or ``boson stars'' could even mimic BHs~\citep{Liebling:2012fv}.

Quite remarkably, very light bosonic particles can also dramatically influence the dynamics of rotating BHs, specially when the BH horizon scale is of the order of the Compton wavelength of the boson. Then, rotating BHs can become unstable against the production of light bosonic particles due to a energy-extraction process known as BH superradiance, akin to the Penrose process and the Blandford-Znajek process~\citep{Penrose:1971uk,ZS,Blandford:1977ds,Brito:2015oca}. This process drives an exponential growth of the field in the BH exterior, while spinning the BH down, forming a dense bound state or ``cloud'' around the rotating BH. 
This mechanism leads to several observable consequences, affecting the mass and spin of SMBHs, as well as their images made possible by the EHT and the ngEHT instruments.

The goal of this section is to discuss how the EHT and the ngEHT can be used to study the existence of ultralight bosons. We will consider three main observables: in Sec.~\ref{sec:spin} we first discuss how precise measurements of the mass and spin of astrophysical BHs allow the exclusion of minimally coupled bosons. 
Section~\ref{sec:direct} is devoted to the discussion of several direct gravitational signatures that can be used to constrain or detect the existence of ultralight bosons. 
Namely, in Sec.~\ref{sec:superevol} we discuss how the ngEHT could detect ultralight bosons through the direct observation of the long-term evolution of the superradiant instability and then present examples where the existence of ultralight bosons could lead to geometries which can differ from Kerr.
In particular, we discuss the so-called Kerr BHs with bosonic hair in Sec.~\ref{sec:BHwithHair}, whereas in Sec.~\ref{sec:GAPRA} we discuss how the oscillating metric perturbations induced by superradiant clouds made of real bosons could be detected using the photon ring autocorrelation. 
We then also mention compact bosonic self-gravitating configurations in Sec.~\ref{sec:bosonstar} which are discussed in more detail in Sec.~\ref{sec:central-brightness-depression}.
Section~\ref{sec:stars} discusses how the motion of S-stars around Sgr A* can constrain the total mass of the cloud in a certain mass window. 
In Sec.~\ref{sec:polarization} we then discuss how for axion-like particles which interact with photons the formation of an axion cloud can lead to periodic oscillations of the orientation of the linear polarization of photons.
Finally, in Sec.~\ref{sec:todo} we close with some open issues.

\subsection{The theory}\label{sec:theory}
Our starting point is the generic Lagrangian density for massive, minimally coupled bosons:
\begin{equation}
\begin{aligned}
\mathcal{L}= \frac{R}{16\pi}
&- \frac{1}{2} \nabla^\mu a \nabla_\mu a - V(a)- \frac{1}{4}B^{\mu\nu}B_{\mu\nu} - \frac{1}{2}m_V^2X_\nu X^\nu + \mathcal{L}_{\rm EH}(H)\\
&- \frac{m_T^2}{4}\left(H^{\mu\nu}H_{\mu\nu} - H^2\right)\,,\label{Lphoton}
\end{aligned}
\end{equation}
where $\nabla_{\mu}$ is the covariant derivative. 
For axion-like particles $a$, the potential is $V(a) = m_a^2 f_a^2\left[1- \cos (a/f_a)\right]$, where $m_a$ is the axion mass, $f_a$ is the decay constant characterizing some high energy scale. 
For a small self-interaction, i.e.\ $a/f_a\ll 1$, the potential $V(a) \approx m_a^2 a^2/2$ simply becomes a mass term. 

The theory above also includes a possible new vector field $X^\mu$ of mass $m_V$ and corresponding tensor $B^{\mu\nu} \equiv \nabla^\mu X^\nu - \nabla^\nu X^\mu$, and a tensor field
$H^{\mu\nu}$ of mass $m_T$. The theory above only considers the Fierz-Pauli term for massive tensors, and describes a broad class of nonlinear theories expanded around vacuum Kerr background~\citep{Brito:2013wya,Brito:2015oca,Brito:2020lup}.  From now on, we will collectively refer to the mass parameter of these new fields as $m_b=(m_a,m_V,m_T)$. Note that the physical mass of the boson is $\hbar m_b$.

The corresponding equation of motion for such massive fields can be studied in a fixed Kerr background, as long as backreaction effects are small, which is the case for bosonic dark matter or for bosons produced from superradiance alone~\citep[see, e.g.,][]{Brito:2014wla,Herdeiro:2017phl}.  
A convenient procedure is to separate angular variables~\footnote{For generic BH spins, separation of variables has only been achieved for massive scalar and vector fields~\citep{Brill:1972xj,Dolan:2007mj,Frolov:2018ezx,Dolan:2018dqv}. For massive tensors no such separation is known, but the full problem, involving a system of elliptic partial differential equations, has recently been solved in~\cite{Dias:2023ynv}.}, expanding the wavefunction $\Psi$ describing a boson of spin $s$ in a given set of angular functions carrying two indices. 
One of them, $l$, specifies the total angular momentum.
The other, namely the azimuthal number, $m$, is associated with the projection of the angular momentum along the $z-$axis~\citep{Brito:2015oca}. 
A Fourier analysis of the resulting ``radial'' equation, i.e., assuming a general time dependence $\sim e^{-i\omega t}$ results in a single ordinary differential equation, which is in fact an eigenvalue problem for $\omega$.

For any bosonic field $\Psi$, there is an unstable mode which grows exponentially in time as $\Psi \sim e^{\omega_I t}$, where the instability rates, $\omega_I$, can be found in~\citet{Brito:2015oca} and have the general dependence~\citep{Detweiler:1980uk,Cardoso:2005vk,Dolan:2007mj,Pani:2012bp,Baryakhtar:2017ngi,Cardoso:2018tly,Dolan:2018dqv,Baumann:2019eav,Brito:2020lup}:
\begin{equation}
\label{eq:timescale}
\omega_I \sim \Upsilon_{Sl m}\alpha^{4l+5+S}\left(m\,\Omega_{\rm BH} - m_b\right)\,,
\end{equation}
where $S=-s,\,-s+1,\,\ldots,\,s-1,\, s$ is the spin projection along the $z$-axis and the $\Upsilon_{sl m}$ coefficients can be found in~\citet{Brito:2015oca}. In the above, $\Omega_{\rm BH}$ is the angular velocity of the BH and $\alpha \equiv M m_b$ is the gravitational coupling. The only exception to the scaling~\eqref{eq:timescale} is a special dipole mode that exists for massive spin-2 fields~\citep{Brito:2013wya,Dias:2023ynv}. This special mode scales as $\omega^{\rm dipole}_I \propto \alpha^3 \left(\Omega_{\rm BH}-\omega_R\right)$, where $\omega_R \sim 0.73~m_b$, and has an instability timescale much shorter that any other superradiant mode. Precise values of $\omega^{\rm dipole}_I$ for generic BH spins and values of $\alpha$ can be found in~\cite{Dias:2023ynv}. 

The instability occurs only for large enough rotations, a clear sign of its superradiant nature. 
Owing to the results above, the modes of a boson field around a spinning BH are amplified if the BH angular velocity at the horizon is larger than the angular phase velocity of the incident wave, i.e., $m\,\Omega_{\rm BH} > m_b$. 
These modes populate the BH over a volume with radius comparable to the Compton wavelength $1/m_b$.

Superradiance is most effective for highly spinning BHs and when the boson Compton wavelength is comparable to the BH gravitational radius $r_g\equiv M$~\citep{Brito:2015oca}.
For a minimally coupled scalar, the growth is dominated by a dipolar mode which can grow on a timescale:
\begin{equation}
\label{eq:inst_scale}
\tau=\frac{1}{\omega_I}\sim \frac{M}{10^6 M_{\odot}}\,\rm yr\, ,
\end{equation}
where $M$ the BH mass. This is the shortest possible instability timescale and requires large BH spins and a gravitational coupling $\alpha\sim 0.42$, or:
\begin{equation}
m_b\sim 5.6\times 10^{-17}\left(\frac{10^6 M_\odot}{M}\right)\,{\rm eV}\,.\label{eq:mass_super}
\end{equation}
The instability deposits the BH rotational energy into the boson field, which forms a bosonic structure outside the horizon, co-rotating with the BH, and with a length scale $\sim 1/(Mm_b^2)$.

We should note that additional interaction terms can be added to action~\eqref{Lphoton}, such as couplings of the axion to photons. In the presence of self-interactions or additional couplings, the exponential growth due to the superradiant instability could be terminated once the field value becomes sufficiently large~\citep{Yoshino:2012kn,Fukuda:2019ewf, Baryakhtar:2020gao,Ikeda:2018nhb,East:2022ppo,Omiya:2022gwu,Spieksma:2023vwl,Chen:2023vkq}. On the other hand, the presence of additional interactions can also lead to unique signatures, as we discuss in more detail in Sec.~\ref{sec:polarization}.

\subsection{Constraints from black-hole spin measurements}\label{sec:spin}
\begin{figure*}
    \centering
    \includegraphics[width=0.8\textwidth]{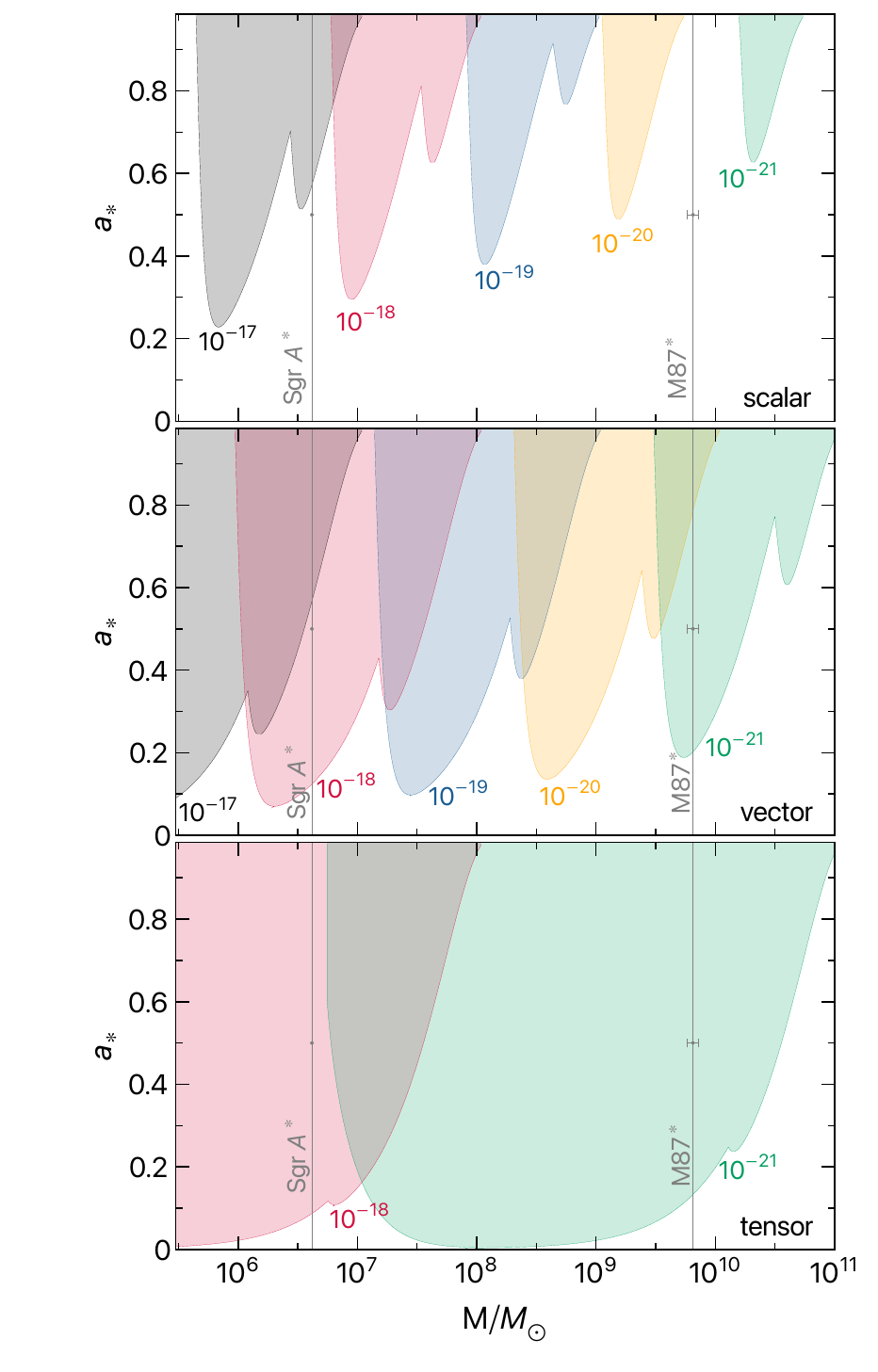}
   \caption{Exclusion regions in the BH spin-mass diagram obtained from the superradiant instability of Kerr BHs against massive bosonic fields for the two most unstable modes.
The top, middle, and bottom panels refer to scalar, vector and tensor fields, respectively. 
For each mass of the field (reported in units of eV), the separatrix corresponds to an instability time scale equal to the Salpeter time $\tau_{\rm Salpeter} \approx 4.5\times 10^7 {\rm\, yr\,}$, i.e., inside each colored region the instability timescale would be shorter than $\tau_{\rm Salpeter}$. For illustration we consider bosons with masses ranging from $10^{-21}\,$eV to $10^{-17}\,$eV. For the massive tensor case we only show two masses to minimize clutter in the figure.
The gray lines and error bars denote the measured mass of Sgr A*~\citep{GRAVITY:2021xju} and M87*~\citep{EventHorizonTelescope:2019dse}.
}
\label{fig:BHspin}
\end{figure*}

From the discussion of the previous section we can infer for example that a BH of mass $\sim 10^{10} M_{\odot}$ like M87* can be superradiantly unstable for ultralight bosons of masses $\sim 10^{-21}\,$eV from Equation~\eqref{eq:mass_super}.
This ultralight boson mass is close to the range of ``fuzzy'' dark matter \citep{Davoudiasl:2019nlo}, which is relevant for  dynamics on galactic scales~\citep{Ferreira:2020fam}.

The instability time scale can be extremely short compared to typical astrophysical time scales and, therefore, relevant for astrophysical BHs (cf.~Equation~\ref{eq:inst_scale}).  
The instability removes rotational energy from astrophysical BHs, and hence a robust observable for this effect concerns the BH ``Regge'' plane: if a bosonic field of a certain mass exists, BHs with certain masses/spins will slow down on short timescales and should not be seen, up to observational uncertainties. 
For very weakly self-interacting bosons, the process depends primarily on the mass and spin of both the BH and the fundamental boson.
By requiring the predicted instability timescale to be smaller than the typical accretion timescale (which tends to instead spin-up the BH), one can then draw regions in the parameter space where highly spinning BHs should not reside, if bosons within the appropriate mass range exist in nature~\citep{Arvanitaki:2009fg,Brito:2013wya,Brito:2014wla,Davoudiasl:2019nlo}. 
This is illustrated in Figure~\ref{fig:BHspin} for scalar, vector and tensor fields, where we show exclusion regions in the BH spin-mass diagram for bosons with masses ranging from $10^{-21}~$eV to $10^{-17}~$eV.

\begin{figure}[htb!]
    \centering
    \includegraphics[width=0.9\textwidth]{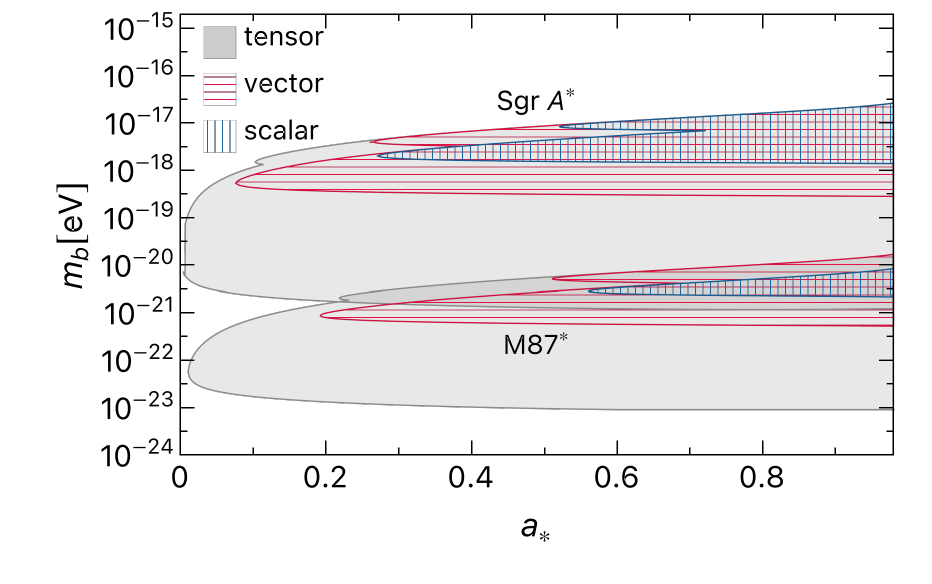}
    \caption{Exclusion regions for the two most unstable modes as a function of the spin of Sgr A* and M87* when fixing Sgr A*'s mass to $M \simeq 4\times 10^{6}M_{\odot}$~\citep{GRAVITY:2021xju} and M87*'s mass to $M \simeq 6.5\times 10^{9}M_{\odot}$~\citep{EventHorizonTelescope:2019dse}. As in Figure~\ref{fig:BHspin}, the separatrices correspond to an instability time scale equal to the Salpeter time $\tau_{\rm Salpeter} \simeq 4.5\times 10^7 ~{\rm yr}$.
    }
    \label{fig:BHspin_M87}
\end{figure}
Such a spin-down effect allows use of BH spin measurements to constrain the existence of ultralight bosons (see~\citet{Brito:2015oca} for a review of current constraints). Measurements of BH mass and spin with ngEHT are discussed in detail in Sec.~\ref{sec:Mass_spin}.
Given that the BH mass is generally much better constrained than the BH spin, the uncertainty in the measurement of the BH mass has a minimal impact on these constraints.
To illustrate the possible constraints that could be obtained from Sgr A* and M87* given their measured mass, in Figure~\ref{fig:BHspin_M87} we show the exclusion regions as a function of the BH spin. In particular, Figure~\ref{fig:BHspin_M87} shows that obtaining a lower limit on its spin is enough to place some constraints on a range of boson masses~\citep[with the specific boson mass range constraint dependent on the magnitude of the measured BH spin;][]{Davoudiasl:2019nlo}. 
For example, from Figure~\ref{fig:BHspin_M87} one can see that for a conservative spin measurement of M87* with $a_{*}\gtrsim 0.5$ \citep{Cruz-Osorio:2021cob}, one could exclude scalar fields with masses around $\sim 3\times 10^{-21}\,$eV. On the other hand, a non-zero spin for Sgr A* would constrain bosons with masses around $\sim 3\times 10^{-18}\,$eV. Given the shorter instability timescale for vector and tensor fields compared to scalar fields, constraints for vector and tensor fields are stronger. This is especially true for tensor fields due to the very short instability timescale of the special dipole mode~\citep{Dias:2023ynv}. For example, measuring the spin of M87* to be $a_{*}\gtrsim 0.2$ would be enough to impose constraints for vector fields with masses $\sim 10^{-21}\,$eV, whereas for tensor fields the measurement of a non-zero spin in Sgr A* and M87* would be enough to exclude massive tensor fields with masses in the whole range from $\sim 10^{-23}\,$eV up to $\sim 10^{-17}\,$eV~\citep{Dias:2023ynv}. Boson masses of this order-of-magnitude are particularly interesting since this mass range is so far mostly unconstrained~\citep{Brito:2015oca,Ferreira:2020fam}.

A key ingredient in the spin-down calculation is the timescale on which astrophysical processes spin up a BH. If a BH is accreting at the Eddington limit, the characteristic timescale for significant spin-up is largely independent of the BH mass and is given by the Salpeter time $\tau_{\rm Salpeter}\approx4.5\times 10^7$ yr~\citep{Shankar:2007zg}. Super-Eddington accretion could reduce this timescale further. On the other hand, EHT observations have indicated that both M87* and Sgr A* are currently significantly sub-Eddington $\dot M/\dot M_{\rm Edd}\sim2\times10^{-5}$ and $\dot M/\dot M_{\rm Edd}\sim10^{-9}$, respectively~\citep{Kuo:2014pqa,EventHorizonTelescope:2019pgp,EventHorizonTelescope:2022xnr}, which would increase the spin-up times accordingly. The impact of the time scale shifts the location of the constraints to the lower mass side\footnote{For sufficiently short astrophysical spin-up timescales no constraints on ultralight bosons can be derived for any measured spin. Given the expected limits from the Salpeter time, it is expected that constraints can be derived assuming sufficiently large spins are measured.}. As the spin up time increases, lighter masses can be ruled out. The power law dependence on the timescale, however, is weak and has an exponent that can be obtained from Equation~\eqref{eq:timescale}.

Finally, we point out that while ruling out ultralight bosons with spin measurements of BHs is relatively straightforward, making a discovery of ultralight bosons via this effect is much more challenging, although technically possible. If one had a reasonable estimate of the initial spin distribution of BHs or believed that the spins of BHs were generally large, and if one measured the masses and spins of a large population of BHs, then one would see a characteristic dependence between the maximum spin of a BH and its mass if an ultralight boson existed in the relevant mass range~\citep[see, e.g.,][where this possibility was studied in the context of BH spin measurements obtained via GW observations of stellar-origin and massive BH binaries]{Arvanitaki:2016qwi,Brito:2017zvb,Ng:2019jsx}. In practice, however, the discovery of ultralight bosons using this method would likely be prone to very large modelling uncertainties.

\subsection{Direct gravitational effects}\label{sec:direct}

\subsubsection{Observing the superradiant instability evolution}\label{sec:superevol}

We now turn to considering the possibility that superradiant evolution within a given timescale may be observable~\citep{Roy:2019esk,Creci:2020mfg,Roy:2021uye,Chen:2022nbb}. 
VLBI techniques have recently allowed the observation of the image of the dark shadow surrounding M87*~\citep{EventHorizonTelescope:2019dse} and Sgr A*~\citep{EventHorizonTelescope:2022xnr} by a global network of radio telescopes. This has opened up the possibility to test physics in the strong gravity regime via BH imaging. As mentioned above, one such interesting phenomenon in strong gravity is the exponential growth of ultralight bosons near a BH via superradiance. This leads to rapid extraction of spin and energy from the BH, which could be detectable as a change in the shadow of the BH image.

To further discuss this phenomenon, we consider the evolution of minimally coupled, massive bosonic fields, as in Equation~\eqref{Lphoton}. We assume that the backreaction due to the presence of the boson fields is negligible on the Kerr background and that the propagation of the boson fields acts as a perturbation over the background Kerr metric. As was mentioned, such a hypothesis is justified by the fact that the total mass of the boson cloud adds up to a fraction of the BH mass that is distributed over a large volume $\sim M^3/m_b^6$, with $m_b$ the mass of the generic boson field, therefore exerting negligible distortion on the background spacetime~\citep{Brito:2015oca,Roy:2021uye}.

\begin{figure}[htb!]
\centering
    \includegraphics[width=0.43\textwidth]{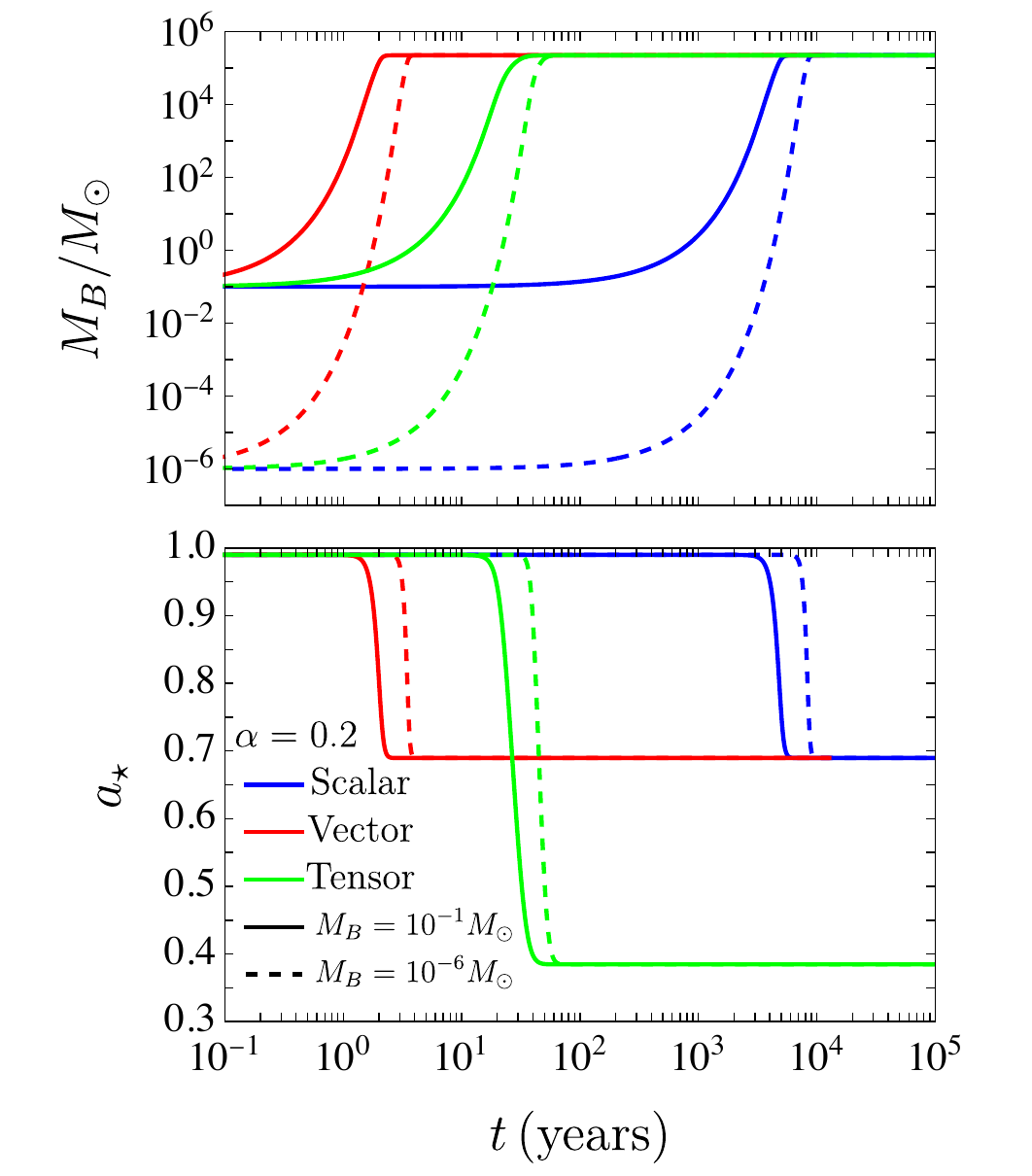}
    \includegraphics[width=0.47\textwidth]{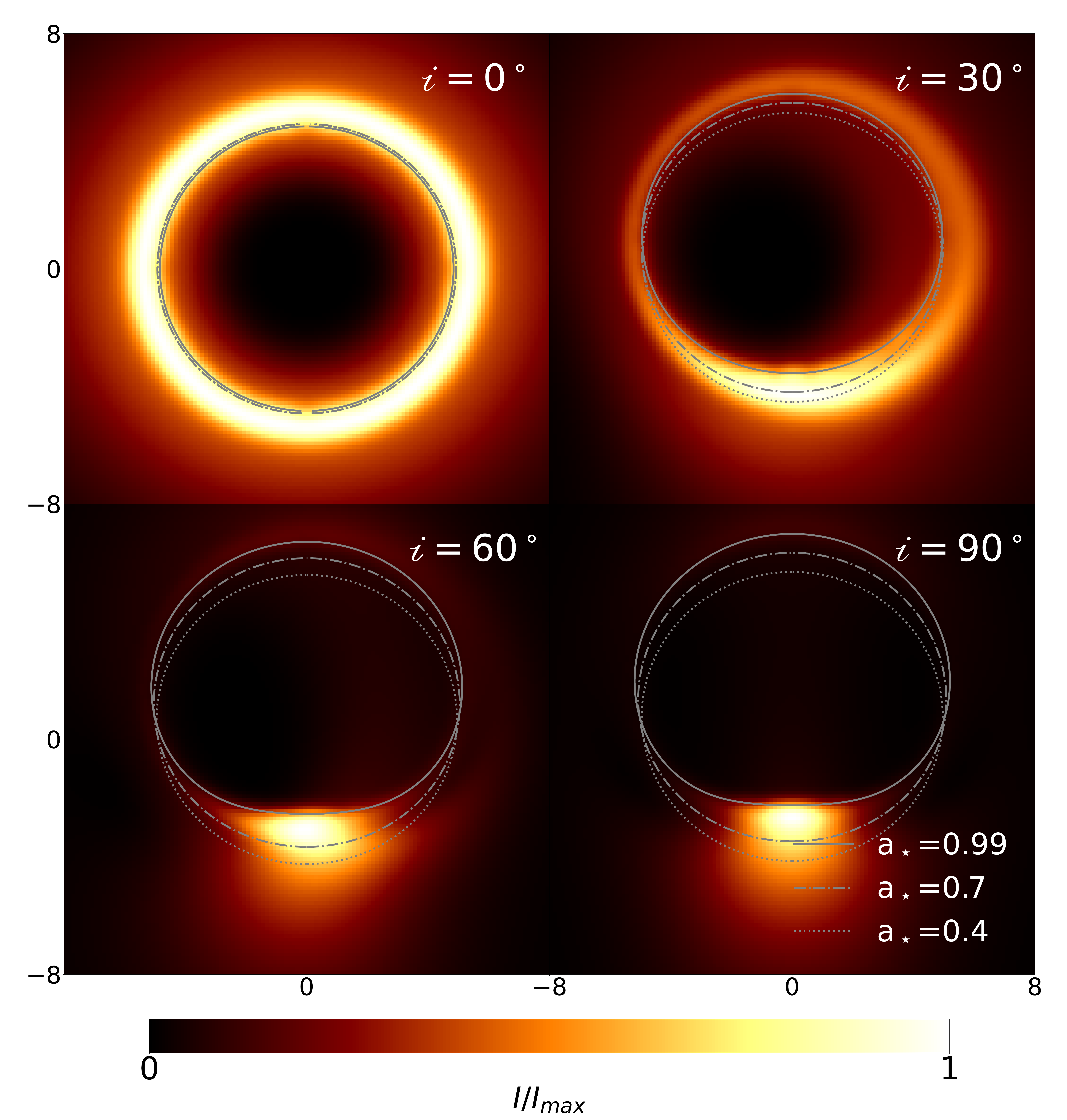}
    \caption{{Left panel:} the evolution of the mass of the boson cloud $M_B$ (top) and of the dimensionless BH spin parameter $a_{\scriptscriptstyle *}$ (bottom),
    as a function of time in years for (initial) $\alpha = 0.2$ and for different choices of the bosonic nature: scalar field (red), vector field (blue), tensor field (green), with the set of quantum numbers as given in the text. The initial mass of the boson cloud is $M_B = 10^{-1}\,M_\odot$ (solid line) and $M_B = 10^{-6}\,M_\odot$ (dashed line). We assume an initial BH mass $M = 4.3\times 10^6\,M_\odot$ and initial spin $a_{*} = 0.99$.
    {Right panel:} evolution of shadow contours (gray lines) during different stages of superradiance for a vector with initial $\alpha = 0.2$ and a BH viewed at different inclination angles. The background depicts the intensity map with an initial value of $a_{\scriptscriptstyle *} = 0.99$.
    The coordinate origin is taken to be the BH location and the axes are specified in units of the initial gravitational radius.}
    \label{fig:BHspin_M87se}
\end{figure}

To better assess this, we have solved the equations describing the evolution of the occupancy number of the boson cloud, $N(t)=M_B/m_b$, where $M_B$ is the mass of the cloud, the dimensionless BH spin parameter $a_{\scriptscriptstyle *}$ due to superradiance, and the mass change of the BH. We have neglected the change in the mass of the BH due to accretion, since we work on timescales that are shorter than the Salpeter timescale.
The exponent of $\alpha$ in Equation~\eqref{eq:timescale} predicts that some of the modes experiencing the fastest superradiant evolution have $l=1,\, S=0$ (scalar field), $l = 1,\, S=-1$ (vector field), or $l = 2,\, S=-2$ (quadrupole mode of a tensor field). For illustration purposes we do not include the evolution of the massive tensor field dipolar mode, which has a much shorter instability timescale~\citep{Dias:2023ynv}. Other than this special dipolar mode, the spectrum of the unstable modes for a massive vector field generally lead to the shortest timescales~\citep{Pani:2012bp,Baryakhtar:2017ngi,Cardoso:2018tly,Dolan:2018dqv}. For example, the timescale for the superradiance evolution associated with a scalar field with $\alpha = 0.2$ around a BH of mass $M\sim 10^{6}~M_\odot$ is $t \sim \mathcal{O}(10^4~{\rm years})$, while repeating the estimate for a vector field with the same parameters reduces the timescale to $t \sim \mathcal{O}(10{\rm~years})$.

Figure~\ref{fig:BHspin_M87se} shows the evolution of the mass of the boson cloud $M_B$ (top left) and $a_{\scriptscriptstyle *}$ (bottom left) due to superradiance caused by a scalar field (red), a vector field (blue) and a quadrupole mode of the tensor field (green), assuming the mass of Sgr A* as $M = 4.3\times 10^6 ~M_\odot$ with initial spin parameter $a_{*} = 0.99$, for an initial coupling $\alpha = 0.2$, corresponding to the mass $m_b \approx 6 \times 10^{-18}$~eV, for two different choices of the initial mass of the boson cloud: $M_B = 10^{-1}~M_\odot$ (solid line) and $M_B = 10^{-6}~M_\odot$ (dashed line). For this example, the change in the BH mass is $\sim 3\%$ for the scalar and vector fields, and $\sim 5\%$ for the tensor field. The initial value of the boson cloud mass only affects the timescale (and not the relative change in the BH mass) since the evolution of both the spin parameter and the BH mass are proportional to the total number of bosons. 

At this point, we note that SMBHs are very old, hence the probability of observing superradiance in one of the observations is vanishingly small, since the instability should have saturated by now. There are, nevertheless, circumstances where this might happen, if, for example, a bosenova suddenly leads to cloud destruction~\citep{Yoshino:2012kn}, and the exponential extraction of the BH spin restarts. A varying boson mass predicted in quintessence models~\citep{Tsujikawa:2013fta} that recently entered the superradiant region is also possible.

Examples of evolution of the shadow contours are shown on the right panel of Figure~\ref{fig:BHspin_M87se}.  
The center of the shadow is clearly separated from the BH located at the origin, with the distance increasing with $a_{\scriptscriptstyle *}$ and the inclination angle $i$. 
Thus the drift of the shadow center, which evolves in the axis perpendicular to the spin projection, can be a smoking gun of superradiant evolution at large $i$~\citep{Chen:2022nbb}. For example, at $i = 60^\circ$, a $10~\mu$as drift is possible for both vector and tensor fields. On the other hand, using photon ring autocorrelation~\citep{Hadar:2020fda}, one can record yearly variations in the azimuthal lapse, $\delta_0$, which is sensitive to the BH spin \citep{Gralla:2019drh} at low $i$. These two ways are thus complementary and both benefit from long-duration observation times.

\subsubsection{Black holes with synchronised bosonic hair}\label{sec:BHwithHair}

The growth of a bosonic field around a Kerr BH due to superradiance may or may not lead to an equilibrium state. If the bosonic field is complex, the BH-bosonic field system can form true stationary configurations: {\it BHs with synchronised bosonic ``hair''}~\citep{Herdeiro:2014goa,Herdeiro:2015gia,Herdeiro:2016tmi}. Such BH equilibrium configurations exist with both scalar and vector (Proca) fields, under a general synchronisation mechanism \citep[see][for more details and generalizations]{Herdeiro:2015tia,Delgado:2019prc,Delgado:2020hwr}.

Focusing on the scalar case, i.e., Equation~\eqref{Lphoton} with $X_{\mu}=0$, $H_{\mu\nu}=0$ and $V(a)=m_a^2|a|^2/2$, but now with the field $a$ complex and $(\nabla a)^2$ also becoming $|\nabla a|^2$, one can construct fully non-linear BH solutions in GR minimally coupled to a complex massive bosonic field $a$.
Although the metric is assumed to be stationary and axially-symmetric, the full solution is not, due to the explicit time dependence assumed in the harmonic ansatz for the scalar $a\sim e^{i(k\varphi-wt)}$, where $k$ and $w$ are respectively the azimuthal harmonic integer and the field frequency~\citep{Herdeiro:2015gia}. This allows one to evade some well-known \textit{no-hair theorems}~\citep{Herdeiro:2015waa}. Although some of the solutions can display quite unusual BH shadow shapes and very distinct gravitational lensing signatures~\citep{Cunha:2015yba, Cunha:2019ikd,Cunha:2018acu,Vincent:2016sjq}, one of the regions in the domain of existence with most potential to describe astrophysical and viable solutions lies in the proximity to the Kerr limit. In that region one may find configurations that might be formed within astrophysical timescales.

The growth timescale of a scalar field due to superradiance is extremely sensitive to the resonance of the BH mass scale $M$ with the Compton wavelength of the ultralight particles. In addition, BHs with bosonic hair are also not absolutely stable and can suffer from their own superradiant instabilities~\citep{Ganchev:2017uuo,Degollado:2018ypf}. 
One possibility
is that M87* started as a Kerr BH and grew scalar hair within an astrophysical timescale (e.g., $\lesssim 0.1\%$ of the Hubble time~\citep{Degollado:2018ypf}), transforming into a BH with bosonic hair state that is effectively stable to its own superradiant instabilities over cosmological timescales~\citep{Ganchev:2017uuo,Degollado:2018ypf}, provided that one restricts to the interval:
\begin{equation}
\label{range}
m_a \, M_{\rm M87}\in [0.1,\,0.3] \qquad \Longrightarrow \qquad m_a \in 1{\rm -}3 \times 10^{-20} ~{\rm eV} \,.
\end{equation}
Within this range, each hairy BH solution can be identified by a two-parameter set of values $\{p,\,M \, m_a\}$, where $p=1-M_H/M$ measures the fraction of the spacetime mass stored in the bosonic hair. 
Here $M_H$ denotes the Komar mass of the horizon and $M$ denotes the total (ADM) mass. 
The parameter, $p$, satisfies $0\leqslant p \leqslant 1$, interpolating between vacuum Kerr BHs in the test limit of $a$ for $(p=0)$, and horizonless boson stars for $(p=1)$. Fully dynamical numerical evolution of complex \textit{vector} fields growing from vacuum Kerr by superradiance~\citep{East:2017ovw} suggest a maximal possible value of $p\sim 9\%$~\citep{Herdeiro:2017phl}. If the process is approximately conservative, then an upper limit of $p \sim 10\%$ should exist regardless of the spin of the bosonic field~\citep{Herdeiro:2021znw}.

The shadow areal radius $\mathcal{R}$ of BHs with bosonic scalar hair can be compared with the Kerr case, for the same total mass $M$. Such a comparison is appropriate if most of the scalar hair is spread over a length scale of $\sim 10~M$, which is typically the case for the hairy BH solutions under consideration here. Under these circumstances, signature effects of the scalar field can be expected to be subdominant on far-away measurements of the total mass $M$, using, e.g., stellar orbital motions over length scales $\gg M$. Assuming $i=17^{\circ}$ for M87*, the relative shadow deviation $\delta \mathcal{R}$ depends very weakly on $M\,m_a$ and is accurately parameterized by a function of $p$ alone~\citep{Cunha:2019ikd}:
\begin{equation}
\hspace{1cm}
\delta \mathcal{R}(p)\equiv 1-\frac{\mathcal{R}_{\rm hairy}}{\mathcal{R}_{\rm Kerr}}\simeq p + p(p-1)A\ ,\qquad\textrm{with}~A\simeq 0.111159 \,.
\label{eq-psi}
\end{equation}
Since $\mathcal{R}_{\rm Kerr}\simeq 19\,\mu$as, the detection of bosonic scalar hair close to the upper limit of $p\sim 10\%$ requires an ngEHT angular resolution close to $1.7\,\mu$as, whereas a finer measurement of $p$ below $1\%$ would require resolutions smaller than $\sim 0.1~\mu$as. Constraining the value of $p$ for M87* via ngEHT observations would directly measure how much mass can exist in a bosonic field in equilibrium with the SMBH.

Naively, one might expect similar results to also hold for BHs with synchronised vector (rather than scalar) hair. However, as was reported recently in~\cite{Sengo:2022jif}, there are some regions in the solution space of BHs with vector hair that could still be compatible with the EHT observations of both M87* and Sgr A*, despite containing a significant portion of the mass stored in the hair, e.g., $p\sim 40\%$. This would include solutions just below $p \sim 29\%$, i.e., the thermodynamic upper limit for configurations that might have grown from Kerr via superradiance. Indeed, as we increase the coupling $M\,m_b$ of the BH solution with vector hair, for a fixed value of $p$ it would become increasingly more difficult for it to be ruled out by EHT (or ngEHT) observations. This is a feature that is in contrast to the scalar case. 
%

\subsubsection{Photon ring astrometry for real bosons}\label{sec:GAPRA}
We next focus on superradiant clouds made of real bosons. The coherently oscillating features of their wavefunctions generate periodic metric perturbations. On long timescales, these metric perturbations dissipate energy from the cloud in the form of potentially detectable gravitational waves, for example, in the LISA band for a cloud outside Sgr A*~\citep{Brito:2015oca}. On the other hand, locally, these metric perturbations also modify the photon geodesics propagating from the BH horizon scale towards us, providing potential signals for EHT and ngEHT~\citep{Chen:2022kzv}.  

More explicitly, metric perturbations around a Kerr background $g_{\mu\nu}^{\rm K}$ can be written as $g_{\mu\nu} \simeq g_{\mu\nu}^{\rm K} + \epsilon \, h_{\mu\nu}$, where $\epsilon \ll 1$ controls the perturbative expansion, and $h_{\mu\nu}$ represents the metric perturbations generated from a bosonic cloud. In this metric, photon geodesics also undergo small perturbations compared to the geodesics in a Kerr background $x_{(0)}^\mu$, i.e.,  $x^\mu \simeq x_{(0)}^\mu + \epsilon \, x_{(1)}^\mu$, where $x_{(1)}^\mu$ is the deviation to the background geodesic calculated using $h_{\mu\nu}$ and the Kerr metric. 

\begin{figure}[htb]
    \centering
      \includegraphics[width=0.52\textwidth]{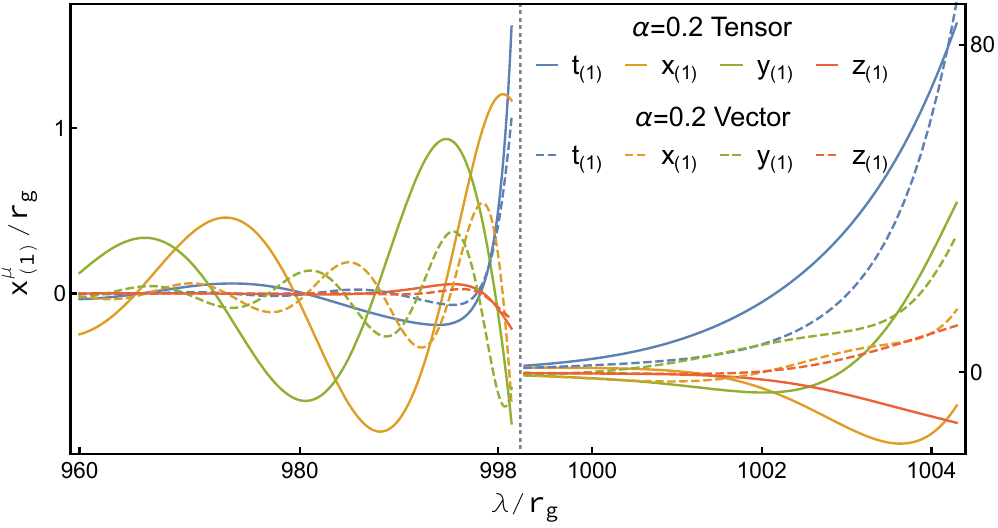}\qquad
     \includegraphics[width=0.4\textwidth]{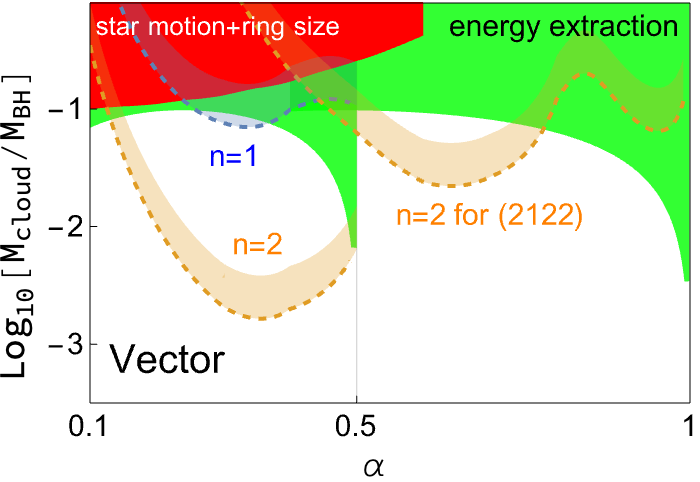}
    \caption{Left panel: examples of deviations from the Kerr background photon geodesics as a function of the affine parameter, $\lambda$, generated by bosonic clouds with $\alpha = 0.2$.
    The initial values are $\lambda_{\rm 0} = 0,\, r_{\rm 0} = 10^3\, r_{\rm g}$, $i=17^\circ$, and $a_* = 0.94$.
    The gray vertical line shows the time at which the unperturbed orbit $x_{(0)}^\mu$ crosses the BH equatorial plane for the first time.
    Right panel: prospects for constraints on the total mass of a vector cloud using photon ring autocorrelations. We show both the ground state $(S,l,m)=(-1,1,1)$ with $\alpha<0.5$ and a higher mode $(S,l,m)=(-1,2,2)$.
    The constraint bands range from a conservative criterium based on ngEHT's spatial resolution $\sim 10~\mu$as to an optimistic criterium based on the intrinsic azimuthal correlation length of the accretion flow $\ell_\phi \approx 4.3^\circ$.
    Constraints from a joint observation of motion of stars and EHT ring size measurements~\citep{Sengo:2022jif} are shown in red, and theoretical bounds on the maximum superradiant extraction for $M_B/M$~\citep{Herdeiro:2021znw} are shown in green.}
    \label{fig:GAPRA}
\end{figure}

The left panel of Figure~\ref{fig:GAPRA} shows examples of the deviation $x_{(1)}^\mu$ using Cartesian Kerr-Schild coordinates $(t, x, y, z)$, for a massive tensor and a vector cloud with $\alpha = 0.2$ in the hydrogenic-like ground state~\citep{Chen:2022kzv}. We take the initial point at $i=17^\circ$ and the BH spin to be $a_{*} = 0.94$ as benchmark values to be consistent with M87*~\citep{EventHorizonTelescope:2019dse}. The evolution of the deviation can be divided in two stages: an oscillatory stage due to the time-varying energy-momentum tensors of the bosonic clouds and a second stage where the deviation grows exponentially when reaching the nearly-critical orbits of the photon ring. This second stage is caused by the instability of the photon ring orbit and is separated from the oscillatory stage by the time $x_{(0)}^\mu$ it first crosses the BH's equatorial plane. Therefore, the deviation for lensed photons significantly surpasses that of direct emission, along with the local plasma dynamics.

To detect small geodesic deviations, one sensitive probe is to use the photon ring autocorrelation discussed in Sec.~\ref{sec:time dependent observables}, which can measure both the time delay and azimuthal lapse from sources emitting in the BH's equatorial plane. The right panel of Figure~\ref{fig:GAPRA} shows the prospects to constrain the total mass of a vector cloud as a function of $\alpha$. The constraint is computed by requiring that the oscillation of the azimuthal lapse generated by the vector cloud is larger than the Gaussian smearing width due to a finite spatial resolution $\sim 10~\mu$as or the correlation length of the accretion flow $\ell_\phi \approx 4.3^\circ$~\citep{Hadar:2020fda}. Two other constraints are shown for comparison, including a joint observation of the EHT ring size measurement and motion of stars, as discussed in Sec.\,\ref{sec:BHwithHair}, and a theoretical bound on the maximum mass a boson cloud can reach due to superradiance, assuming that there is no angular momentum supplement to the BH~\citep{Herdeiro:2021znw}.

For a massive tensor cloud that couples to electromagnetic photons directly, constraints for $M_B$ will be more stringent~\citep{Chen:2022kzv}, for which the $n=1$ photon ring can already constrain a previously unexplored region of the parameter space. In addition, one can also probe their existence if the massive tensor is dark matter and forms a soliton core outside Sgr A*.

On the other hand, a scalar cloud generates time delays more efficiently than spatial deflections, which can in principle also be detected using time domain correlations. Compared to the azimuthal lapse, time delays are more difficult to detect, due to the large correlation time~\citep{Hadar:2020fda}. However, a nearby point-like source such as a hotspot or a pulsar can strongly boost these searches with a significantly better time resolution~\citep{Chesler:2020gtw}. These sources can also play important roles to look for $n=2$ peaks in the autocorrelation.

The oscillation amplitude of the azimuthal lapse is more significant in the inner region of the critical curve due to a stronger photon ring instability in this region~\citep{Chen:2022kzv}. Thus, the improvements in spatial resolution and dynamic range of the ngEHT are crucial to resolve these fine structures in the expected oscillation pattern. Searches for the $n=2$ ring, which can probe larger regions of the parameter space for both vector and tensor clouds, can also be boosted with an increase in baseline coverage.

\subsubsection{Boson and Proca stars}\label{sec:bosonstar}
For complex bosons, the hairy BH solutions described in Sec.~\ref{sec:BHwithHair} connect smoothly to horizonless, self-gravitating structures known as boson or Proca stars, when the field is a massive scalar or vector, respectively.  These are interesting BH mimickers and, in the context of EHT observations, the appearance of such solutions has been explored \citep[e.g.,][]{Vincent2016,Olivares:2018abq,Herdeiro:2021lwl}. 
While it is shown that under some circumstances these objects produce ring-like structures similar to BH images, in other cases they can produce images with a bright core that are qualitatively different, and therefore easily distinguishable. These two distinct behaviors, as well as the size of the ring, depend on a number of factors, such as the compactness of the stars, the particular boson field model used to construct the stars, or the inclination angle at which observations are made. In principle, knowledge of these properties could make it possible to set bounds on the star parameters that are compatible with present and future observations \citep{Olivares:2018abq,Herdeiro:2021lwl}. For further details we refer to Sec.~\ref{sec:central-brightness-depression} where the capabilities of the ngEHT to discriminate these objects are discussed in the context of horizon physics.
  
\subsubsection{Motion of S-stars}
\label{sec:stars}
Besides constraints from lensing signals from the Galactic Center by the ngEHT, one should also take into account that observations of S-stars orbiting Sgr A* may impose interesting constraints on putative boson clouds that could be present around Sgr A*. 
S-stars have been largely monitored and studied in the past decade using both astrometry and spectroscopy, with particular attention paid to the orbit of the star S2. The latter is one of the closest stars to the Galactic Center, reaching a minimum distance of $120$ AU ($\sim 1200$ Schwarzschild radius) from the central mass. 
The high precision of the data collected independently by both the GRAVITY Collaboration \citep{gravity2019} and the UCLA Galactic Center group \citep{ghez2008} enabled constraint of both the mass of the central object $M \sim 4.3 \times 10^6 ~ M_{\odot}$ and the Galactic Center distance $D \sim 8.3$~kpc. 
The hypothesis that Sgr A* is in fact a SMBH has been supported by the direct measurement of both gravitational redshift and the Schwarzschild precession value of $\Delta \omega = 12.1''$ per revolution in the orbit of S2~\citep{GRAVITY:2018ofz,GRAVITY:2020gka, do2019}. Data collected for S2 have also been used to test the presence of an extended mass within its apocenter, with particular attention paid to spherically symmetric dark matter density distributions~\citep[see, e.g.,][]{Lacroix:2018zmg,Bar:2019pnz,Heissel:2021pcw}). 
The GRAVITY Collaboration \citep{GRAVITY:2021xju} provided the current $1~\sigma$ upper bound of $0.1 \%$ of $M$ (equivalent to $\delta M \sim 4000 ~M_{\odot}$) on the dark mass around Sgr A* using the motion of four S-stars (S2, S29, S38, S55).

In the context of ultralight scalar fields, a bosonic structure around a SMBH has an impact on stellar orbits~\citep{Cardoso:2011xi, Fujita:2016yav,Ferreira:2017pth,Boskovic:2018rub,gravity2019scalar,DellaMonica:2023dcw,DellaMonica:2022kow}. The presence of a scalar cloud may affect the orbital elements of S2 in a way that is potentially detectable by the GRAVITY interferometer, since the in-plane precession it induces in the orbit is competitive with the first Post-Newtonian (PN) correction, i.e., with the Schwarzschild precession. Describing the scalar cloud by two parameters, the fractional mass $\Lambda = M_B/M$ and its dimensionless mass coupling constant $\alpha$, the largest variations in the orbital elements of S2 \citep{gravity2019scalar} are expected for:
\begin{eqnarray}
\label{range_S2}
0.001 \lesssim \alpha \lesssim 0.05 \, ,
\end{eqnarray}
corresponding to an effective scalar field mass of $10^{-20} {\rm ~ eV} \lesssim m_a \lesssim 10^{-18} \rm ~ eV$ and an effective peak position of $1.2 \times 10^4 ~ r_{\rm g} \lesssim R_{\rm peak} \lesssim 3 \times 10^6 ~ r_{\rm g}$.
The latter is in fact comparable with S2's orbital range of $3 \times 10^3 ~ r_{\rm g} \lesssim r_{\rm S2} \lesssim 5 \times 10^4 ~ r_{\rm g}$, meaning that the scalar cloud has a larger impact on the dynamics of S2 if the star crosses regions of space where the scalar density is higher. 
We note that for Sgr A* and for ultralight scalar fields with masses in the range~\eqref{range_S2}, the superradiant instability timescale is in general longer than a Hubble timescale (see Figure~\ref{fig:BHspin}), except for values of $\alpha$ close to the upper end of that limit, hence the scalar cloud must be formed by means of a different process. 
A recent analysis of the astrometry and the radial velocity of S$2$ showed no substantial evidence for a scalar cloud structure with mass coupling roughly in the range~\eqref{range_S2}. 
The fractional mass of the cloud can be constrained to be $\Lambda \lesssim 10^{-3}$ at the $3\,\sigma$ confidence level, corresponding to $0.1 \% $ of the central mass, setting a strong bound on possible bosonic structures around Sgr A*~\citep{GRAVITY:2023cjt}. 

\subsection{Polarimetric measurements for axion-induced birefringence}\label{sec:polarization}
We have so far focused on bosons that interact only gravitationally or interact only very weakly through possible additional couplings. However, from a particle physics perspective, there are strong reasons to include couplings to all of the Standard Model particles.
In fact, the introduction of an axion was originally made specifically in order to couple with gluon sector, so as to explain the smallness of the electric dipole moment of the neutron~\citep{Peccei:1977hh}. 
More broadly, one could then expect all axion-like particles to have nontrivial couplings to all other fields. 
In such theories, self-interactions and/or couplings to other particles are generically present and those can change the evolution of the superradiant instability. For example, the superradiance process is expected to be highly suppressed once the self-interactions of the axions become dominant. In the case of axions with a cosine potential $V(a) = m_{a}^2\, f_{}^2 \, \cos (a/f_a)$ in Equation~\eqref{Lphoton}, the self-interactions from the expansion of the potential will lead to emission of scalar radiation towards infinity and transitions into decaying modes. Thus one would expect that in some regimes superradiance and energy loss can balance each other and that the axion cloud enters a coherently oscillating state where the maximum field value of the cloud, $a_{\rm max}$, located at the equatorial plane of the Kerr BH at a radius $\sim 2\, r_g/\alpha^2$, can saturate at around $a_{\rm max}\sim f_a$ \citep{Yoshino:2012kn, Baryakhtar:2020gao}.  
For $a_{\rm max}$ below $10^{15}~$GeV, the total mass of the axion cloud is less than $1\%$ of the BH mass for $\alpha > 0.1$. 
Constraints from spin measurements as discussed in Sec.~\ref{sec:spin} do not apply any more since the extraction of the rotation energy can be considerably slowed down~\citep{Baryakhtar:2020gao}. However, one can still expect to see signatures of the ultralight axion if it couples to the visible sector, for example through axionic couplings to photons: %
\begin{equation}
\mathcal{L}_{\rm int} = g_{a\gamma} a F_{\mu\nu} \tilde{F}^{\mu\nu} / 2\,,
\label{eq:aEM}
\end{equation}
where $F^{\mu\nu} \equiv \nabla^\mu A^\nu - \nabla^\nu A^\mu$ is the field strength tensor of photons, $\tilde{F}^{\mu\nu}$ is its dual, and $g_{a\gamma}$ is the axion-photon coupling.
Once this coupling is turned on, the oscillating axion field
can periodically rotate the EVPA of the linearly polarised emissions \citep{Carroll:1989vb,Harari:1992ea}. 
The period of such oscillations is approximately the inverse of the axion mass and may be written as:
\begin{equation}
\label{eq:oscillation_period}
    T \simeq \frac{2\pi}{m_a} = 4\times 10^{5}\left(\frac{10^{-20}~{\rm eV}}{m_a}\right)~\textrm{seconds},
\end{equation}
with small corrections dependent on $\alpha$ \citep{Dolan:2007mj,Brito:2015oca}.
Using the equations of covariant radiative transfer in a local frame to take into account the plasma and curved spacetime effects, the axion effect in a local frame is equivalent to adding a term in parallel with the plasma-induced Faraday rotation $\rho_{V}^{\textrm{FR}}$, i.e., $\rho_{V} = \rho_{V}^{\textrm{FR}} - 2 g_{a\gamma}\, {\rm d} a/{\rm d} s$, where $s$ is the proper time and the second term is proportional to the gradient of axion field along the line-of-sight \citep{Chen:2022oad,Chen:2021lvo}. 
The shift of the EVPA is independent of the photon frequency, which can be distinguished from the astrophysical Faraday rotation. Since the superradiantly-grown axion cloud has angular momentum, the variations of the EVPA behave as a propagating wave along the photon ring for a nearly face-on BH \citep{Chen:2019fsq}.

\begin{figure*}[htb!]
    \centering
    \includegraphics[width=0.44\textwidth]{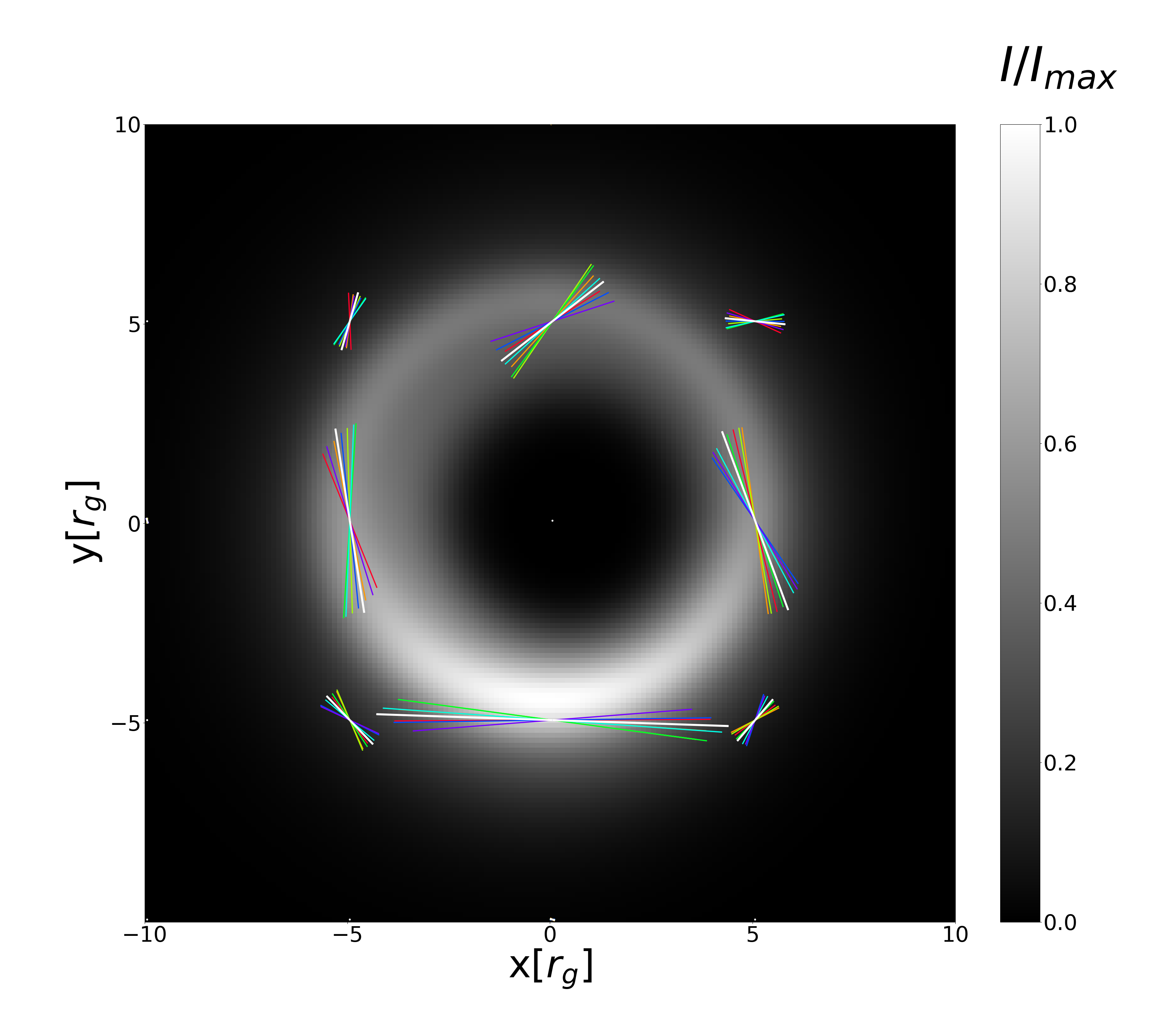}
    \includegraphics[width=0.55\textwidth]{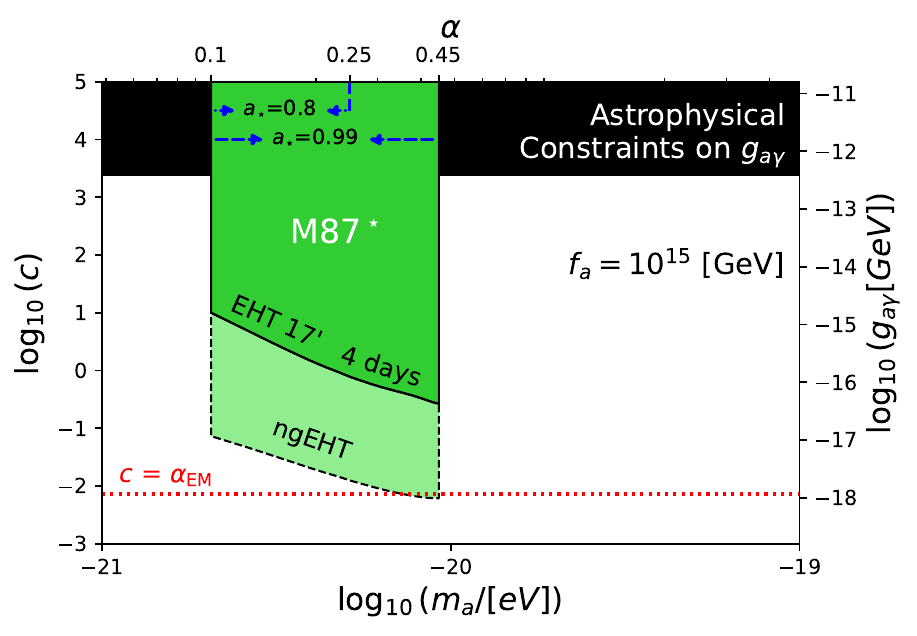}
    \caption{{Left panel:} illustration of a covariant radiative transfer simulation (\texttt{ipole}) of the polarised emission from a Kerr BH surrounded by an axion cloud. Different colors on the EVPA quivers, which range from red through to purple, represent the time variation of the EVPA in the presence of the axion-photon coupling. White quivers are the EVPAs when the axion field is absent. The intensity scale is normalized so that the brightest pixel is unity.
    {Right panel:} the upper limit on the axion-photon 
coupling \citep{Chen:2021lvo}, characterized by $c \equiv 2 \pi g_{a \gamma}\,f_a$, derived from the EHT polarimetric observations of SMBH M87$^\star$ \citep{EventHorizonTelescope:2021bee} and  prospect for ngEHT \citep{Chen:2022oad}. 
The bounds from other astrophysical observations assuming $f_a = 10^{15}~$GeV are shown for comparison. 
   }
\label{figaxion}
\end{figure*}

An illustration of the EVPA variations around M87* is shown in the left panel of Figure~\ref{figaxion}, where different colors of the quiver lines represent separate oscillation phases within one period.
Using the four days of polarimetric measurements of M87* in 2017 \citep{EventHorizonTelescope:2021bee}, one can already constrain the dimensionless axion-photon coupling $c \equiv 2 \pi g_{a \gamma}\, f_a$ to previously unexplored regions~\citep{Chen:2021lvo}, which is shown on the right panel of Figure~\ref{figaxion}.
The upper bound on the axion mass window is determined by the spin of the BH via the superradiant condition and two examples of spins are shown. The lower bound is required to have a superradiant rate timescale much shorter than the age of the Universe.

The fidelity of mm/sub-mm VLBI polarization maps is limited primarily by the ability to reconstruct unknown station-based calibration factors, e.g., complex station gains and polarimetric leakages. Closure traces are polarimetric closure quantities defined on station quadrangles which are (by construction) insensitive to linear station-based corruptions of the coherency matrix, extending the more familiar closure phases and closure amplitudes \citep{2020ApJ...898....9B}. Conjugate closure trace products (CCTPs) are combinations of closure traces that are sensitive exclusively to polarization, being unity otherwise, and are therefore providing direct, non-imaging evidence for complex polarization structures \citep[see, e.g., Figure 13 of][]{EventHorizonTelescope:2021bee}. 
The time-variable EVPA evolution produced by light axion clouds about M87* and Sgr A* results in time-variable excursions of the CCTP phase.  
Estimates based on the 2017 M87* EHT observations suggest that over a single observing campaign, the EHT can easily constrain the degree of axion-induced EVPA rotations to $2^\circ$ for $\alpha \simeq 0.4$ \citep{Wang:2023eip}, similar to the limits presented in \citet{Chen:2021lvo}.  
With its considerably larger bandwidth and number of available quadrangles, the ngEHT should be able to improve these limits by nearly two orders of magnitude, detecting EVPA fluctuations as small as $0.05^\circ$ using CCTP phases alone \citep{Wang:2023eip}.

The robustness of the superradiant phase is essential for the existence of a saturated axion cloud and thus the constraints. The gravitational perturbation is hard to prevent the build-up process since the rotating SMBH dominates the gravitational potential. 
The parametric instability of the axion to photons \citep{Ikeda:2018nhb, Boskovic:2018lkj,Spieksma:2023vwl} will be kinetically suppressed by the effective photon mass in the presence of plasma that is much higher than the axion mass. The conversion of axion and magnetic fields to electric fields that can heat the plasma is also highly suppressed.

For the ngEHT, one can increase the sensitivity by correlating the different data sets of the EVPA variations by a factor of the square root of the number of data sets \citep{Chen:2022oad}. Simultaneous correlation of EVPA variations at three different frequencies can increase the statistics or falsify suspicious signals. Correlations of azimuthal EVPA at different radii from the BH center require the angular resolution to be $\sim 10~\mu$as and the dynamic range of linear polarization to be above $100$. More sequential observations and higher time cadence also bring more statistics. 
On the other hand, emissions that reach the sky plane simultaneously and originate from different emission points where the axion field oscillates at different phases can potentially wash out the axion-induced EVPA oscillation, especially for lensed photons that orbit around the BH several times. At 86~GHz these lensed photons contribute less and the amplitude of the EVPA oscillations becomes larger. EVPAs observed at larger radii from the photon ring without the contamination of lensed photons should produce larger and more robust predicted signals.

The time variability of the accretion flow will lead to EVPA variations at each point in the sky plane. To suppress such turbulent behavior, \citet{Chen:2021lvo} introduced the differential EVPA in the time domain, the difference between two sequential observations in the data, in order to test axion-photon couplings. The price to pay is a suppressing factor of $\sin{\left(\omega t_{\textrm{int}}/2\right)}$, where  $t_{\textrm{int}}$ is the interval time between two sequential observations. For longer axion oscillation periods it provides a smaller value, which suppresses the sensitivity at lower axion masses on the right panel of Figure~\ref{figaxion}. 
On the other hand, such time variability of the EVPA can be modeled as a kind of noise using GRMHD simulations or extracting from a turbulent field \citep{Lee:2020pvs}. Thus one can expect to get rid of the suppression factor in the future with the ngEHT.

In the right panel of Figure~\ref{figaxion}, we also give the prospective sensitivity of the ngEHT to the axion-photon coupling \citep{Chen:2022oad}. There are two aspects contributing to the enhancement of the sensitivity: the removal of the suppression factor, $\sin{\left(\omega t_{\textrm{int}}/2\right)}$, after a more sophisticated understanding of the accretion flow's variations, and an increase in number of data sets. Reasonable expectations include ten times the observation time, five different radii from the BH center for the EVPA, and three different frequencies. We also take into account a larger radial wave-function in the equatorial plane of the BH. 
In the high mass region, the lowest possible value of $c_{\rm min} = \alpha_{\rm EM}$ can potentially be covered, where $\alpha_{\rm EM}$ is the electromagnetic fine-structure constant. 

Non-superradiantly-produced axions could also exist around SMBHs as a component of dark matter, contributing to birefringence signals as well. Field profiles of axion dark matter could be described by soliton cores in the centers of galaxies~\citep{Yuan:2020xui} or gravitational atoms bound to the SMBHs~\citep{Gan:2023swl}. The excellent angular resolution of the ngEHT will again help to resolve the coherent oscillating EVPA variations from different points in the sky plane.

\subsection{Open issues}\label{sec:todo}
In this section we discuss a number of open issues related to other interactions and how astrophysical environments can affect the superradiant instability and affect some of the observables mentioned above.

\subsubsection{Other types of interactions}

In addition to the axion-photon coupling discussed in Sec.~\ref{sec:polarization}, various types of ultralight bosons can interact differently with standard model particles, leading to corresponding oscillatory parameters in the standard model. Testing variations of fundamental constants in the strong gravity region, such as near a SMBH, can therefore test the existence of ultralight bosons.  
A CP-even scalar, often called a dilaton, can couple to the Lagrangian density of the standard model. The two main consequences of this are shifts in $\alpha_{\rm EM}$ from $\phi \,F_{\mu\nu}F^{\mu\nu}/\Lambda_\gamma$, and in the mass of a fundamental fermion, $\psi$, from $\phi\,m_\psi \bar{\psi}\, \psi/\Lambda_\psi$, where $\phi$ is the dilaton field, $m_\psi$ is the fermion's rest mass, and $\Lambda_\gamma$ and $\Lambda_\psi$ are the cut-off energy scales of the two couplings.

Measuring the fine-structure constant $\alpha_{\rm EM}$ usually requires spectroscopic observations. For example, comparing absorption lines from S-stars around Sgr A* can test variations of $\alpha_{\rm EM}$ between the star and the Earth \citep{Hees:2020gda} and constrain the ultralight boson's couplings \citep{Yuan:2022nmu}. The fluorescent iron K$\alpha$ line from X-ray observations can also potentially constrain its value near the innermost stable circular orbit (ISCO) \citep{Bambi:2013mha}. 
The frequency bands of the EHT and the ngEHT can, in principle, observe molecular lines, but they can only exist far away from the SMBH with lower temperatures compared to the horizon-scale plasma.

Similarly, spectroscopic observations can constrain the ratio between the electron and proton mass \citep{Murphy:2008yy}, which is influenced by the oscillations of the dilaton. In addition, the value of the electron-to-proton mass ratio can be set freely in GRMHD simulations, which could be used to study potential imprints of dilatonic couplings in the plasma.

A vector boson cloud that kinetically mixes with electromagnetic photons through couplings of the form $\epsilon F_{\mu\nu} X^{\mu\nu}$ can heat up the plasma \citep{Caputo:2021efm}, where $\epsilon$ is the kinetic mixing coefficient. Predictions of the phenomenological consequences require detailed studies, for example, through GRMHD simulations. In addition, the dense vector cloud may produce time-oscillating electromagnetic signatures, with frequencies much larger than the vector boson mass, arising from Compton-like scattering of photons with the electrons in the plasma~\citep{Caputo:2021efm}.

Compared with terrestrial searches for ultralight boson dark matter, the signals from a bosonic cloud can be much more significant due to the large field value that can be attained in the cloud. Using a Newtonian approximation and hydrogenic  wavefunctions, the field values for a scalar and a vector cloud at the superradiant ground state are related to the total cloud mass by 
\be\begin{split} \frac{ M_{B}}{M_{\rm BH}}
\approx \left\{\begin{array}{ll}
0.5\% \times \left(\dfrac{{\phi_0}}{10^{16}~ {\rm GeV}}\right)^2\, \left(\dfrac{0.4}{\alpha}\right)^4\,,\\
 0.8\% \times \left(\dfrac{X_0}{10^{17}~ {\rm GeV}}\right)^2\, \left(\dfrac{0.4}{\alpha}\right)^4\,,
\end{array}\right. \label{eq:MC}
\end{split}\ee
where $\phi_0$ and $X_0$ are the maximal field values of a scalar and a vector cloud, respectively. Thus a cloud with $10\%$ mass of the BH can have a significant field value up to the grand unification scale.  
Therefore we expect that the ngEHT could be used to constrain or detect the different couplings mentioned above.
However, further work is needed to fully assess the detectability of these type of interactions.

Finally, significant field values and their interactions with matter fields can lead to an enhanced production of matter within a cloud. A pertinent example is the interaction between a scalar cloud and Standard Model neutrinos~\citep{Chen:2023vkq}. 
The substantial scalar field value within the cloud facilitates the generation of a significant number of neutrinos parametrically, akin to the mechanism observed during the preheating phase in the early universe~\citep{Greene:1998nh, Greene:2000ew}. Furthermore, spacetime variations in bosonic fields contribute to the acceleration of neutrinos~\citep{Chen:2023vkq}.
The extended oscillation period of bosons surrounding M87* results in cyclic neutrino fluxes characterized by an angular preference, particularly within the polar angle region. 
Given the potential of the ngEHT to identify a diverse range of SMBHs and measure their masses and spins, joint observations involving the ngEHT and high-energy neutrino observatories stand to effectively constrain a wider spectrum of scalar masses. Notably, the joint observation approach also enables the identification of astrophysical neutrino backgrounds originating from blazars~\citep{Kovalev:2023crn}.

\subsubsection{Parametric instability of axion clouds}
An important question is whether we understand well enough how couplings of ultralight bosons with Standard Model particles affect the evolution of the superradiant instability and how taking all the effects into account can affect the observations. For example, \citet{Ikeda:2018nhb} considered as a starting point the action~\eqref{Lphoton} with $V(a)=m_{a}^2 \,|a|^2/2$, describing a real massive (pseudo)scalar field with axionic couplings to the electromagnetic field in Equation~\eqref{eq:aEM}.
They found that above some critical value for the coupling constant $g_{a\gamma}$, axionic clouds around BHs can transfer an important fraction of their energy to the electromagnetic field due the development of a parametric instability. Their results can be translated into a maximum mass $M_{B}$ that the cloud can reach: 
\begin{equation}
\frac{M_B}{M}\sim \left(\frac{0.1}{M\mu}\right)^4\left(\frac{2\times 10^{-17}{\rm~ GeV}^{-1}}{g_{a\gamma}}\right)^2 \,.
\end{equation}
An important question is therefore how much of this phenomenology is important and whether plasma couplings can quench the instability due to the modified dispersion of electromagnetic waves in a plasma medium. This was partially studied in \citet{Sen:2018cjt} and \citet{Boskovic:2018lkj}. Considering the case of a cold plasma, the plasma frequency is given by:
\begin{equation}
    \omega_{p}=\sqrt{\frac{4 \pi n_{e} e^{2}}{m_{e}}} \approx 10^{-13}\left(\frac{n_{e}}{10^{-4} \mathrm{~cm}^{-3}}\right)^{1 / 2} ~\mathrm{eV} \,,
\end{equation}
with $m_e$, $e$ and $n_{e}$ denoting the mass, charge and number density of the free electrons, respectively.  
Taking $n_e \sim 10^{4}$ -- $10^{8} \mathrm{~cm}^{-3}$ from EHT observations of M87* \citep{EventHorizonTelescope:2021srq}, photons attain an effective mass of around $10^{-9}$ -- $10^{-7.5}$~eV, which is much higher than the relevant axion masses $\sim 10^{-21}$~eV. Thus, one should expect the parametric instability to be highly suppressed. 

There are, however, some caveats to this simplistic argument that warrant further study. For example, \citet{Sen:2018cjt} showed that there can still be a small instability window at large couplings $g_{a\gamma}$ when $m_a$ is much smaller that the plasma frequency. In addition, these analysis were done using a flat-space approximation and treating the interaction of the electromagnetic field and the plasma as simply giving the photon an effective mass, which might not be a good approximation to describe these systems \citep[see, e.g.,][]{Cardoso:2020nst,Cannizzaro:2020uap,Cannizzaro:2021zbp}. It is also known that non-linear plasma effects can open a transmission window in plasmas \citep{Cardoso:2020nst} which could therefore allow the parametric instability to develop. Understanding how all these effects affect the superradiant instability is a challenging task, but one which is important to pursue.

\subsubsection{Plasma heat-up from axion clouds in a magnetic field}
In addition to the parametric instability discussed above, another important question is whether, in the presence of a background magnetic field and axion-photon couplings, the axion cloud can sufficiently heat up electrons/positrons/ions so as to terminate the superradiance process before the BH spins down or the field saturates the self-interaction limit. A similar process happens, for example, due to the kinetic mixing of dark photons with (normal) photons \citep{Caputo:2021efm}. 

To understand this problem using a flat-space approximation, it is useful to write the equations of motion for both the transverse electromagnetic field $A_T$ and the axion field $a$ in the presence of a background magnetic field $B_0$ in terms of Fourier modes:
\begin{equation}
\left\{\vec{k}^{2}-\omega^{2}+ \left(\begin{array}{cc}
\Omega_{p}^{2} & \omega g_{a \gamma} B_0 \\
\omega g_{a \gamma} B_0 & m^2
\end{array}\right)\right\}\left(\begin{array}{c}
A_{T} \\
a
\end{array}\right)=0 \,, \label{eomAa}
\end{equation}
where $ \Omega_{p}^{2} \equiv \dfrac{\omega_{p}^{2}}{1+\frac{i \nu}{\omega}}$ and:
\begin{equation}
    \nu=\frac{4 \sqrt{2 \pi} \alpha^{2} n_{e}}{3 m_{e}^{1 / 2} T_{e}^{3 / 2}} \log \Lambda_{C} \approx 3 \times 10^{-21} \mathrm{eV}\left(\frac{n_{e}}{0.04~ \mathrm{cm}^{-3}}\right)\left(\frac{8000~ \mathrm{K}}{T_{e}}\right)^{3 / 2} \,,
\end{equation}
is the electron-ion collision rate contributing to the friction in the plasma. 
We can then solve Equation~(\ref{eomAa}) to obtain the dissipation rate of the axion field in the plasma:
\begin{equation}
\gamma_{a}^{I} = \frac{g_{a \gamma}^2 \, B_0^2\, \nu}{2 \omega_p^2} \,.
\end{equation}
Again taking EHT observations of M87* as an example, where it was found that $n_{e} \sim 10^{4}$--$10^{8}~\mathrm{cm}^{-3}$, $B_0 \sim 1\textrm{--}30$~G and $T_e \sim (1\textrm{--}12) \times 10^{10}$~K \citep{EventHorizonTelescope:2021srq}, the dissipation rate of the axion field due to the heating up of the plasma is roughly:
\begin{equation}
\gamma_a^I \sim 10^{-53} \left( \frac{g_{a \gamma}}{10^{-12}~ {\rm GeV}^{-1}}  \right)^2  ~{\rm eV}\,,
\end{equation}
in the non-relativistic limit $|\vec{k}| \ll \omega$ (in this limit the result applies to the longitudinal mode of the photon as well). This is too small to stop superradiance. Whilst we employed a flat-space approximation, one expects that strong gravity effects will not change this picture considerably, although further studies are necessary to  assess this.

%% file: Tests_GR_Kerr.tex
\section{Tests of GR and the Kerr hypothesis with the ngEHT\label{sec:Tests_GR_Kerr}}

BHs in GR unavoidably harbor unphysical curvature singularities~\citep{Penrose:1964wq,Hawking1966,Penrose:1969pc,Hawking:1970zqf,Penrose:1972ui,Hawking:1973uf,Senovilla:1998oua,Senovilla:2014gza}.
The question is therefore not \textit{if} GR and the Kerr hypothesis break down, but rather \textit{how} and \textit{where} this break down occurs.

There is consensus that trans-Planckian curvature scales are governed by an as of yet unknown quantum theory of gravity which modifies the core of a BH~\citep{Ashtekar:2005cj,Bojowald:2007ky,2018NatPh..14..770E}.
However, quantum-gravity effects may affect the structure of BH spacetimes at much lower curvature scales and even structure outside the horizon~\citep{Frolov:1979tu,Frolov:1981mz,Hajicek:1986hn,Hajicek:2000mi,Hajicek:2002ny,Ambrus:2005nm,Almheiri:2012rt,Barcelo:2014npa,Rovelli:2014cta,Barcelo:2014cla,Haggard:2016ibp,Barcelo:2015noa,Giddings:2016btb,Giddings:2019jwy,Bacchini:2021fig,Eichhorn:2022bbn}.
Furthermore, matter (itself potentially non-minimally coupled) within GR may form exotic compact, supermassive objects which could be found at the cores of galaxies (see Sec.~\ref{sec:Ultralight_fields}).
Finally, classical modifications of GR, i.e., so-called modified gravity theories, are being investigated (some motivated by questions in cosmology) and may contain BH solutions that differ from Kerr~\citep[see Sec.~\ref{sec:non-minimal} as well as][for reviews]{Berti:2015itd,Cardoso:2019rvt}.
In view of these many distinct theoretical possibilities, an observational program which pushes as far as possible into the strong-gravity/near-horizon regime is called for, in order to place constraints on theoretical proposals or even discover a deviation from GR.

\noindent\paragraph{\bf The astrophysics challenge\\}
A critical challenge for this program is set by the astrophysical environment of a SMBH: the Kerr hypothesis already technically breaks down for astrophysical BHs, since the Kerr solution is strictly a vacuum solution~\citep{Kerr:1963ud,Newman:1965tw,Newman:1965my,Bekenstein:1971hc,Robinson:1975bv,Chrusciel:2012jk}, 
yet SMBHs exist within an astrophysical environment typically composed of an accretion disk and relativistic jet outflows. 
This astrophysical environment is only partially understood, introducing a systematic uncertainty in the comparison between data and theory. Within GR a large body of literature is already devoted to an ever improving understanding of the accretion disk physics and the resulting observational images. 
Beyond GR it is a key outstanding challenge to achieve a commensurate level of control over the systematic uncertainties introduced by a BH's astrophysical environment.

\noindent\paragraph{\bf Searching for a breakdown of GR vs testing modified gravity\\}
There are two questions that we aim to answer with ngEHT observations. First, we ask whether GR fails, and second, we ask whether a modified theory of gravity performs better at explaining the observational data.

Within GR, there are clear predictions for the image of a BH (depending on the accretion disk properties). Exploring whether the observational data agrees with this expectation tests whether or not GR fails for this observation. Knowledge of theoretical predictions beyond GR is unnecessary for such a test.

Beyond GR, predictions for the image of a BH (or a horizonless spacetime) exist in part.
Where they exist, it is possible to ask whether the given theory performs better than GR at explaining the data.
It is our aim to ask this question as comprehensively as possible.
Controlling the systematic uncertainties arising from a BH's astrophysical environment is essential to accomplish this task.
Thus, we aim at completing the following program.

\noindent\paragraph{\bf Confronting ngEHT observations with theoretical predictions\\}
To compare predictions of modified theories of gravity with those of GR, we have to complete the following steps: (i) investigate the theoretical properties of objects/spacetimes beyond GR, in particular their shadows, (ii) examine images of these objects/spacetimes as they appear when illuminated by an accretion disk, and (iii) account for the capabilities of the ngEHT array.

\noindent \newline We summarize progress on this program in several scenarios beyond GR in Sec.~\ref{sec:physics-scenarios-beyond-GR} and find that step (i) has already been completed for many objects/spacetimes beyond GR, where the critical curve (shadow boundary) has been calculated. This includes: gravity with Chern-Simons terms \citep{Amarilla:2010zq,Okounkova_2018}, Gauss-Bonnet gravity \citep{Guo:2020zmf,Konoplya:2020bxa,Kumar:2020owy,Wei:2020ght}, gravity coupled to nonlinear electrodynamics \citep{Allahyari:2019jqz}, scalar-vector-tensor gravity \citep{Moffat:2015kva,Wang:2018prk}, tensor-vector gravity \citep{Vetsov:2018mld}, braneworld settings \citep{Amarilla:2011fx,Eiroa:2017uuq}, Kaluza-Klein BHs \citep{Amarilla:2013sj}, BHs inspired by asymptotically safe gravity \citep{Held:2019xde} and Loop Quantum gravity \citep{Liu:2020ola}, regular BHs \citep{Abdujabbarov:2016hnw,Amir:2016cen,Tsukamoto:2017fxq,Stuchlik:2019uvf}, naked singularities~\citep{Abdikamalov:2019ztb,Kumar:2020yem} and wormholes~\citep{Gyulchev:2018fmd,Brahma:2020eos,Bouhmadi-Lopez:2021zwt}.
See also~\citet{Perlick:2021aok,Vagnozzi:2022moj}.
These constraints have been derived through comparing the diameter of the critical curve obtained from theoretical considerations with the observational measurements of the diameter of the shadow image. We refer to these as \textit{projected constraints}: they provide an estimate of the parameter ranges that an (ng)EHT array may constrain, once simulated images go beyond the critical curve and account for the presence of an accretion disk.

In fact, step (ii) has been partially achieved for some settings beyond GR, where most studies are limited to simple models of accretion disks~\citep{Shaikh:2018lcc,Liu:2020vkh,Zeng:2020dco,Bauer:2021atk,Dong:2021yss,Eichhorn:2021etc,Gyulchev:2021dvt,Daas:2022iid} and a more limited number of covariant MHD simulations~\citep{Mizuno:2018lxz, Olivares:2018abq, Roder:2022hqn,Chatterjee2023b,Chatterjee2023a} are available.
Step (iii) has only been achieved for very few settings~\citep{Vincent:2020dij,Eichhorn:2022fcl}.
Therefore, different theoretical scenarios are in different stages of development when it comes to readiness for comparison with actual observational data.

We then take a different approach and condense all these beyond-GR scenarios down to three general signatures which can be parametrically constrained.
In Sec.~\ref{sec:beyond-GR-science-cases} we discuss the resulting science cases with respect to the capabilities of the ngEHT.

\noindent\paragraph{\bf Relations between ngEHT and other observational constraints\\}
There are several observational programs that are currently making significant progress in constraining deviations from GR, most notably GW measurements~\citep{LIGOScientific:2017zic,LIGOScientific:2017bnn,LIGOScientific:2017vwq,LIGOScientific:2019zcs,LIGOScientific:2021nrg,LIGOScientific:2021sio}.
Their relation to ngEHT constraints is subject to two considerations.
First, because GWs arise in the dynamical regime of a gravity theory, they can be challenging to simulate beyond GR, such that a waveform catalogue that enables tests of modified-gravity theories is not yet available. In contrast, the ngEHT accesses the (mostly) stationary regime of the spacetime, where one only requires (spinning) stationary solutions of modified-gravity theories. Second, current constraints on deviations from GR from GW observations do not necessarily apply to SMBHs: unless BH uniqueness theorems hold, constraints on the metric of roughly solar-size BHs do not constrain the metric of SMBHs.

\subsection{Scenarios for physics beyond GR}
\label{sec:physics-scenarios-beyond-GR}

In Sec.~\ref{sec:specificspacetimes}, we turn to specific spacetimes and compact objects.
We discuss various scenarios which: arise from quantum fluctuations, are motivated by quantum gravity, or feature violations of cosmic censorship.

Here, we review physics scenarios beyond GR and the resulting violations of the Kerr hypothesis, according to how it is circumvented. The ngEHT may also provide insight into particle physics beyond the Standard Model. Such new-physics effects typically involve minimally coupled exotic matter, and were discussed separately in Sec.~\ref{sec:Ultralight_fields}. In Sec.~\ref{sec:non-minimal} we discuss classical and quantum theories of modified gravity, including non-minimally coupled matter fields and effective field theory. In Sec.~\ref{sec:specificspacetimes} we turn to specific spacetimes and compact objects: we discuss various scenarios which arise from quantum fluctuations, are motivated by quantum gravity, or feature violations of cosmic censorship.

In each subsection, we briefly summarize the theoretical motivation and observational status of the physics scenario beyond GR. We then summarize the state-of-the-art with respect to VLBI imaging, to point out where additional work is necessary to achieve readiness for future comparison with data. This includes, in particular, the question of whether potential ngEHT signatures have been studied in idealized scenarios or in more realistic settings which take into account astrophysical uncertainties and limitations in instrument capabilities.

We remain agnostic as to the theoretical viability of these scenarios. Instead, we take the point of view that observational constraints should guide the search for physics beyond GR.

\subsubsection{Modified gravity}
\label{sec:non-minimal}

The assumption that AGNs are supermassive Kerr BHs rests on the field equations of GR, although the Kerr metric is also a solution of some theories beyond GR, e.g., if the action contains no Riemann invariants~\citep{Psaltis:2007cw,Barausse:2008xv}. 
The EFEs can be taken to follow from a theorem by~\cite{Lovelock:1971yv}. 
The theorem, in turn, relies on several assumptions and physics beyond GR, which may lead to alternative paradigms for AGNs and can be classified by how the theorem is circumvented \citep[see, e.g.,][]{Pani2013PhRvD,Berti:2015itd}. 
In particular, the theorem assumes:
(i) spacetime is a differentiable manifold endowed with a metric,
(ii) four spacetime dimensions,
(iii) diffeomorphism symmetry, (iv) a local action principle,
(v) equations of motion with (at most) second-order time derivatives,
(vi) no gravitational degrees of freedom beyond the massless graviton, and
(vii) no non-minimal coupling.

Assumption (i) is violated in many (although not all) approaches to quantum gravity. For instance, this gives rise to the fuzzball paradigm, see Sec.~\ref{sec:specificspacetimes} in string theory. Nevertheless, many approaches to quantum gravity work with an effective spacetime metric, that is constructed at a phenomenological level, e.g., in Loop Quantum Gravity \citep{Ashtekar:2005qt}, or non-commutative spacetimes\footnote{See~\citep{Vagnozzi:2022moj} for \textit{projected constraints} on non-commutativity derived by confronting calculations of the critical curve with observations}. For BHs, this effective metric is typically regular (see Sec.~\ref{sec:specificspacetimes}).

In general, it is expected that at sufficiently low curvature scales, quantum effects can be captured in an effective action, even if the UV theory goes beyond the spacetime setting.
Within approximations, this effective action is (partially) known, e.g., in string theory \citep{Gibbons1988,Garfinkle1991,Sen1992},
asymptotically safe quantum gravity \citep[e.g.,][]{Bonanno:2020bil} and causal dynamical triangulations~\citep[e.g.,][]{Loll:2019rdj}.
At this level, quantum and classical theories can be treated on the same footing.\footnote{One may expect that quantum-gravity effects are generically tied to higher curvature scales than classical modifications of GR. However, this expectation has been challenged in, e.g.,~\citet{Frolov:1979tu,Frolov:1981mz,Hajicek:1986hn,Hajicek:2000mi,Hajicek:2002ny,Ambrus:2005nm,Almheiri:2012rt,Barcelo:2014npa,Rovelli:2014cta,Barcelo:2014cla,Haggard:2016ibp,Barcelo:2015noa,Giddings:2016btb,Giddings:2019jwy,Bacchini:2021fig,Eichhorn:2022bbn}. Equating the scale of quantum gravity to the Planck scale is, in fact, a simple dimensional estimate which does not use any information whatsoever on the gravitational dynamics. It may thus not do justice to the actual theory of quantum gravity which may, e.g., (i) not follow naturalness arguments and have a scale different from the Planck scale, or (ii) have several dynamical scales.}

In the following, we review how different modifications violate the assumptions of Lovelock's theorem and discuss the state of maturity of different studies of BH shadows in these various settings.

In settings with extra dimensions, the solutions with an event horizon~\citep[see][for a review]{Emparan:2008eg}, can be projected onto four-dimensional spacetime, in which they appear as four-dimensional BHs with a modified line element. Modifications to the critical curve have been investigated~\citep{Amarilla:2011fx,Papnoi:2014aaa,Singh:2017vfr,Amir:2017slq,Long:2019nox} and \textit{projected constraints} have been obtained using EHT results~\citep{Banerjee:2019nnj,Vagnozzi:2019apd}.

Projected constraints on theories with violations of Lorentz-invariance \citep[or more generally diffeomorphism symmetry, e.g.,][]{Barausse:2011pu} have been obtained in~\citet{Ding:2019mal}, \citet{Khodadi:2020gns}, \citet{Vernieri2012PhRvD}, and \citet{Sotiriou2014PhRvD}.
Moreover, they have been found to exhibit a nested structure for trapping horizons as a function of the frequency of the photons that follow the null geodesics~\citep{Carballo-Rubio:2021wjq}.
It is an open question as to whether the same statement holds for the photon sphere.
If that is indeed the case, tests of achromaticity of the emission ring could constrain violations of Lorentz invariance in gravity.
Such Lorentz-invariance violations are already strongly constrained by other observations \citep[e.g.,][]{EmirGumrukcuoglu:2017cfa, Gupta:2021vdj}.\newline

In a four-dimensional diffeomorphism-invariant setting, one can circumvent the Lovelock theorem by modifying the action.
A restriction to second-order field equations ensures the absence of potential (Ostrogradski) ghost instabilities and significantly constrains the allowed set of interactions.
The converse is not true: if the equations are higher than second order, the theory does not automatically have ghost instabilities.

General theories of ghost-free interactions between gravity and a scalar, so-called scalar-tensor theories~\citep{Horndeski:1974wa, Deffayet:2011gz}, between gravity and a vector, so-called vector-tensor theories~\citep{Heisenberg:2014rta, deRham:2020yet}, and between gravity and a second (potentially massive) tensor mode, so-called bimetric or massive gravity theories~\citep{deRham:2010kj, Hassan:2011zd}, have been constructed with second-order equations of motion.
Theories with higher-order equations of motion but without Ostrogradski instability include infinite-derivative gravity~\citep{Biswas:2005qr, Modesto:2011kw, Biswas:2011ar}. 
Theories with higher-order equations of motion include, e.g., Stelle's higher derivative gravity \citep{Stelle:1977ry}, in which spherically-symmetric BH solutions have been found numerically~\citep{Lu:2015cqa, Podolsky:2019gro}. In this theory, BHs with small enough horizon suffer from a classical long-wavelength instability~\citep{Brito:2013wya,Held:2022abx}.

Such theories with higher-order equations of motion can also be interpreted as the leading-order terms in an effective field theory (EFT)~\citep{Endlich:2017tqa, Cardoso:2018ptl, deRham:2020ejn}, which may, for instance, arise from integrating out quantum fluctuations in a quantum theory of gravity.
An EFT includes interactions order-by-order and is only valid up to a cutoff scale at which interactions of yet higher order can no longer be neglected. 
In the EFT (some of) the additional degrees of freedom, associated instabilities, and the new BH branches, may thus be an artefact of extrapolating the EFT beyond its regime of validity.

The existence of a (perturbative) EFT parameter can facilitate the order-by-order extension of some of the properties/theorems of GR to larger classes of modified-gravity theories, e.g.,~\cite{Xie:2021bur} for the example of circularity. In these settings, more image features are likely to be shared between GR and the beyond-GR theory.

\noindent\paragraph{\bf Black-hole uniqueness and ngEHT prospects\\}
Beyond GR, BH uniqueness may be violated~\citep{Sotiriou:2011dz}. Kerr BHs remain a solution of a subclass of these theories~\citep[see, e.g.,][]{BenAchour:2018dap, Motohashi:2019sen}. Generically, if curvature invariants built from the Riemann tensor are absent in the action, it is straightforward to show that solutions of GR remain solutions~\citep{Psaltis:2007cw,Barausse:2008xv}. This includes a large class of nonlocal theories of gravity~\citep{Li:2015bqa}. 
Such theories can therefore be more challenging to constrain with the EHT.
Other modifications deform Kerr BHs~\citep{Yunes:2009hc, Yunes:2011we, Yagi:2012ya, Herdeiro:2014goa, Herdeiro:2015waa, Silva:2017uqg, Ayzenberg:2018jip} and/or admit additional BH branches~\citep{Lu:2015cqa, Podolsky:2019gro}.

If BH uniqueness holds, all astrophysical constraints on modifications of GR at different scales \citep{Yunes:2013dva, Will:2014kxa, Berti:2015itd} should be accounted for. For modifications which are tied to local curvature scales, the EHT and ngEHT probe a regime which is already constrained by GW and Solar System observations~\citep{Glampedakis:2021oie}.
If BH uniqueness is violated, the ngEHT is uniquely positioned to test modifications of GR on scales associated with AGNs.

\noindent\paragraph{\bf Spherically-symmetric black holes
and VLBI observations\\}
Most of the known explicit BH solutions in modified gravity are restricted to spherical symmetry.
For many of these, spherically-symmetric deviations in the critical curve have been quantified~\citep{
Ayzenberg:2018jip,
Allahyari:2019jqz,
Konoplya:2019goy,
Konoplya:2019fpy,
Shaikh:2019fpu,
Zhu:2019ura,
Islam:2020xmy,
Khodadi:2020jij,
Konoplya:2020bxa,
Kumar:2020sag,
Guo:2020zmf%
};
see also~\citet{Perlick:2021aok} for a recent review.

In spherical symmetry, there is a degeneracy between modified-gravity effects and the mass-to-distance ratio of a given BH~\citep{Kocherlakota:2022jnz}. 
A second independent mass measurement is required to break this degeneracy.
Moreover, all deviations in the critical curve are degenerate across all the different modified theories of gravity~\citep{EventHorizonTelescope:2021dqv, Vagnozzi:2022moj}. Breaking at least some of this degeneracy is achievable using finite-order lensing features such as photon rings~\citep[e.g.,][]{Wielgus:2021peu,Kocherlakota2023}.

A much smaller set of theories has been investigated by the use of disk models and resulting photon rings, which includes theories in which a scalar field couples to the Gauss-Bonnet or the Pontryagin invariant~\citep{Zeng:2020dco,Bauer:2021atk}, as well as Stelle's higher-derivative gravity~\citep{Daas:2022iid}.
Modifications of the emission spectrum and the Blandford-Znajek process have been investigated by \cite{Liu:2020vkh} and \cite{Dong:2021yss}, respectively.
Even less is known about full (GR)MHD simulations which have been performed for Einstein gravity coupled to additional matter fields in~\cite{Mizuno:2018lxz} and \citet{Roder:2022hqn}.
More recent (GR)MHD simulations of spinning (axisymmetric) BHs have been performed in \cite{Chatterjee2023b,Chatterjee2023a}.
In general, degeneracy is expected between modifications of the spacetime and properties of the astrophysical environment, making such studies critical. In particular, it may be the case that the \textit{projected constraints} are overly optimistic. Marginalizing over astrophysical parameters may significantly weaken the constraints~\citep{Cardenas-Avendano:2019pec}.

\noindent\paragraph{\bf Spinning black holes and VLBI observations\\}
Spinning BHs beyond GR have been constructed with the Janis-Newman algorithm \citep{Janis:1968zz}.
See~\citet{Capozziello:2009jg, Modesto:2010rv, Bambi:2013ufa, Azreg-Ainou:2014pra, Toshmatov:2014nya, Kumar:2020owy, Wei:2020ght} for examples.
We note that it is not generally expected that the Janis-Newman algorithm is valid in modified-gravity theories~\citep{Drake:1998gf,Hansen:2013owa}.

In some theories, BH solutions have been explicitly extended to the case of slow spin~\citep{Yunes:2009hc,Cano:2019ore}. 
Some axisymmetric solutions have been numerically constructed for arbitrary spin~\citep{Kleihaus:2011tg,Kleihaus:2015aje,Fernandes:2022gde} but full analytical solutions are not yet known. 
In the few cases in which spin has been explored, there is no hidden (Carter-like) constant of motion \citep{Owen:2021eez}. Moreover, some theories can break reflection symmetry about the equatorial plane~\citep{Cardoso:2018ptl,Cano:2019ore,Chen:2021ryb}, and others can break circularity~\citep{BenAchour:2020fgy}, cf.,~Sec.~\ref{sec:parameterized}. 
It remains an important open question as to how far (some of) these properties can be tied to generic image features.

\noindent\paragraph{\bf Open challenges\\}
The open challenges to further quantify the constraining power of the ngEHT for modified gravity theories are as follows.
\begin{itemize}
\setlength\itemsep{0.5em}
    \item[\textbullet] Go beyond spherical symmetry in solutions to modified theories of gravity and account for spin -- this will allow breaking of degeneracy in the Kerr mass and potentially even between different modifications.
    \item[\textbullet] Determine which modified-gravity theories break BH uniqueness and admit (stable) non-Kerr BH branches -- this identifies a class of theories for which Solar System and GW constraints need not apply.
    \item[\textbullet] Go beyond calculations of the critical curve and instead simulate images, using simple disk models in a first step, and full GRMHD simulations, ideally including polarization and frequency dependence, in a second step.
    \item[\textbullet] Understand degeneracies between modifications of GR and properties of a BH's astrophysical environment.
\end{itemize}

\subsubsection{Specific spacetimes and compact objects beyond GR}
\label{sec:specificspacetimes}

In addition to the explicit BH solutions in modified gravity (see Sec.~\ref{sec:non-minimal}), several classes of spacetimes have been motivated by theoretical considerations both in and beyond GR.

\noindent\paragraph{\bf Exotic compact objects from semi-classical physics\\}
Gravastars \citep{Mazur:2001fv, Mazur:2004fk, Cattoen:2005he, Chirenti:2007mk}
are expected to form if quantum fluctuations  become so sizable during gravitational collapse that they trigger a phase transition to spacetime regions with an effective stress-energy tensor which violates energy conditions. The viability of such a formation mechanism is debated in the literature~\citep{Chen:2017pkl}. 
After formation, the resulting stationary spacetime is effectively described by several radial shells with distinct effective stress-energy tensors. 

In the simplest effective gravastar geometries~\citep{Visser:2003ge}, an exterior asymptotically flat vacuum solution to GR is glued to an interior de-Sitter patch. The interior and the exterior are connected by a thin shell of energy-condition violating matter which facilitates smooth matching. More recently, gravastars with AdS interior have been constructed~\citep{Danielsson:2017riq}.

Spinning gravastars have been constructed in the slow-rotation limit by perturbing non-rotating spacetimes~\citep{Cardoso:2007az, Pani:2015tga, Danielsson:2017pvl} and are subject to an ergoregion instability~\citep{Friedman:1978wla}.

Semi-classical relativistic stars have also been proposed as a possible endpoint of gravitational collapse, being supported by the most elementary form of quantum pressure provided by gravitational vacuum polarization~\citep{Visser:2008rtf,Barcelo:2009tpa,Carballo-Rubio:2017tlh}. As with gravastars, proposals for formation mechanisms~\citep{Barcelo:2007yk,Barcelo:2015noa} are only partially understood.

Contrary to gravastars, which have a vacuum core in which the equation of state is $p=-\rho$, semi-classical relativistic stars are formed by a delicate energetic balance between matter and vacuum polarization, and can thus have different equations of state~\citep{Ho:2017vgi,Arrechea:2021pvg,Arrechea:2021xkp}. Due to technical limitations in our knowledge of quantum field theory on rotating backgrounds~\citep{Ottewill:2000qh,Zilberman:2022aum,Zilberman:2022iij}, spinning solutions are still to be obtained.

\noindent\paragraph{\bf Regular black holes\\}
In some approaches to quantum gravity \citep{Donoghue:1993eb, Reuter:1996cp}, quantum fluctuations have been found to weaken the gravitational force, lead to geodesic defocusing, and thus provide a natural mechanism for singularity resolution. 
Thus, regular BHs, both spinning  \citep[e.g.,][]{Reuter:2010xb,Bambi:2013ufa,Neves:2014aba,Toshmatov:2014nya,Azreg-Ainou:2014pra,Ghosh:2014hea,Brahma:2020eos,Mazza:2021rgq,Eichhorn:2021iwq,Franzin:2022wai} and non-spinning \citep[e.g.,][]{Bonanno:2000ep, Nicolini:2008aj, Gambini:2013ooa,Ashtekar:2018lag, Platania:2019kyx,Bodendorfer:2019cyv,Bodendorfer:2019nvy,Bodendorfer:2019jay}, have been constructed or motivated in multiple approaches to quantum gravity.\footnote{This is possible within an \textit{effective} metric setting even in quantum-gravity approaches where spacetime is not a smooth manifold at high curvature scales~\citep[see, e.g.,][]{Ashtekar:2005qt, Modesto:2004xx, Modesto:2005zm, Hossenfelder:2009fc, Gambini:2013ooa, Rovelli:2014cta, Ashtekar:2018lag}.}

On a more phenomenological level, regular BHs have been put forward in, e.g., \citet{Dymnikova:1992ux} and \citet{Hayward:2005gi}, and reviewed in \citet{Ansoldi:2008jw}.
They also arise as 
solutions to GR coupled to non-linear electrodynamics~\citep{Ayon-Beato:1998hmi}.
They can also be regarded as a useful paradigm for beyond-GR-spacetimes, given that they resolve one (of several) problematic aspects of BHs in GR, namely the curvature singularity (though not necessarily the Cauchy horizon). There are still open questions regarding their dynamical evolution due to the exponential mass inflation instability~\citep{Brown:2011tv,Frolov:2017rjz,Carballo-Rubio:2018pmi,Carballo-Rubio:2021bpr,Barcelo:2022gii}, which is generic but for a specific set of geometries~\citep{Carballo-Rubio:2022kad}.

Generically, regular BHs come with (at least) one free parameter that determines the curvature scale at which deviations from Kerr become sizable \citep[e.g.,][]{Held:2019xde,Contreras:2019cmf,Kumar:2019pjp,Li:2020drn,Kumar:2019ohr,Kumar:2020ltt,Kumar:2020yem,Eichhorn:2022oma}. 
Parameter values of order $1\%$ in these models 
lead to $\mathcal{O}(0.1\%)$ effects at the horizon and correspondingly somewhat smaller effects at the photon sphere and on the shadow \citep[see, e.g.,][]{EventHorizonTelescope:2021dqv,Eichhorn:2022oma}. 
Such parameter values arise if the scale of new physics corresponds to curvature radii of the order of the gravitational radius, and hence to scales far
above the Planck scale. This could be, for instance, a consequence of the back-reaction associated with mass inflation~\citep{Barcelo:2022gii,Carballo-Rubio:2022nuj}.

Key image features of regular BHs~\citep[see, e.g.,][]{Abdujabbarov:2016hnw,Eichhorn:2021etc,Eichhorn:2021iwq,Lima:2021las,Islam:2021ful,Eichhorn:2022oma} are: (i) at fixed mass, the diameter of the critical curve, as well as all photon rings of regular BHs, is smaller than for a Kerr BH,  
(ii) the relative photon-ring separation increases, (iii) at finite spin (and non-face-on inclination), the photon-ring shape deviates from Kerr and can, in non-circular\footnote{The term ``circular'' here refers to a technical property of the spacetime.} regular BHs, become non-reflection-symmetric about the horizontal image axis and feature cusps and dents.

Property (i) is testable with the ngEHT, if an additional, independent mass measurement is used, e.g., from stellar orbits. In fact, the EHT observation of M87* already provides a first constraint on a regular BH with particularly large deviations from Kerr \citep{Eichhorn:2022oma}, while other regular BHs remain unconstrained \citep{EventHorizonTelescope:2021dqv}.
Property (ii) may be testable for some regular BHs \citep{Eichhorn:2022oma}, where estimates show that the separation between $n=1$ and $n=2$ photon rings may reach several $\mu$as (for a shadow diameter corresponding to that of M87*), cf.,~Sec.~\ref{sec:multiring}.
Superresolution techniques may be particularly useful to constrain such regular BHs. Properties (iii) and (iv) are likely beyond the reach of ngEHT, although studies accounting for finite resolution have not been conducted yet.

In addition, the overall lensing structure of these spacetimes differs from Kerr, therefore a spacetime tomography approach~\citep{Broderick:2005my,Broderick:2005jj,Tiede:2020jgo,Wong:2020ziu} is also likely to provide constraints on regular BHs.
Studies of localized emission regions around regular BHs are, however, currently in their infancy \citep{Eichhorn:2022oma}. Further, there may be polarization signatures \citep{Liu:2022ruc}, which is also not yet well-explored.

\noindent\paragraph{\bf Fuzzballs\\}
In string theory, BHs are argued to 
be replaced by an exotic compact object (ECO), i.e., a fuzzball. 
Fuzzballs have been motivated by the individual microstates of fluctuating BH geometries in string theory~\citep{Mathur:2005zp, Hertog:2017vod,  Mayerson:2020tpn}. For a specific class of five-dimensional extremal BHs, the relevant microstates have been shown to successfully reproduce the Bekenstein-Hawking entropy~\citep{Strominger:1996sh, Callan:1996dv}. Some individual microstate geometries turn out to be regular and horizonless~\citep{Balasubramanian:2008da}. The fuzzball paradigm conjectures that classical BH geometries are the coarse-grained superposition of these individual microstates. In view of classical instabilities~\citep{Keir:2016azt,Cardoso:2014sna,Eperon:2016cdd,Marolf:2016nwu}, stability remains an open question. Uncharged, non-supersymmetric, and four-dimensional microstate geometries of direct relevance for astrophysical BHs are challenging to construct~\citep[see][for a recent construction of charged, four-dimensional, and non-supersymmetric microstate geometries]{Bah:2021jno}. The latter break reflection symmetry about the equatorial plane which suggests an asymmetry in their shadows. Moreover, these geometries do not exhibit an ergosphere, which poses a challenge with regard to a viable jet-launching mechanism.

Ray-traced images of the celestial sphere in charged solutions of supergravity interpreted as effective fuzzball geometries have been computed in~\citet{Bacchini:2021fig}. When a deformation parameter is tuned towards the extremal BH limit, geodesics that enter the interior of the effective fuzzball geometry encounter increasingly large curvature close to timelike surfaces of infinite redshift. The growing redshift (between an emission region at high curvature and an observer at asymptotic infinity) has been argued to increasingly darken the appearance of these effective fuzzball geometries~\citep{Bacchini:2021fig}. Quantifying this conjectured darkening could provide a route to constrain the deformation parameter.

\noindent\paragraph{\bf Non-hidden wormholes and naked singularities\\}
We refer to non-hidden wormholes as spacetimes with two (or more) asymptotically flat regions which are not separated by a horizon. Similarly, we refer to naked singularities as asymptotically flat spacetimes with a curvature singularity which is not hidden by a horizon.
Such objects can even occur as vacuum solutions of GR: for instance, the Kerr solution itself can describe a naked singularity if the spin parameter exceeds its extremal value, i.e., $|a_{*}|>1$. 
Whether or not naked singularities can dynamically form is connected to cosmic censorship~\citep{Wald:1997wa}.
Their physical viability is questioned by severe theoretical shortcomings such as geodesic incompleteness and/or closed timelike curves~\citep{Morris:1988cz}.
Nevertheless, they provide a useful testing ground to understand the capabilities of an ngEHT array. Non-hidden wormholes also arise from regular BHs, at sufficiently large values of the regularization parameter. In the limit of near-critical spin, even Planck-scale modifications of GR can result in a horizonless spacetime~\citep{Eichhorn:2022bbn}.

Images of naked singularities strongly depend on whether or not they feature a photon sphere. If they do, they can cast a critical curve
that may be similar to that of a BH and the respective deviations in the critical curve have been quantified~\citep{Abdikamalov:2019ztb,Kumar:2020yem,Bao2023}.
If they do not possess a photon sphere, there may still be strong-lensing effects, but image features typically appear more distinct from BHs~\citep{Shaikh:2018lcc}.
Accretion physics around naked singularities has been investigated in \cite{Joshi:2011zm,Joshi:2013dva}, finding potentially detectable differences in the high-frequency tail of the emission spectrum.
Accretion disks around naked singularities have been investigated in~\cite{Gyulchev:2019tvk,Gyulchev:2020cvo}, \cite{Shaikh:2019hbm}, and \cite{Deliyski:2023gik}.

In distinction to naked singularities, non-hidden wormholes need not contain curvature singularities. In contrast to BHs, they do not feature event horizons, irrespective of whether or not they have a photon sphere. 
Discriminating between wormholes and BHs via VLBI imaging is still not straightforward, at least at present angular resolutions. 
Indeed, it has been shown that many wormholes have unstable photon orbits and are able to cast critical curves very similar to BHs, even though in most models the critical curves have a smaller size than those of BHs with the same ADM mass \citep{Bambi:2013nla,Gyulchev:2018fmd,Amir:2018pcu,Brahma:2020eos,Bouhmadi-Lopez:2021zwt}. 
One promising signature to probe wormholes from their images is based on the fact that the observers may see photons falling into the mouth of their side, then reflected either from the throat or from the potential barrier near the photon sphere on the other side of the throat. Even the photons directly from the other side could be detectable.
Typically, the combination of these effects would form multiple light rings inside the dark spot of the image \citep{Shaikh:2019jfr,Wang:2020emr,Wielgus:2020uqz,Peng:2021osd}, see also \cite{Delijski:2022jjj} for the effect of polarization.
After blurring due to imperfect resolution, these effects appear as an overall enhancement of intensity within the dark spot \citep{Ohgami:2015nra,Ohgami:2016iqm,Paul:2019trt,Vincent:2020dij,Guerrero:2021ues,Eichhorn:2022fcl}.
This can be used to distinguish the images of wormholes from those of BHs with an ngEHT array, even though the EHT may not achieve this.

\subsubsection{New-physics effects in light propagation and matter dynamics}
\label{sec:propagation}
Most studies to-date of settings beyond GR account for changes in the spacetime, but use the standard form of the geodesic equation for ray tracing and, where considered, GR dynamics for the accretion disk. Going beyond GR, the geodesic equation can also, in principle, be modified. 
For instance, it has been known since the seminal work of \citet{Drummond:1979pp} that quantum effects can modify the propagation of light: quantum fluctuations of charged matter fields, most importantly the electron, give rise to the Euler-Heisenberg effective action, which exhibits terms that couple the electromagnetic field strength-tensor to the Ricci scalar, Ricci tensor and Riemann tensor. Even though for GR BHs, the latter term is negligible; for BHs beyond GR, Ricci-flatness need not hold and all three terms may be present. \citet{Drummond:1979pp} found that gravitational lensing in the Schwarzschild metric becomes polarization dependent through such terms, although of an unmeasurably small amount for the Solar System -- the case may be different for BHs, although a quick estimate of the type of term, $F^2\, R_{\mu\nu\kappa\lambda}R^{\mu\nu\kappa\lambda} \frac{\hbar^2}{m_e^2\, c^2}$, indicates that it remains a tiny correction to classical electrodynamics even close to the horizon of a supermassive BH. 
On the one hand, this particular example, which arises within Quantum Electrodynamics on a classical GR background, serves as a reminder that quantum effects, may in principle modify the propagation of light through a given geometry. Therefore, working with the standard geodesic equation, as is done in most of the literature, may turn out to be a restriction on the physics beyond GR. On the other hand, the example highlights that within quantum field theory on a GR background, such effects are likely completely negligible. However, there are settings, such as, e.g., regular BHs with large values of the deviation parameter, in which the new-physics scale is rather low. In such settings, the new physics may also affect the propagation of matter. Assuming the standard null geodesic equation when calculating the image of such BHs may thus amount to a certain amount of tuning, in that the new-physics scales in the geometry and the matter sector differ from each other.

Besides quantum effects, classical modifications that involve non-minimal coupling between the electromagnetic field strength and curvature terms may lead to deviations in the geodesic equation. In \citet{Bertolami:2008zh} it has been shown, however, that coupling an arbitrary function of the curvature scalar to the matter Lagrangian only leads to modifications of the massive geodesic equation, not the massless one. Such terms therefore affect the dynamics of the accretion disk, but not the propagation of photons. As a second example, \citet{Allahyari:2020jkn} consider a Horndeski coupling between photons and curvature and perform a parametric estimation of effects on the BH shadow, from which they constrain the corresponding coupling.

To the best of our knowledge, a systematic inclusion of new physics in the propagation of photons and the dynamics of the accretion disk has not yet been attempted. Such investigations must of course also account for experimental and observational constraints from various other settings, including astrophysical observations as well as laboratory tests.

Violations of Lorentz symmetry resulting in a modified dispersion relation for photons could in principle lead to potentially observable signatures when images at different frequencies are available. This can be quantified considering a generic modified dispersion relation~\citep{Colladay:1998fq,Liberati:2012th}:
\begin{equation}
E^2=(cp)^2+f^{(n)}\frac{(cp)^n}{M_{\rm P}^{n-2}},\qquad n>2.    
\end{equation}
The group velocity $c_{\rm g}$ is given by:
\begin{equation}
c_{\rm g}=\frac{\partial E}{\partial p}=c\left(\frac{cp}{E}+\frac{n}{2}\frac{f^{(n)}}{M_{\rm P}^{n-2}}\frac{(cp)^{n-1}}{E}\right)=c_{\rm p}\left(1+\frac{n}{2}f^{(n)}\frac{(cp)^{n-2}}{M_{\rm P}^{n-2}}\right),
\end{equation}
where we have defined the phase velocity $c_{\rm p}=c^2p/E$.
For two values of momenta $p_1$ and $p_2$, there is a displacement of the photon sphere:
\begin{equation}
\Delta r_{\rm ph}=3GM\left|\frac{1}{c_{\rm g, 1}^2}-\frac{1}{c_{\rm g, 2}^2}\right|.
\end{equation}
We can estimate the order of magnitude of the constraints that can be placed on $|f^{(n)}|$, for a given value of $n>2$, by equating the displacement above with the angular resolution of ngEHT (there will be a numerical difference between the radius of the photon sphere and the apparent size of the photon ring, but both have the same scaling with the group velocity). For frequencies of 230 and 345 GHz, respectively, the deviation in the speed of propagation from the relativistic speed of light is of the order $10^{-23}$, which does not lead to a detectable effect in the achromaticity of the photon ring. 
In fact, such modified dispersion relations for photons are already tightly constrained (with the leading-order term constrained to be transplanckian) with other observations \citep{Addazi:2021xuf}.

\subsection{Science cases}
\label{sec:beyond-GR-science-cases}

As summarized above, many of the compact objects motivated in various beyond-GR scenarios share qualitatively common features among each other that distinguish their shadow images from the Kerr case, at least in principle. Here, we delineate a program to test the following generic features.
\begin{itemize}
    \item[\textbullet] First, horizonless objects such as boson stars, gravastars, fuzzballs as well as horizonless regular spacetimes, non-hidden wormholes and naked singularities, may mimic the \textit{central brightness depression} of a BH. However, the central brightness depression is typically less pronounced than for a BH. Accordingly, a high dynamic range enables a better distinction of BHs from these horizonless spacetimes, cf., Sec. ~\ref{sec:central-brightness-depression}.
    \item[\textbullet]
    Second, BHs beyond GR (both singular ones that may occur in modified-gravity theories and regular ones that may be motivated by quantum gravity) often feature deformations of the $n=1$ (and higher-order) photon ring, some of which are non-degenerate with the spin. Thus, a high-enough resolution and confident extraction of the lensed emission
    may constrain \textit{parametric deviations} from Kerr spacetime, cf., Sec.~\ref{sec:parameterized}.
    \item[\textbullet]
    Third, all BHs cast an \textit{inner shadow}\footnote{Diffuse foreground emission can obscure the inner shadow.}, i.e., a central dark image region which is bounded by the direct image of the horizon, cf., Sec.~\ref{sec:inner_shadow}. Just like the photon ring(s), the inner shadow can be deformed in shape and size in scenarios beyond GR.
    \item[\textbullet]
    Fourth, some alternative spacetimes can lead to a significantly larger separation between different photon rings (in comparison to the Kerr spacetime). This applies to horizonless spacetimes and the respective occurrence of inner photon rings. It also applies to a potential distinction between the $n=1$ and the $n=2$ photon rings. Observational searches of \textit{multi-ring structures}, cf.,  Sec.~\ref{sec:multiring}, can thus constrain deviations from GR, even if the $n=2$ photon ring of the Kerr spacetime remains unresolvable.
\end{itemize}
This motivates testing for such image features without theoretical bias and, where possible, in systematic parameterizations beyond GR. Below, we summarize the current status of this effort. It is important to keep in mind that testing GR with spacetime images is a fast-evolving field. Thus, the above list of signatures may not be exhaustive and further promising image features may be added to the ngEHT effort in the future.

\subsubsection{Horizonless spacetimes and their central brightness depression}
\label{sec:central-brightness-depression}

In GR, the defining characteristic of a BH is its horizon, a one-way membrane that can only be crossed inwards. This behavior is a paradigmatic illustration of the strength of the gravitational interaction in its nonlinear regime, and is therefore of essential importance as a test of GR and the Kerr hypothesis.

While this short introduction does not aim at capturing the important technical details in the mathematical definition of horizons, it is important to stress that there are several definitions that have differing motivation and scope. The prevailing definition, both for historical and popularity reasons, is the notion of an event horizon~\citep{Penrose:1969pc,Hawking:1976ra,Hawking:1973uf}. This definition has nevertheless several important drawbacks, which includes its global nature and the associated impossibility of observing an event horizon in any physical experiment~\citep{Visser:2014zqa}. This has motivated the search for quasi-local definitions of horizons, such as apparent, dynamical or trapping horizons, which can be coincident depending on the situation~\citep{Hayward:1993wb,Ashtekar:2004cn,Gourgoulhon:2008pu}.

In general, any horizonless spacetime can thus be modelled by coefficients that quantify (i) absorption ($\kappa$), (ii) reflection ($\Gamma$), (iii) re-emission ($\tilde{\Gamma}$), and (iv) transmission ($\tau$) of the central region~\citep[cf.,][]{Carballo-Rubio:2022bgh}. These coefficients satisfy a sum rule $\kappa+\Gamma+\tilde{\Gamma}+\tau=1$ coming from energy conservation. In spherical symmetry, we can also define an effective radius $R$ of the central object. These parameters take different values depending on the model being considered, and are associated with different image features as we will discuss in more detail below. We already know from GR that total absorption ($\kappa=1$) results in a central brightness depression. The more dominant the absorption gets, the closer the horizonless object will mimic a BH. Thus, no experiment can exclude the possibility of horizonless spacetimes. 
However, the ngEHT will tighten experimental constraints on these coefficients. Such quantitative constraints are crucial since they can
exclude beyond-GR scenarios in which the resulting horizonless objects are predicted to exceed these constraints.

In this section, we will describe the image features associated with each of these coefficients $\Gamma$, $\tilde{\Gamma}$ and $\tau$, as well as analyzing their observability. The brightness of these features is linearly proportional to the corresponding coefficient, but their characteristics are a function of the effective radius of the horizonless object. The linearity of the problem implies that we can discuss these features independently, focusing on one of them at a time without loss of generality.

\begin{itemize}
\setlength\itemsep{0.5em}
    \item[\textbullet] The effective radius $R$ can be constrained to be below $R\lesssim 3\,r_{\rm g}$ if the ngEHT finds clear evidence of a photon ring (see Sec.~\ref{sec:Light_ring} for a detailed discussion).

    \item[\textbullet] Reflection on a physical surface has been studied, and compared with EHT images for some values of the coefficients $R$ and $\Gamma$, in~\citet{SgrA_PaperVI} and \citet{Carballo-Rubio:2022bgh}, the latter providing an extensive exploration of the parameter space, and a quantitative analysis of the observability of these features by EHT and a tentative ngEHT configuration based on the corresponding values of angular resolution and dynamic range. The horizonless spacetime metric in that case is a spherically symmetric shell that reflects a fraction $\Gamma$ of all incoming rays. For ultracompact objects, simulated images show an inner set of photon rings, cf., upper panels in Figure~\ref{fig:ht1}, which cannot be resolved by the EHT as the application of a Gaussian filter shows (lower panels). On the other hand, improvements in image dynamic range and angular resolution such as those expected to be achievable by the ngEHT can noticeably change the situation, leading to a constraint $\Gamma\lesssim 10^{-1}$, at least for $i=0^\circ$ (see Figure~\ref{fig:ht2}).

    \begin{figure}[ht]
    \centering
    \includegraphics[width=0.99\textwidth]{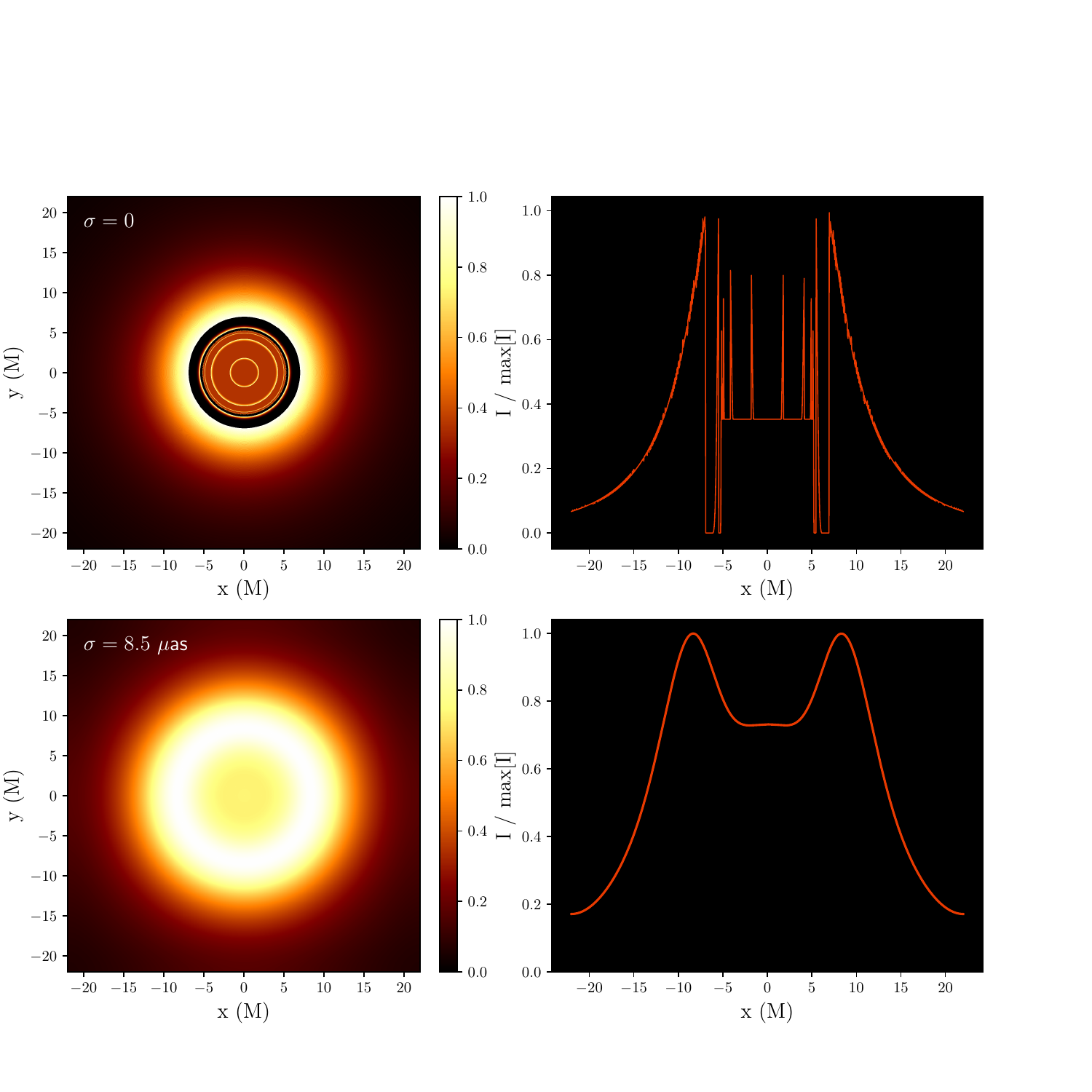}
    \caption{Images of spacetimes where specular reflection takes place, but with partial absorption ($\Gamma=0.5$) and an intrinsic brightness ($\eta=10^{-2}$) included. We take an inclination $i=0^{\circ}$, $\epsilon=10^{-3}$, without filter (top row) and a Gaussian filter with the EHT angular resolution of $20\ \mu\mbox{as}$ (bottom row). We see that these values of $\eta$ change appreciably the structure of the central depression in brightness.
    \label{fig:ht1}}
    \end{figure}

    \begin{figure}[ht]
    \centering
    \includegraphics[width=0.99\textwidth]{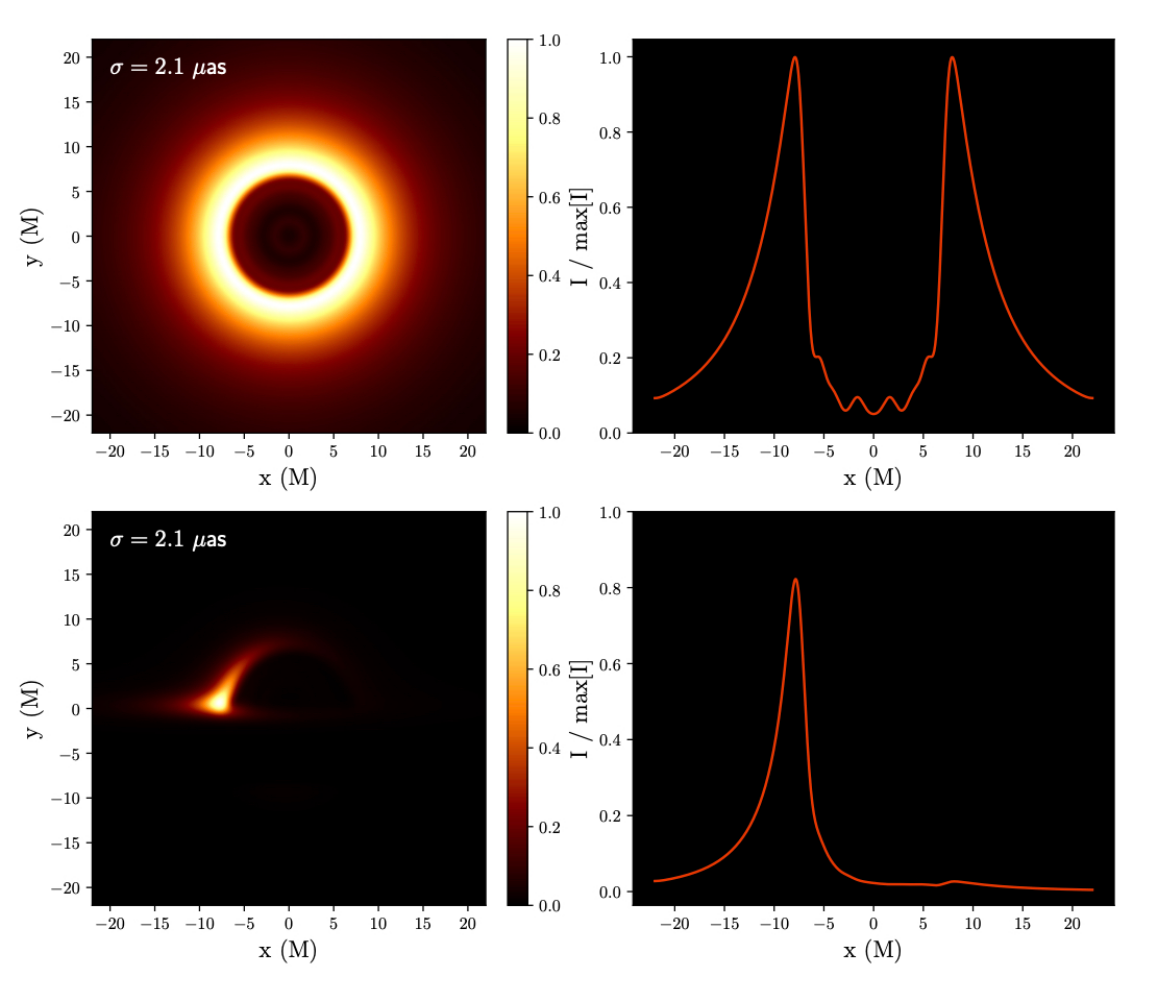}
    \caption{Results of applying a Gaussian filter with an angular resolution of $5\ \mu\mbox{as}$ for $\epsilon=10^{-3}$, $\Gamma=0.5$, $\eta=10^{-3}$, $i=0^{\circ}$ (top row) and $i=85^{\circ}$ (bottom row). 
    The optimistic value of angular resolution of $5\ \mu\mbox{as}$ can pick up the innermost structure of the simulated image. However, higher inclination angles make it more difficult to discern the features associated with the existence of a surface.
    \label{fig:ht2}}
    \end{figure}
    
    \item[\textbullet] Re-emission from a physical surface has been also studied in \citet{Carballo-Rubio:2022bgh} within the framework of the same spherically symmetric model and using the same tools to assess the observability of the corresponding features, that take the form of a central region with uniform brightness. As shown in Figure~\ref{fig:ht1}, the size of the novel features associated with re-emission can be large enough that even the angular resolution of EHT is enough to pick up these features. In fact, for the ideal situation of $i=0^\circ$, it is possible to constrain the re-emission channel ($\eta\lesssim 10^{-3}$), but not the specular reflection channel. On the other hand, improvements in image dynamic range and angular resolution such as those expected to be achievable by the ngEHT can greatly improve the situation, leading to more stringent constraints on the re-emission channel ($\eta\lesssim 10^{-4}$) and specular reflection channel ($\Gamma\lesssim 10^{-1}$), at least for $i=0^\circ$ (see Figure~\ref{fig:ht2}).
    
    In addition, infrared observations constrain re-emission from a physical surface. For instance, Sgr A* and M87* cannot have a physical surface that is in equilibrium with the surrounding accretion environment~\citep{Broderick:2005xa,Broderick:2007ek,Narayan:2008bv,Broderick:2009ph,Broderick:2015tda}. However, gravitational lensing may prevent sufficiently compact objects from reaching equilibrium~\citep{Lu:2017vdx,Cardoso:2017njb,Cardoso:2019rvt}.
    Furthermore, more complete descriptions of the behavior of the surface, including rotation~\citep{Zulianello:2020cmx} and absorption~\citep{Carballo-Rubio:2018jzw,Carballo-Rubio:2022imz}, can have an important impact on the features of the re-emitted radiation and delay reaching equilibrium, respectively. 
    Together with available lower bounds on $R$~\citep{Carballo-Rubio:2018vin}, such considerations can reduce the allowed parameter space (see also Sec.~4 in~\cite{SgrA_PaperVI} for Sgr A*).

    \item[\textbullet] The case without surface and with full transmission was recently investigated in \citet{Eichhorn:2022fcl}, including a quantitative analysis of the capabilities of a tentative ngEHT configuration. The horizonless spacetime metric in that case is an overspun, regular BH with Planck-sized deviations from the Kerr spacetime. Simulated images show an inner set of photon rings (upper panel in Figure~\ref{fig:horizonless}) which cannot be resolved by the EHT (lower left panel). A ten-telescope extension of the EHT, specified in the appendix of \citet{Eichhorn:2022fcl}, using a multifrequency reconstruction at 230 GHz and 345 GHz is capable of: (i) distinguishing the horizonless spacetime from the Kerr spacetime with the same disk model by a difference in the central brightness depression by a factor of about 40, (ii) showing non-concentric intensity contours in the shadow region, indicating that the ngEHT may be on the brink of resolving the inner photon rings, see right lower panel in Figure~\ref{fig:horizonless}. It is an intriguing open question as to whether superresolution techniques could resolve the inner photon rings.
    
    Less compact horizonless objects, in particular boson stars, have been analyzed in \citet{Vincent2016,Olivares:2018abq}. The enhancement of a central low-density region by gravitational lensing could in principle produce an image with a central brightness depression comparable in size to the shadow of a Kerr BH of the same mass, and with a similar morphology (see Figure~\ref{fig:kerr_vs_boson}). Numerical simulations of accreting boson stars have shown that for the family of solutions with minimal coupling $V(\Phi) \propto |\Phi|^2$ this effect is only present for the unstable members, and that for such cases the predicted size of the dark region is always smaller than that of the shadow of a Kerr BH with the same mass. For the parameters of Sgr A*, this difference is $\gtrsim 15~\mu$as and therefore distinguishable with present EHT capabilities \citep{Olivares:2018abq}. Nevertheless, the distinction could becomes more challenging for other ECOs.
    For instance, semi-analytical calculations for the case of Proca stars predict that stable members of the family can produce central brightness depressions which overlap in size with those in Kerr BH images \citep{Herdeiro:2021lwl} (see Figure \ref{fig:Schw-vs-Proca}). In this case, more precise tests are required to distinguish between a ring produced by MHD effects and a true photon ring produced by the capture of photons by an event horizon. These tests may include looking for deviations from circularity in the shape of the dark region due to the observing angle.

In fact, if lensing is weak (as it is the case for the Proca star shown in Figure \ref{fig:Schw-vs-Proca}), head-on views would produce more circular dark regions, while near-edge-on views would show ellipsoidal, elongated shapes \citep{Herdeiro:2021lwl}. 
Our ability to distinguish each case would improve with the maximum resolution achievable by the ngEHT, with the possibility of resolving the thin photon ring being of particular importance. Observations at different 
wavelengths would also play a crucial role. In fact, the apparent location of a true photon
ring should be achromatic, while that of a ring of stalled plasma would depend on the optical depth of the accretion flow at different frequencies.
Other features which distinguish images of simulated accretion onto boson stars from BH images include a smaller asymmetry due to Doppler beaming and the absence of relativistic jets \citep{Olivares:2018abq}. In part, the reason for this is that simulations have been performed only for nonrotating mini boson stars, where, for instance, the Blandford-Znajek mechanism cannot operate \citep{Blandford:1977ds}. Although it has been suggested that rotating boson stars with minimal coupling are likely unstable, this is not necessarily the case for Proca stars or for some cases with self interaction \citep{Sanchis-Gual:2019ljs}.
However, GRMHD simulations of 
surfaceless ECOs are still relatively rare in the literature, and additional studies may be
required to explore the conditions and level of confidence with which these objects can be distinguished from BHs.

\begin{figure}
    \centering
    \includegraphics[width=\linewidth]{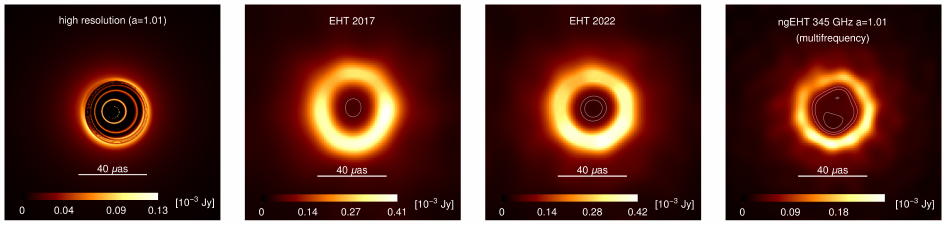}
    \caption{We compare the simulated high-resolution image (far-left panel) of a horizonless spacetime (generated by overspinning a regular BH to $a_*=1.01$) and its reconstructions using the ehtim toolkit~\citep{Chael:2018oym}, as seen by: (i) the 2017 EHT array (middle-left panel), (ii) the 2022 EHT array (middle-right panel), (iii) a multifrequency observation at 230 GHz and 345 GHz of a potential ngEHT array with ten additional telescopes (far-right panel). We also show (for the three reconstructed images) contour lines at $0.035$~Jy, $0.05$~Jy, and $0.065$~Jy (whenever they exist) to visualize the structure of the central brightness depression. See also~\citet{Eichhorn:2022fcl}.}
    \label{fig:horizonless}
\end{figure}

\begin{figure}
 \centering
\includegraphics[height=5.1cm]{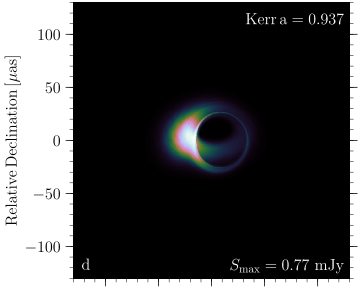}
\includegraphics[height=5.1cm]{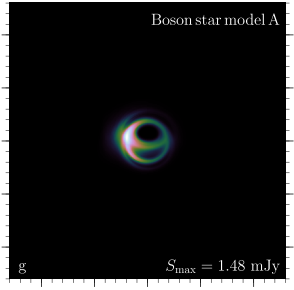}
\caption{Ray-traced images from GRMHD simulations of accretion onto a Kerr BH (left) and a boson star (right) with the mass and distance of Sgr A*. Despite the different size, these simulations show that under some circumstances, a horizonless, surfaces ECO can mimic the morphology of a BH
image by a combination of GRMHD and lensing effects. Figures taken from \citet{Olivares:2018abq}.
}
\label{fig:kerr_vs_boson}
\end{figure}

\begin{figure}
\centering
\includegraphics[height=8cm]{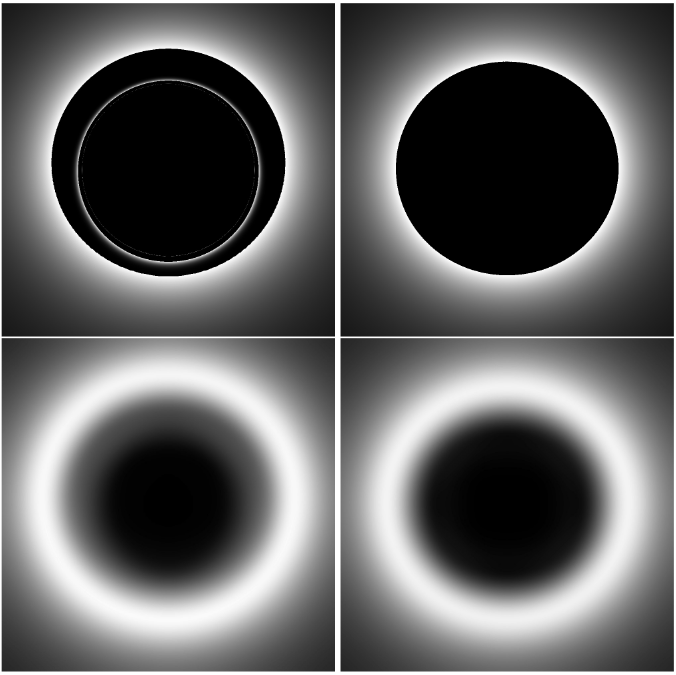}
\caption{Ray-traced images of a Schwarzshild BH (left) and a Proca star (right)
surrounded by thin accretion disks terminating at the location predicted by the spacetime properties. The lower panels are blurred by a Gaussian kernel, highlighting the possible degeneracy when observing at a single frequency without resolving the thin photon ring. 
Figure reproduced with permission from~\cite{Herdeiro:2021lwl}.
}
\label{fig:Schw-vs-Proca}
\end{figure}

\end{itemize}

\subsubsection{Parametric tests of the Kerr paradigm}
\label{sec:parameterized}

A complementary approach to the study of specific spacetimes beyond GR is to parameterize deviations from the Kerr spacetime as generally as possible \citep{1979GReGr..10...79B,Collins:2004ex,Vigeland:2009pr,Vigeland:2011ji,2011PhRvD..83l4015J,2013PhRvD..88d4002J,Cardoso:2014rha,2016PhRvD..93f4015K,2020EPJC...80..405G,Kocherlakota2020,Delaporte:2022acp}.
Because these parameterized spacetimes lack a Lagrangian origin and can be made as general as possible (given assumptions about the symmetries of the spacetime), they in principle provide theory-agnostic tests of the Kerr spacetime, under the assumption that a metric adequately captures all relevant gravitational degrees of freedom. 
Non-metric theories may thus fall outside the parameterized BH spacetimes in their current form. In practice, a comprehensive test of the underlying parameter spaces is difficult because of their high dimensionality.
The most general form of a parameterized spacetime contains several free functions of the spacetime coordinates. Assuming series expansions and truncating at finite order reduces this freedom to a finite set of free parameters. Explicit tests rely on (a) choosing specific functions (e.g., regular BHs can be embedded in parameterizations through specific choices of functions) or (b) working at finite order in the series expansion.

Parameterizations typically make assumptions about the spacetime and its symmetry properties:
\begin{itemize}
    \item[\textbullet] The most general parameterizations to date assume only axisymmetry and stationarity \citep{Delaporte:2022acp}.
    \item[\textbullet] In addition to axisymmetry and stationarity, one may assume circularity~\citep{Xie:2021bur}, which is an isometry that imposes conditions on the Ricci tensor, resulting in parameterizations with five \citep{2016PhRvD..93f4015K} or four free functions \citep{Papapetrou:1966zz}.
    \item[\textbullet] In addition to circularity, one may demand the existence of a Killing tensor \citep{1979GReGr..10...79B, Vigeland:2011ji, 2013PhRvD..88d4002J},  which implies a conserved quantity of the geodesic motion, generalizing the Carter constant and guaranteeing integrability. Beyond technical simplicity, there is no fundamental reason why theories beyond GR must satisfy circularity nor admit a generalized Carter constant.
    \item[\textbullet] Reflection symmetry about the equatorial plane can be preserved or broken~\citep{Chen:2021ryb}.
    The existence of a Killing tensor guarantees the vertical symmetry of the critical curve on the image plane, even if the spacetime is reflection asymmetric~\citep{Cunha:2018uzc,Chen:2020aix}.
\end{itemize}

\noindent One can utilize these spacetimes, model the surrounding accretion flow and then simulate an intensity image to see how these parameterized models can cast shadow images by solving the geodesic equations numerically. There have been various studies utilizing these spacetimes to investigate deviations from the Kerr metric \citep{EventHorizonTelescope:2020qrl, Younsi2016, Younsi:2021dxe, Volkel:2020xlc, Mizuno:2018lxz,Kocherlakota2020,Kocherlakota2022,Nampalliwar:2021oqr,Ayzenberg:2022twz,Nampalliwar:2022smp,EventHorizonTelescope:2022xnr}.
\begin{figure}
    \begin{center}
    \includegraphics[width=0.98\linewidth]{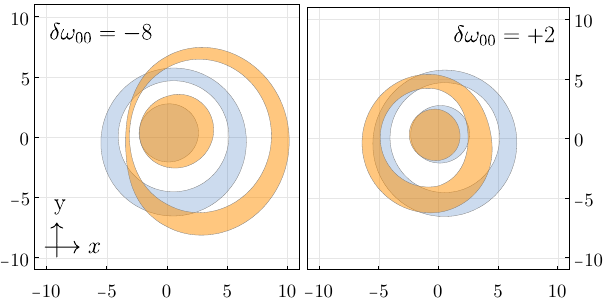}
    \\
    \includegraphics[width=0.98\linewidth]{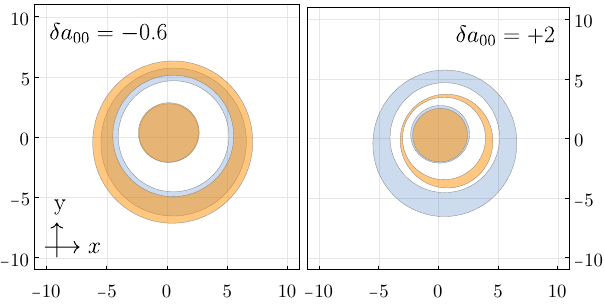}
    \end{center}
    \caption{We show the inner shadow (inner shaded region) and the first lensing band (outer shaded band) of deviations from a Kerr BH with spin $a_{*}=0.9$ viewed at near-face-on inclination of $17^{\circ}$. In each panel, we compare the Kerr case (blue-shading and the same throughout all panels)
    to spacetimes with varying KRZ parameters~\citep{2016PhRvD..93f4015K} (orange-shading) $\delta\omega_{00}$ and $\delta a_{00}$. The parameter $\delta\omega_{00}$ relates to deviations of the asymptotic spin parameter. The parameter $\delta a_{00}$ relates to deviations in the first parameterized post-Newtonian coefficients. See also \citep{Cardenas:2022}.
    }
    \label{fig:LB_KRZ}
\end{figure}
Observables which are most relevant to constrain parameterizations of spacetimes with a horizon are likely the photon rings.
In \citet{Younsi:2021dxe} and \citet{Ayzenberg:2022twz}, photon rings have been used to investigate constraints on parameters of circular metrics with and without Carter-like constants, obtained by analyzing simulated images of spacetimes with accretion disks.

One can also constrain the parameters in these parameterizations with other observations, e.g., GWs \citep{Cardenas-Avendano:2019zxd,Carson:2020iik,Shashank:2021giy}, 
or X-ray data \citep{Cardenas-Avendano:2019zxd,Yu:2021xen}, or even Solar System tests. Thus, if uniqueness theorems hold beyond GR, some parameter values are already too small to be further constrained by the ngEHT. 
However, beyond GR, BH uniqueness theorems do not need to hold and thus SMBHs may well correspond to different solutions from stellar-mass BHs. In this case, only constraints from observing the same object with different techniques are meaningful.
In Figure~\ref{fig:LB_KRZ} we exemplify how the inner shadow and the first ($n=1$) lensing band, i.e., the image region in which all lensed (equatorial) emission must occur, deforming when such constraints are set aside~\citep[cf.,][]{Cardenas:2022}.

\subsubsection{Image signatures of event horizons: the inner shadow}
\label{sec:inner_shadow}

BH can create unique image signatures through their extreme gravitational lensing and event horizon. These signatures are influenced by their surrounding accretion and emission properties. In the high-magnetic-flux MAD state of BH accretion~(\cite{2003ApJ...592.1042I,2003PASJ...55L..69N,2011MNRAS.418L..79T}; see also \cite{1974Ap&SS..28...45B}), which is favored by polarimetric EHT observations of \m87 \citep{EventHorizonTelescope:2021srq}, the magnetic pressure exceeds the gas pressure in the disk near the BH. In time-averaged simulation data, the near-horizon material forms a thin, highly magnetized structure in the equatorial plane.
This thin equatorial structure is the source of most of the observed 230~GHz emission. 

\begin{figure*}[t]
\centering
\includegraphics[width=\textwidth]{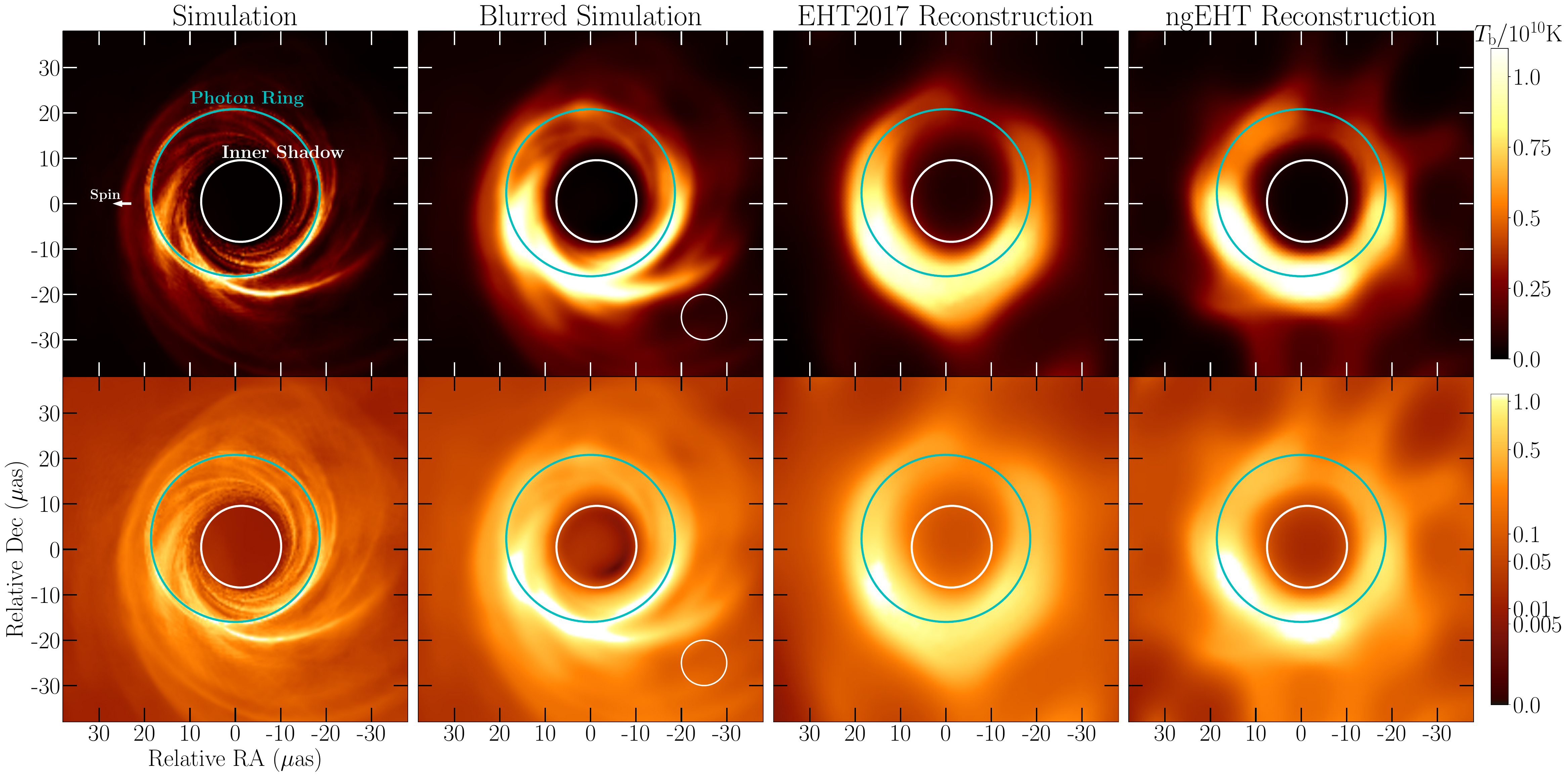}
\caption{Left: example GRMHD snapshot image of \m87 for a MAD accretion model onto a BH with mass $M=6.2\times10^9\,M_\odot$ and spin $a_{*}=0.9375$ \citep{2021ApJ...918....6C}. 
Center left: the same GRMHD snapshot image, convolved with a Butterworth filter with a cutoff frequency of $1/10\,\mu$as.
Center right: reconstruction of the simulation model from synthetic data generated on EHT2017 baselines.
Right: reconstruction of the simulation model from synthetic data generated from a concept ngEHT array. 
The top row shows images in a linear color scale and the bottom row shows the same images in gamma scale.
In all images, the white curve shows the boundary of the inner shadow, while the cyan curve shows the boundary of the shadow. The BH spin vector points to the left (East).  
}
\label{fig:Inner_Shadow_Reconstructions}
\end{figure*}

\citet{2021ApJ...918....6C} showed that MAD simulations of \m87 naturally exhibit a deep flux depression whose edge is contained well within the photon ring and critical curve (see Figure~\ref{fig:Inner_Shadow_Reconstructions}). 
This darkest region, or \textit{inner shadow}, corresponds to rays that terminate on the event horizon before crossing the equatorial plane even once.
This region is bounded by the direct image of the equatorial event horizon. More generally, as long as the emission near a BH is predominantly near the equatorial plane and extends all the way to the horizon, the darkest region in the observed image will correspond to the inner shadow. In addition, because of gravitational redshift, the image brightness falls rapidly near the edge of the inner shadow. As a result, detecting an image feature associated with the inner shadow requires images with high dynamic range.

The inner shadow of a Kerr BH has a significantly different dependence on its parameters than the critical curve \citep{Takahashi2004}. For instance, the critical curve of a Schwarzschild BH is always a circle with radius $\sqrt{27}~r_{\rm g}$, while the inner shadow of a Kerr BH has a size, shape, and relative displacement that depend sensitively upon the viewing inclination relative to the spin axis and the position angle of the projected spin axis on the sky. The photon ring and inner shadow thus provide complementary information. When considered independently, each is subject to degeneracies in its size and shape as a function of BH mass, spin, and viewing angle: these degeneracies can be broken via simultaneous observations of both features. 

In both GRMHD simulations with strong magnetic fields and in semi-analytic, optically thin disk models, the photon ring and the inner shadow are both prominent as observable features. In Figure~\ref{fig:Inner_Shadow_Reconstructions}, we investigate the ability of the EHT and ngEHT arrays to recover the inner shadow feature with simulated image reconstructions of a snapshot image from a radiative GRMHD simulation of \m87 \citep{Chael_2019}. We generate synthetic VLBI data from the 230~GHz simulation image using the EHT baseline coverage on 2017~April~11 \citep{EventHorizonTelescope:2019ths}. We also generate synthetic data using example 230~GHz and 345~GHz ngEHT coverage, assuming a flat spectral index in the underlying source model. This ngEHT concept array used here  \citep{Raymond_2021} adds 12 telescopes to the current EHT, substantially improving the EHT's baseline coverage, angular resolution, and imaging dynamic range. In both cases, we generated synthetic data including thermal noise and completely randomized station phases from atmospheric turbulence, but we did not include the time-variable amplitude gain errors that complicate real EHT imaging \citep{EventHorizonTelescope:2019jan,EventHorizonTelescope:2019ths}.

The second column of Figure~\ref{fig:Inner_Shadow_Reconstructions} shows the simulation image blurred to half of the nominal ngEHT resolution at 230~GHz (using a circular Gaussian blurring kernel of $10\,\mu$as FWHM). The remaining columns show the EHT and ngEHT reconstructions using synthetic data. Both reconstructions were performed using the \texttt{eht-imaging} library \citep{Chael:2018oym}; the settings used in imaging the 2017 data were the same as those used in \texttt{eht-imaging} in the first publication of the M87 results in \citet{EventHorizonTelescope:2019ths}. While the EHT2017 reconstruction shows a central brightness depression, its size and brightness contrast cannot be strongly constrained or associated with the inner shadow. However, the increased baseline coverage of the ngEHT array significantly increases the dynamic range, and the image reconstruction recovers better the position and size of the high-dynamic-range inner shadow that is visible in the simulation image blurred to the equivalent resolution.

This imaging test is idealized. We neglect realistic station amplitude gains and polarimetric leakage factors that complicate image inversion from EHT data. However, \m87 is weakly polarized, making accurate recovery of the total intensity image possible with no leakage correction \citep{EventHorizonTelescope:2019ths,EventHorizonTelescope:2021bee}, and image reconstruction of EHT data with even very large amplitude gain factors is possible with a relatively small degradation of the reconstruction quality using \texttt{eht-imaging}. In the ngEHT reconstruction, we assume a flat spectral index between 230~GHz and 345~GHz and simply stack the visibility data from simulated ngEHT observations at both frequencies.
A more realistic approach would solve for the spectral index between the two frequencies simultaneously with the image during the fit \citep{Chael_2022}.

This somewhat idealized example demonstrates that the ngEHT array \textit{could} constrain the presence of an inner shadow in \m87 if it is indeed present in the image. In particular, detecting this feature does not require dramatic increases in imaging \textit{resolution} (which, in the absence of a 230~GHz VLBI satellite, is limited by the size of the Earth) but does require increases in the imaging \textit{dynamic range}, which is limited by the sparse baseline coverage of the EHT array. Once its presence is established via imaging, parametric visibility domain modeling may better recover the size and shape of the inner shadow to higher accuracy than is possible from imaging alone \citep[e.g.,][]{EventHorizonTelescope:2019ggy}.

\begin{figure*}[t]
\centering
\includegraphics[width=0.95\textwidth]{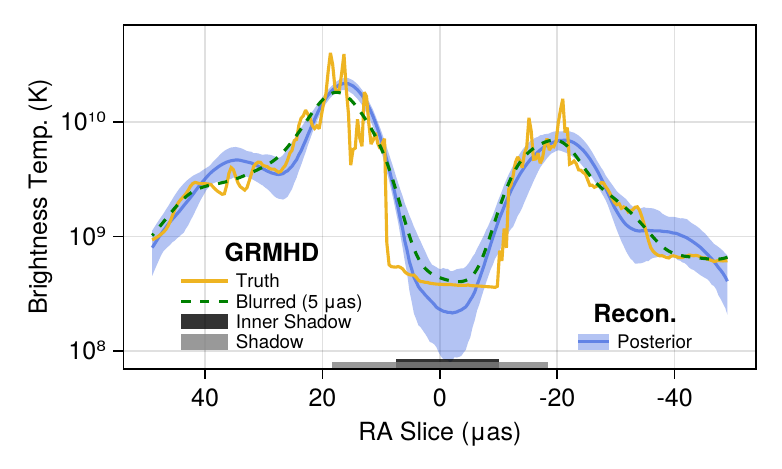}
\caption{Simulation and reconstruction brightness cross sections along the East-West axis. The yellow curve shows the 1D brightness profile of the \m87 GRMHD simulation snapshot in Figure~\ref{fig:Inner_Shadow_Reconstructions}. The dashed green curve shows the 1D brightness profile from the same simulation after convolution with a Gaussian with a FWHM of $5~\mu$as. The solid blue line shows the median brightness profile extracted from the \texttt{Comrade} ngEHT reconstructions. The band shows the 99\% posterior credible interval for the 1D brightness profile. The gray band along the bottom shows the region interior to the BH shadow, and the black band shows the inner shadow region. 
}
\label{fig:Inner_Shadow_Slices}
\end{figure*}

In Figure~\ref{fig:Inner_Shadow_Slices}, we show cross sections of the simulation and reconstructed images. The reconstructions are from the Julia \citep{Julia-2017} Bayesian VLBI imaging package \texttt{Comrade} \citep{Comrade}. \texttt{Comrade}'s imaging approach is similar to \citet{2020ApJ...898....9B, Pesce2021}, and fits a rasterized grid of pixels to the data. For our image model we used a $16\times 16$ raster with a $100~\mu{\rm as}$ field of view. A flat Dirichlet prior was chosen for the raster fluxes and we fit to visibility amplitudes and closure phases. The simulation image features faint foreground emission from the approaching relativistic jet, which lies in front of the bulk of the emission in the equatorial plane and provides a finite brightness ``floor'' inside the inner shadow. With the addition of new sites and short interferometric baselines, the ngEHT achieves the dynamic range necessary to identify the brightness depression associated with the inner shadow, which extends to levels 
$10^{2}$ 
dimmer than the peak of the emission for this simulation.

\begin{figure*}[t]
\centering
\includegraphics[width=0.99\textwidth]{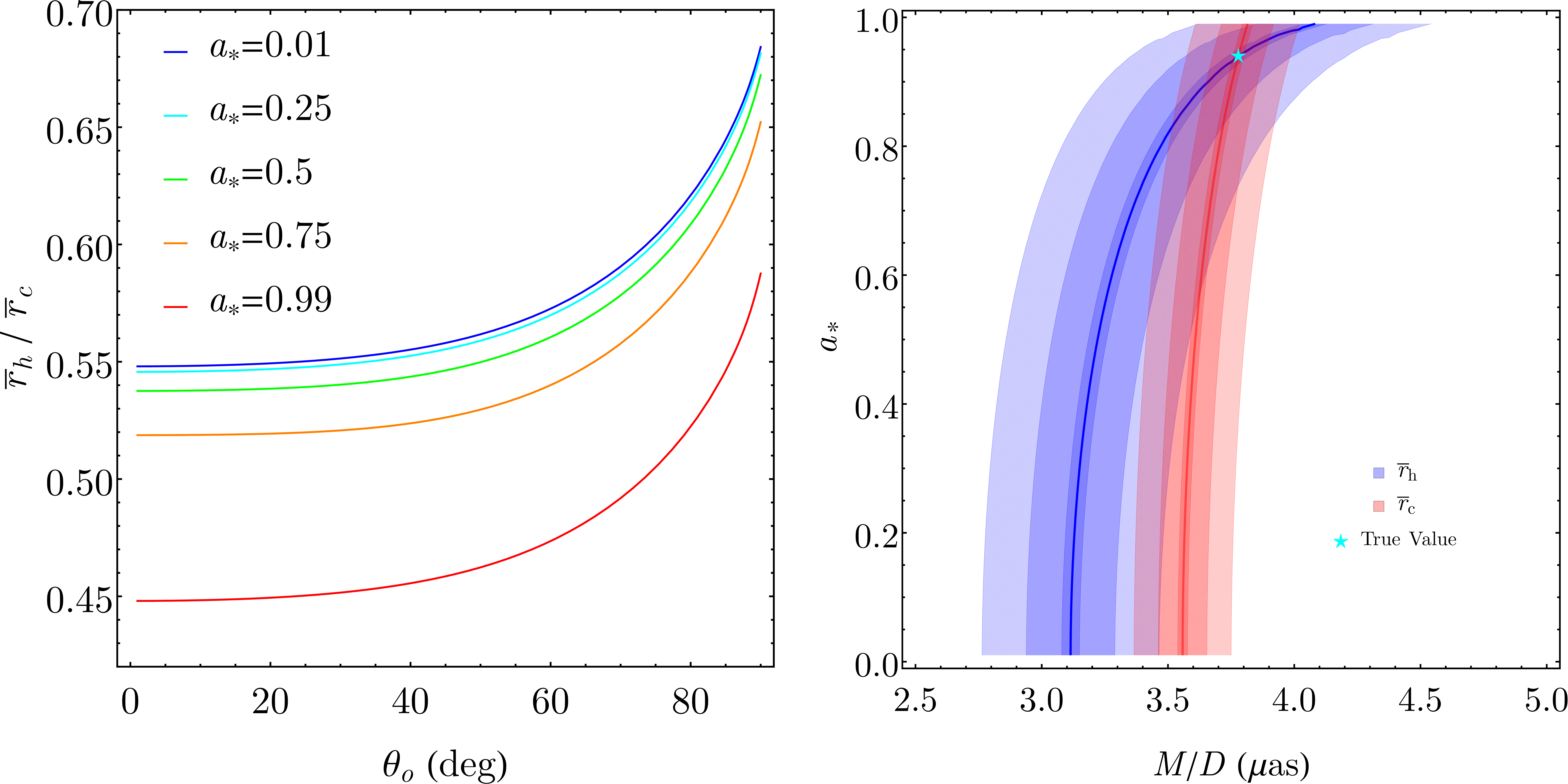}
\caption{Left: the ratio of the mean radius of the lensed horizon $\bar{r}_{\rm h}$ to the mean radius of the critical curve $\bar{r}_{\rm c}$. In the low-inclination case, $\bar{r}_{\rm h}/\bar{r}_{\rm c}$ shrinks from $\approx\!55$\% at zero spin to $\approx\!45$\% at maximal spin.
Right: simultaneous constraints on the BH mass-to-distance ratio $M/D$ and spin $a_{*}$ enabled by measuring the mean radius of the lensed horizon (blue, $\bar{r}_{\rm h}$) and critical curve (red $\bar{r}_{\rm c}$), when the inclination is fixed to $17^\circ$, as is appropriate for \m87 \citep{Mertens2016,CraigWalker:2018vam}.
Without fixing the mass, multiple values of $a_{*}$ provide the same result for the size of each feature, but combining a measurement of both features breaks this degeneracy.
The shaded regions show regions corresponding to $0.1$, $0.5$, and $1\,\mu$as errors on the radius measurement. 
The input mass scale and spin are $M/D=3.78\,\mu$as and $a_{*}=0.94$.
}
\label{fig:SpinMassConstraint}
\end{figure*}

Figure~\ref{fig:SpinMassConstraint} shows how measurements of both the mean radius of the inner shadow  ($\bar{r}_{\rm h}$) and of the critical curve ($\bar{r}_{\rm c}$) could be used to estimate BH parameters. In the low-inclination case, the ratio $\bar{r}_{\rm h}/\bar{r}_{\rm c}$ shrinks from $\approx\!55$\% at zero spin to $\approx\!45$\% at maximal spin. For $\theta_{\rm o}\lesssim30^\circ$, $\bar{r}_{\rm h}/\bar{r}_{\rm c}$ is approximately independent of the inclination, providing a pathway to measuring the BH spin. Importantly, measuring $\bar{r}_{\rm h}/\bar{r}_{\rm c}$ for an astrophysical BH would \textit{not} require accurate measurements of the BH mass $M$ or distance $D$.

Figure~\ref{fig:SpinMassConstraint} demonstrates how a simultaneous measurement of the radius of the critical curve and the lensed horizon could be used to constrain the mass and spin in \m87 when the inclination is fixed at $\theta_{\rm o}=17^\circ$ \citep{Mertens2016}.
These simultaneous constraints are analogous to those discussed in \citet{2022ApJ...927....6B}, which considers constraints from measuring multiple lensed images from a single face-on emitting ring.
The blue line shows the space of mass-to-distance ratios $M/D$ and spins $a_{*}$ that give the same mean lensed horizon radius for an image of \m87; the red line shows the same for the critical curve.
The red and blue lines intersect in only one location corresponding to the input BH mass $M/D=3.78~\mu$as and spin $a_{*}=0.94$.
The shaded bands around the intersecting lines show absolute errors in the radius measurements of $0.1,\,0.5,$ and $1~\mu$as.
Given a reported EHT radius measurement uncertainty of $1.5~\mu$as from geometric modeling of the EHT 2017 data in  \citet{EventHorizonTelescope:2019ggy}, measurements of the ring and inner shadow radius and centroid locations at $\lesssim\!1~\mu$as precision may be feasible with the ngEHT.
In addition to reducing uncertainty in the image size measurement itself, precisely constraining both features will depend on reducing systematic uncertainty in the relationship between the gravitational features and images from a set of plausible astrophysical models \citep[e.g.,][]{EventHorizonTelescope:2019ggy}.

\subsubsection{Resolvable multi-ring structures}
\label{sec:multiring}

\begin{figure}
    \begin{center}
    \includegraphics[width=\linewidth]{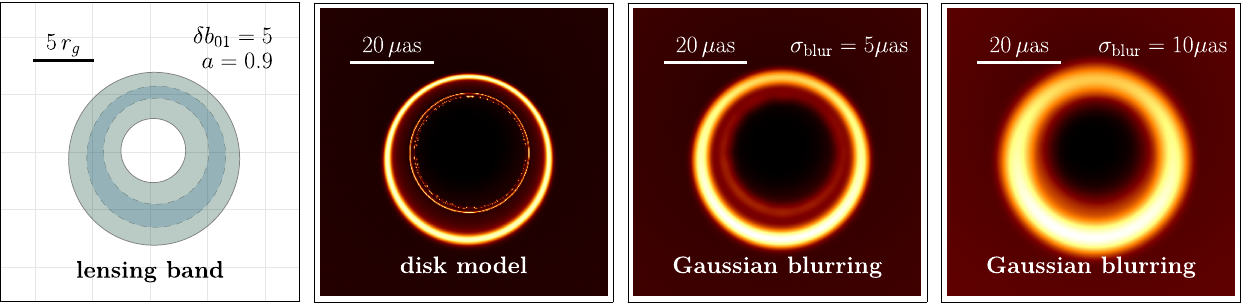}
    \end{center}
    \caption{Multi-ring structures for specific circular deviations of Kerr spacetime ($b_{01}=5$ in the KRZ parameterization~\citep{Konoplya:2016jvv}) at fixed spin parameter of $a_{*}=0.9\,M$ and viewed at the inclination of M87*. From left to right we show (i) the $n=1$ lensing band (for Kerr spacetime in blue-dashed and with deviation in green-continuous), cf.~\cite{Cardenas:2022}; (ii) the resulting image in this spacetime assuming a specific disk model, cf.~\cite{Eichhorn:2022fcl}; (iii) the same image but blurred with a Gaussian kernel with $\sigma_\text{blur}=5\,\mu\text{as}$; and (iv) with $\sigma_\text{blur}=10\,\mu\text{as}$. (For the translation of $r_g=M$ to the overall image scale in $\mu\text{as}$, we determine the maximum diameter of the convex hull of all image pixels with at least half of the average intensity per pixel.)
    \label{fig:LB_multiring_KRZ}
    }
\end{figure}

\begin{figure}
    \begin{center}
    \includegraphics[width=\linewidth]{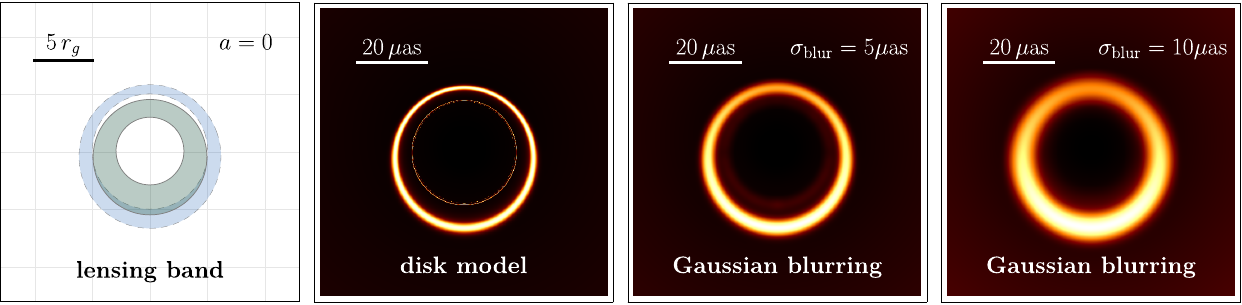}
    \end{center}
    \caption{As in Figure~\ref{fig:LB_multiring_KRZ} but for a non-spinning ($a_{*}=0$) regular BH with exponential falloff in the mass function as in~\citep{Simpson:2019mud}. The deviation is chosen near-critical such that if it is further increased, the regular spacetime would transition from a BH to a horizonless object. 
    \label{fig:LB_multiring_regBH}
    }
\end{figure}

In any given spacetime, the $n^\text{th}$-order photon rings (for $n>0$) 
necessarily lie
within finite lensing-band regions in the image plane, irrespective of the radial location of the emission region. With astrophysical priors on the emission region, these finite lensing-band regions can only become more restrictive. 
Moreover, each lensing-band region of order $n$ contains all higher-order ($m>n$) ones and thus provides a theoretical upper limit on the maximum radial separation of successive photon rings in the image plane (cf., Sec.~\ref{sec:Light_ring} for a discussion of the ngEHT capabilities of resolving photon rings in the Kerr spacetime). 

Some of the possible deviations from Kerr (Sec.~\ref{sec:parameterized}) can significantly broaden the $n=1$ lensing-band region~\citep[see][and Figure~\ref{fig:LB_multiring_KRZ} for an example]{Cardenas:2022}. For any such deviations, higher-order photon rings could become resolvable at an appreciably lower resolution than for the Kerr spacetime. This entails that the respective scenarios for new physics could be constrained by the non-observation of such multi-ring structures, even if the ngEHT resolution is insufficient to resolve multiple photon rings in Kerr spacetime. One such scenario to be constrained by multi-ring structures are regular BHs with an exponential falloff function (see Figure~\ref{fig:LB_multiring_regBH}).

Constraints on new physics from searches for multi-ring structures naturally benefit from increased (effective) resolution. However, they also benefit from an increase in dynamic range. This is because higher-order rings are less bright, as can be seen in the middle-right panels in Figures~\ref{fig:LB_multiring_KRZ}--\ref{fig:LB_multiring_regBH}, where the secondary ring appears far lower in intensity than the primary one.

\subsection{ngEHT challenges for numerical simulations}
\label{sec:numerical}
Numerical simulations played a crucial role in interpreting
the observations of M87* and Sgr A* obtained during the 2017
EHT campaign \citep{Collaboration2019_V}.
Libraries of synthetic images from GRMHD models have enabled
constraints on parameters of the accretion flow and the BH spacetime with increasing accuracy.
These include the BH mass, accretion rate, and
inclination angle. Under a few additional assumptions,
they also provide information on the magnitude of the BH
spin and the typical ratios between electron and ion temperatures \citep[e.g.,][]{Roelofs2023,Chatterjee2023c}.

From the point of view of testing of GR and the
Kerr hypothesis, GRMHD simulations can be used to understand the extent to which deviations from a model can be attributed to the spacetime, as opposed the those attributable to the
accretion flow model \citep{Bronzwaer2020}.
In fact, insights obtained from simulation-based modeling,
such as the deviation of the size of the bright ring with
respect to the actual BH shadow, have been
used by the EHT Collaboration to test the Kerr hypothesis and to study a few specific BH alternatives
\citep{Psaltis2020,EventHorizonTelescope:2021dqv,SgrA_PaperVI,Younsi:2021dxe}.
One of the most interesting prospects of the ngEHT is the possibility of probing the spacetime around supermassive compact objects in even greater detail.
In light of this, we list here what we consider to be some of the most
interesting research avenues that should be addressed using numerical
simulations (not necessarily in order of importance).

\noindent\paragraph{\bf GRMHD-informed semi-analytic models\\}
Synthetic images created from semi-analytic models of the accretion
flow are much faster to generate as compared to those from GRMHD models, 
offering clear advantages when fitting images by sampling over large regions
of the parameter space \citep{palumbo_bayesian_2022}.
This is especially important when considering alternatives to
Kerr BHs, for which the size of the parameter space
increases considerably. 
Semi-analytic models have been
used to explore emission from fluid configurations around
non-Kerr BHs and other compact objects
\citep[see e.g.][]{Vincent2016,Vincent:2020dij,Kocherlakota:2022jnz,bauer_spherical_2022,daas_probing_2022,Ozel2022,Younsi:2021dxe}.
However, in many cases these are limited to spherically or 
axisymmetric static configurations, and leave out much of the physics present in simulations, e.g., turbulence and temporal 
variability.
With increasingly detailed observations enabled through
the ngEHT, more detailed models will help to better constrain the properties of the spacetime and the accretion flow.

Work to characterize variability in GRMHD simulations has been performed previously \citep{georgiev_universal_2022}.
This characterization becomes increasingly relevant in light of the difficulties
of GRMHD models within the EHT library to match the observed variability
of Sgr A*, which could point to the necessity of considering different
accretion models or including additional physical processes \citep{Collaboration2019_V}.

Alternatives to Kerr BHs may come with additional features in the
accretion flow that are essential to include in realistic
semi-analytic models. For instance, models in which the event horizon
is replaced by a surface that exchanges momentum
with electromagnetic radiation
\citep{SgrA_PaperVI}
may be incomplete without considering
a similar exchange of momentum with matter.
It is to be expected that the interaction of the
relativistically-infalling plasma with a hard surface would produce
strong shock waves that could be detectable (notwithstanding redshift of subsequent emission) even when the surface itself
has a very large heat capacity and produces almost no thermal emission.

Another issue to consider is the possibility of images with ring-like
features that do not originate from a BH shadow.
This situation occurs even in several SANE~\citep{2012MNRAS.426.3241N} Kerr models of the EHT
simulation library, where the radius of maximum emission appears to be
related to the position of the ISCO \citep{Bronzwaer2020},
and for models where there is strong
emission coming from the jet base \citep{Collaboration2019_V}.
In non-Kerr spacetimes with a maximum in the rotation velocity profile
of circular geodesics, very clear ring structures may appear
as a result of the suppression of the magnetorotational instability (MRI) \citep[see][and Sec.~\ref{sec:bosonstar}]{Olivares:2018abq}.
This latter effect has already been considered in semi-analytic models of
horizonless compact objects \citep{Herdeiro:2021lwl}
However,
there are several questions that need to be answered, as its
robustness, or whether there are mechanisms
other than the suppression of the MRI that could produce similar
ring-like structures.
A better understanding of these features
would aid in constructing semi-analytic models of accretion onto alternative
compact objects with the correct emission geometry.

\noindent\paragraph{\bf Perturbative deviations from the Kerr metric\\}
General-relativistic hydrodynamic and GRMHD simulations
of accretion in alternative spacetimes are still relatively uncommon.
Examples include non-BH objects such as boson stars
\citep{Meliani2016,Meliani2017a,Olivares:2018abq,teodoro_tidal_2021},
BHs in specific alternative theories
of gravity \citep{Mizuno2018,Fromm2021,Roeder2022,Roeder2022a,Chatterjee2023b}
and BHs in theory-agnostic parameterized metrics
\citep{Nampalliwar2022,Chatterjee2023a}.
Some of these examples were conceived mainly as proofs of principle,
and are either extreme cases that exhibit properties that are easily
distinguishable from those of Kerr BHs, or consider accretion
regimes distinct from that expected around the main EHT targets.

However, observations of M87* and Sgr A* during the EHT 2017 campaign
show a remarkable agreement with the Kerr hypothesis.
This indicates that deviations with respect to the Kerr geometry are
likely small and can be treated as perturbations using the existing
parameterized expansions (see Sec.~\ref{sec:parameterized}).
The search for such deviations would benefit from a systematic study of
the effects that varying the leading order deviation parameters
have on the accretion flow properties and subsequent synthetic images.

Although the different deviation parameters span
a large parameter space, the regions to explore could be reduced
by exploring near to the Kerr models most favored by the EHT scoring
\citep{Collaboration2019_V},
and by making informed choices that account for existinf degeneracies,
as was done by \citep{Nampalliwar2022} for the degeneracy between
the spin and the size of the ISCO.
The parameter space exploration would also benefit from the use of
informed semi-analytic models as described above. 
     
\noindent\paragraph{\bf The structure of the accretion flow onto SMBHs\\}
All of the simulations used for the interpretation of the EHT M87* observations
and most of those used for Sgr A* are variations of the same physical scenario:
a rotation-supported torus in hydrodynamic equilibrium is initialized at a distance
of a few $r_{\rm g}$ from the SMBH, a weak poloidal magnetic loop is added
inside the torus, and the system is slightly perturbed.
Subsequently, the MRI~\citep{Balbus1991}
produces turbulence, which amplifies the magnetic field and facilitates angular momentum transport.
Depending on the dynamical importance of the magnetic field in the saturated state,
models are classified as SANE (less magnetized)
or MAD (more magnetized).

However, large scale simulations of accretion onto Sgr A* fed by stellar winds show
flow patterns that differ from this scenario in several aspects.
Most notably, they show that the MRI is relatively unimportant at several scales.
While at large scales the magnetic fields are weak and passively advected \citep{Ressler:2020tbr},
at horizon scales they accumulate and become dynamically, regulating
accretion in a way similar to MADs \citep{Ressler2020a}.
Due to computational limitations, however, few works have treated the problem using GRMHD \citep{Ressler2020a,lalakos_bridging_2022}.

For a correct interpretation of the ngEHT observations, it is essential to understand
in which regime the accretion flow pattern can be represented by typical tori simulations
and whether this regime is realized by in the environment of the ngEHT targets, or
if a change in the simulation paradigm is needed.
Of course, this consideration also applies to alternative spacetime geometries and compact objects, for which
uncertainties associated with the accretion flow need to be sufficiently understood
before drawing conclusions regarding the spacetime properties or the presence of new fundamental physics.

\noindent\paragraph{\bf Beyond ideal MHD\\}
In general, models based on GRMHD simulations show an excess in variability
when compared to observations at 230 GHz \citep{Collaboration2019_V}.
However, it is expected that non-ideal effects like viscosity and thermal
conduction will lead to a decrease in variability. Their inclusion
is motivated by the fact that plasma in the vicinity of Sgr A* and M87* is
practically collisionless, and these can be used in GRMHD simulations
to describe leading order corrections due to kinetic effects
\citep{chandra_extended_2015,Foucart2017}.
The consequences of these and other non-ideal effects like resistivity
\citep{Ripperda2019} for the source variability and other observable properties still needs to be investigated.

Another research direction that requires physics beyond ideal MHD is
the study of BH magnetospheres.
These regions are difficult to simulate using GRMHD and are more readily described using force-free electrodynamics (FFE).
This commonly leads to the choice of ignoring regions with
very high magnetizations when producing ray-traced images from
GRMHD data.
However, general relativistic simulations
combining the physics of GRMHD and FFE have already been performed in the
context of accretion onto millisecond pulsars \citep{parfrey_general-relativistic_2017}.
These kinds of simulations may become increasingly important in light of accretion scenarios such as those
described in \citet{Blandford2022,Blandford2022a},
in which magnetospheres play a crucial role in the
dynamics and overall flow pattern.

Similar to the discussion regarding the structure of the accretion flow, any attempt to probe gravity and new fundamental physics in the vicinity of BHs must take into account the limitations of the plasma model,
and the pursuit of research lines such as those outlined above is expected to contribute towards helping alleviate these current uncertainties.

\subsection{Outlook}
\label{sec:beyond-GR-outlook}

Many of the known beyond-GR scenarios, (cf., Sec.~\ref{sec:physics-scenarios-beyond-GR}) result in similar signatures for VLBI observations of SMBHs.
Amongst these signatures, we have identified as promising science cases for the ngEHT: (i) a less pronounced central brightness depression (cf., Sec.~\ref{sec:central-brightness-depression}), (ii) deformations of the $n=1$ photon ring (cf., Sec.~\ref{sec:parameterized}), and (iii) potentially resolvable multi-ring structures (cf., Sec.~\ref{sec:multiring}).
There may well be further relevant science cases in the future, given the field of testing GR and other theories of gravity with BH images is now evolving quickly.

We highlight the key role of BH uniqueness theorems: in GR, BH uniqueness holds and implies that the Kerr spacetime can be constrained by a range of different observations of distinct BHs, namely with electromagnetic signals, GWs, and with PN constraints. 
Beyond GR, the assumption of BH uniqueness is a very strong constraint on the theory.
For instance, even in one of the simplest extensions of GR, curvature-squared gravity~\citep{Stelle:1977ry}, there are several spherically-symmetric BH solutions at any given value of the asymptotic mass, albeit only one stable one~\citep{Brito:2013wya, Held:2022abx}.
There are therefore two distinct possibilities regarding the constraints that the ngEHT may impose.
\begin{itemize}
\setlength\itemsep{0.5em}
    \item[\textbullet] If BH uniqueness holds beyond GR, then many beyond-GR spacetimes are already constrained by other observations. Possible ngEHT signatures are then in many cases small and difficult to detect.
    \item[\textbullet] If BH uniqueness does not hold beyond GR, then the ngEHT has the potential to constrain the strong-field regime of SMBHs. Then, ngEHT signatures may be large enough to be detected.
\end{itemize}
There is by now a considerable body of theoretical work on images of spacetimes beyond GR. However, many of these are limited to determining the critical curve, which in itself is not observable. Images with disks have been generated for some spacetimes and enable us to better assess whether these spacetimes may be distinguished from the Kerr spacetime with the ngEHT.

By moving beyond static images, it is likely that much can be learned from a spacetime-tomography approach to spacetimes beyond GR. There is currently a gap in the literature in the sense that systematic investigations of how informative, e.g., time-dependent emission, hot spots and similar features can be, have not yet been conducted.

A further critical gap in the literature concerns the power of superresolution techniques.
In \cite{Broderick:2022tfu},
it has been demonstrated that, under additional assumptions, superresolution techniques allow one to reconstruct the $n=1$ photon ring.
Beyond GR, these techniques may be invaluable in imposing meaningful constraints.

From the phenomenological side, there are several observables for which little is known beyond their Kerr signature. This includes polarization, which is poorly investigated in spacetimes beyond GR. Similarly, the achromaticity of the ring could be a strong way to distinguish the Kerr spacetime from at least some settings beyond GR.
However, studies of images at several frequencies are so far mostly lacking in settings beyond GR.

%% file: Binaries.tex
\section{Exploring binary black holes with ngEHT} \label{sec:introbinaries}
In our dark energy, dark matter driven Universe, structures are expected to build up hierarchically from the merger of smaller scale collapsed haloes, suggesting that larger galaxies assemble by via mergers over cosmic time, potentially also involving primordial BHs as seeds of collapse. 
Since the growth of SMBHs hosted in galactic nuclei is understood to be driven by both accretion and BH-BH mergers, supermassive BH binaries (SMBHBs) are believed to be the natural outcome of galaxy mergers. 
In particular, mergers are expected to be the dominant growth channel for BHs hosted in galaxies that reside in dense environments, especially at high BH masses, the range that is accessible to ngEHT observations \citep{Kulier+2015, Weinberger+2018,RicartePN2018,Pacucci+2020}.
Binary BH mergers, or other strongly dynamical spacetimes such as collapsing configurations, may probe the strong-curvature regime of GR \citep{cardoso_2023}.
SMBBHs in the GW-driven regime are the critical missing piece to the assembly of supermassive BHs. The ngEHT can probe the most astrophysically relevant parameter space (total masses of order $10^8-10^9\,M_{\odot}$) that are challenging for both Pulsar Timing Arrays (PTAs) or LISA. 
Furthermore, multiple detections of binaries at different redshifts (especially with a fortunate simultaneous GW detection) probe cosmology. 
SMBHBs would be very efficient natural multimessenger laboratories for addressing ngEHT science goals, especially those described in Secs.~\ref{sec:Mass_spin}, \ref{sec:Tests_GR_Kerr}, \& \ref{sub-sec:Multi-mess_SMBHB}.

During SMBH mergers, dynamical friction and interactions with the stellar contents and accreting materials draws the two SMBHs to the nucleus of the newly created merger remnant \citep{Merritt_2005}. 
The evolution and fate of the SMBHB is dictated by a range of physical processes that operate to cause it to shrink via the loss of angular momentum in the orbit. 
The environmental interactions that drive the binary down to separations of ${\sim}0.1$--10\,pc are well understood, but the detailed mechanism(s) implicated in causing continued inspiral beyond this point, and in particular down to the sub-parsec scale regime in which GW emission takes over to shrink the binary orbit, still remain unclear \citep[e.g.,][]{Begelman_1980}.  
A number of viable solutions to this long-standing and so-called ``final parsec problem'' \citep{Armitage_2002,Milosavljevic_2003} have been proposed. For instance, interactions with gas in a circumbinary disk and three-body interactions with stars in the innermost regions could all contribute to, and have a significant influence on, the evolutionary timescale for the binary.
Uncovering the details of the physics during this last parsec of evolution and final merger informs the science cases of ongoing and future GW detectors such as PTAs, space-based GW interferometry (e.g., LISA), and other future advanced GW facilities.

Dynamically speaking, the formation and evolution of SMBHBs evolves through three main phases after galaxy mergers \citep{Begelman1980, armitage2002accretion, Colpi2014SSRv}: the pairing stage, the hardening stage, and the gravitational radiation dominated final coalescence phase. 
During the pairing phase, the binary separation is of the order of several $\text{kpc}$, and the two SMBHs migrate inwards towards the center through dynamical friction with gas until a compact binary is formed at separations of a few $\text{pc}$ \citep{Armitage_2002}. 
The second hardening phase involves interaction with the stellar population in the innermost regions and could also be modulated by the presence and availability of gas in the vicinity. 
It is not clear how long it is likely to take SMBHBs to reach the critical separation where angular momentum loss via GW emission starts taking over. 
The presence of gas in the inner regions has been demonstrated to facilitate this stage \cite{armitage2002accretion}. 
In idealized simulations, it appears that mergers and final coalescence can occur within a Hubble time, in agreement with theoretical work modeling the mass assembly history of SMBHs over cosmic time that requires BH mergers to explain the mass distribution of SMBHs as observed today.
The recent observational detection of a SMBHB in UGC 4211 with a 230~pc separation by ALMA opens up a new window into potential detection for candidates in the transition phase between the pairing and hardening stage \cite{Koss+2023}.
The lack of even indirect observational evidence for SMBHBs at the sub-parsec separation phase is impeding progress in this field and it is in this domain that the ngEHT stands to be transformative. 

The remnant evolution of SMBH binaries at the gravitational radiation phase is determined by their emitted GWs. 
The frequency of GWs emitted by SMBHBs at this phase enters into the PTA band, which is about $1~\text{per year}$ to $1~\text{per decade}$. 
Simultaneously, their angular separation assuming they are located at Gpc distances is roughly a few to several tens of $\mu$as, which is within the reach of the ngEHT. 
The combination of multi-band electromagnetic observations and PTA observations forms a multi-band and multi-messenger astrophysical era of SMBHBs.  
For smaller binary separations and shorter orbital periods the orbital decay is GW-driven and may outpace the viscous time scale in the disk, leading to a decoupling of the binary from the disk~\citep[see, e.g.,][]{Gold2019}. 
The transition where this decoupling occurs is obtained by equating the GW time scale with the viscous time scale of the disk, which depends strongly on the geometric thickness. 
For thin disks the decoupling could occur at separations relevant for the ngEHT. 
For geometrically thicker disks, accretion may proceed until smaller separations with mini-disk formation \citep{Paschalidis2021}.

Much like the study of binary stars, which are generally described by their method of discovery and observation, SMBHBs are apt to be found by the ngEHT in one of the following ways: transiting binaries, astrometric binaries, and telescopic binaries. 
We discuss these techniques and the candidates detected by each of these methodologies in more detail later in this section.  
A wide range of SMBHB candidates are detected via multiple techniques and these include periodically variable quasars~\citep{Orazio1457,Charisi:2016fqw} and quasars exhibiting broad emission lines that indicate high recoil velocities ($\geq 1000~\text{km}\, \text{s}^{-1}$) \citep{Eracleous_2012}. 
Among the periodically variable AGN, the low-luminosity AGN exhibit shorter mm-wavelength variation timescales and may be superior targets for the ngEHT as these SMBHBs may be resolved with relative astrometry~\citep{D_Orazio_2018}. 
Recently, the search for SMBHBs amongst quasars with offset broad lines using very long baseline arrays was conducted by~\citet{Breiding2021ApJ}, however, no double radio sources were found to be resolved. One possible reason for their non-detection is that the projected orbital separation lies within the limit of the current observing resolution~\citep{Breiding2021ApJ}. The ngEHT with significantly better resolution has the capacity to resolve potential candidates.

The ngEHT will have a nominal angular resolution of $\sim  15~\mu$as, which translates to a spatial resolution of $\leq 0.13$~pc across redshifts. 
Additionally the adoption of superresolution techniques might help improve this further, by factor of a few for imaging (\citealt{Chael_2016}, \citealt{Akiyama_2017a}, \citealt{Broderick_hybrid}) or substantially more for geometric modeling of simple structures such as displaced but individually unresolved emission regions. 
This means that the ngEHT can therefore \textit{spatially resolve} SMBHBs that have entered their steady-state GW emission phase. 
The orbital period at this stage is typically short (ranging from months to years), which makes it accessible to multi-epoch observations with the ngEHT.  
Furthermore, \citet{DOrazio_2018} estimate that between 1 and 30 sub-parsec SMBHBs should have millimeter flux densities in the $\gtrsim 1$~mJy regime and will hence also be accessible with the ngEHT.

The current best studied candidate SMBHB is the radio source OJ287 ($z = 0.306$). 
This source has been observed for over $120$ years and is a quasi-periodic flaring source. 
There is a well observed double flare structure with $\sim$12 year periodicity (as measured from Earth), or $\sim$9.2 years in the source proper time.
This difference has been attributed to the presence of an as of yet unseen secondary BH with mass $1.5 \times 10^{8} M_{\odot}$ passing through the accretion disk of a much more massive primary, with an estimated mass of $\sim1.835 \times 10^{10} M_{\odot}$ \citep[e.g.,][]{sillan_88, lehto_96, Dey-et-al-2018-b}.
The mass and period of OJ287 imply a Newtonian semi-major axis of 1.16 $\times$ 10$^{4}$ AU, corresponding to $12.4~\mu$as at the source redshift, and with the estimated eccentricity of $\sim$0.65, a maximum primary-secondary separation at apoapsis of $\sim 20.3~\mu$as.
This renders it directly resolvable by the ngEHT at its highest frequency. 
At periapse, the secondary would move at a proper (local) circular velocity of $\sim$8.11 $\times$ 10$^{4}$ km s$^{-1}$, or 0.271 c, corresponding to $\sim$14 $\mu$as yr$^{-1}$ as seen from Earth, slowing down to 1.72 $\times$ 10$^{4}$ km s$^{-1}$, or 0.057 c ($\sim$3 $\mu$as yr$^{-1}$) at apoapse. 
If the secondary BH in OJ287 were visible at 345~GHz, the ngEHT should be able to easily detect its absolute astrometric motion, while the motion of the primary would be $\sim$20 times smaller, and likely not detectable even with a decade-long observing campaign.  
The ISCO radius for OJ287 is $\sim$1100 AU or $\sim$1.2 $\mu$as, and therefore unresolvable by the ngEHT.

An alternate explanation for OJ287 has also been proposed. Detailed investigation of 120 epochs of VLBA observations of this source reveal that viewing angle changes due to a putative precessing (and nutating) jet could cause the morphological changes of the pc-scale jet as well as the radio variability that is observed~\citep{2018MNRAS.478.3199B}. The jet would complete a full orbit in projection in about 22 years, twice the dominant time scale observed in the optical waveband. It is likely that the optical emission is also produced by the jet. While a binary BH model or the Lense-Thirring (LT) effect is required to explain jet precession, the piercing of the accretion disk by a secondary BH is not required in this scenario.
By contrast, the obvious stability of the jet and jet motion do not support the plunging SMBHB interpretation, as it is incompatible with regular disturbances of the accretion disk. \citet{britzen_23} show that the phase of the precession relates to the variability of the Spectral Energy Distribution (SED). 
The precession model for OJ287 is further supported by other, independent observations. \citet{komossa} failed to detect the 2022 outburst predicted by the ``plunging''-model. Instead, OJ287 was at low optical–UV emission levels, declining further into November. 
The predicted thermal bremsstrahlung spectrum was not observed either, at any epoch. Furthermore, the authors estimate a SMBH mass of $10^{8}$ M$_\odot$ for OJ287 and confirm the mass estimate by \citet{2018MNRAS.478.3199B} based on the precession model. 
\citet{yuan} validate the plausible predictions of a jet with precession characteristics in OJ287 based on an archival study of VLBI observations at 2.3, 8.6, 15, and 43 GHz. 
The first GMVA plus ALMA observations reveal a compact and twisted jet extending along the northwest direction, with two bends within the inner 200 ${\rm \mu}$as, resembling a precessing jet in projection \citep{zhao}. 
Recent Space VLBI observations with RadioAstron at 22~GHz with an angular resolution of $\sim 150~\mu$as, or $\sim 40~r_{\rm g}$) spatial resolution, add to this physical picture a high brightness temperature, qualitatively confirming violent processes in the inner part of the source \citep{OJ287-RA-2023}. 
VHE flaring emission in OJ287 has been investigated by, e.g., \citet{lico_22}.

ngEHT observations could resolve the question of which of these clearly conflicting and competing models for OJ287 provide a proper description of the source. 
While OJ287 may be the favorite SMBHB in our own backyard for the ngEHT, our current understanding of the assembly history of BHs suggests that multiple resolvable sources ought to exist. 
At present, there is an ongoing effort to find and characterize additional viable SMBHB candidates besides OJ287.  

We briefly review the possibilities for the detection of SMBHBs via a range of observational techniques. As we have noted above, the ngEHT will have sufficient angular resolution to identify sub-parsec SMBHBs at any redshift, providing a powerful complement to GW observations of galaxy mergers from PTAs and future planned facilities like LISA.
At the present time there are several theoretical uncertainties with predictions of the merger rates stemming from our incomplete understanding of the astrophysical processes that effect coalescence, as well our lack of knowledge about the abundance and masses of initial BH seeds. This translates directly into a lack of secure predictions for the separation distribution for SMBHBs.
This is once again where data from the ngEHT could significantly alter and constrain theoretical models, permitting better-calibrated subsequent predictions. 
While the ngEHT is not a survey instrument, it will nevertheless be able to observe many binary candidate sources by utilizing sub arrays. 
In addition, key synergies with the next-generation VLA (ngVLA) can effectively address the limited field of view of the ngEHT.
%
\subsection{Multiple supermassive black hole systems}
Studies of binary and multiple stellar systems and their evolution have revealed that a significant proportion of stars are in multiples~\cite{Tokovinin1997}, with substantial evidence that some binary stellar systems were likely former triple systems~\citep{Eggleton2017}. 
It is known from planetary systems and close binary stars that therein a third body orbiting the binary impacts the orbit of the inner binary. 
This astrophysical coupling process is referred to as the von Zeipel, Kozai and Lidov mechanism \citep{vonZeipel1910, Kozai1962, Lidov1962}, and it explains a variety of phenomena~\citep{Naoz2016}. 
The mechanism is effective in the merger processes with the changes it can create in the inner orbit under certain conditions \citep{Antonini2017, Stephan2016}. 

A few candidates for multiple SMBH triple systems have been identified in recent years, e.g., SDSS J0849+111 \citep{Pfeifle2019, liu_2019}, J150243.09+111557.3 \citep{Deane2014}, NGC 7733-7734 \citep{Yadav2021}, and SDSS J1056+5516 \citep{Kalfountzou2017}. 
The presence of another SMBH sufficiently close to, and in orbit around, a SMBHB system can affect the orbit of the binary SMBH and accelerate the merger process significantly. 
As a result of the existence of a third SMBH in the vicinity, an oscillation will occur between the orbital eccentricity of the SMBHB in the inner orbit and the angle between the orbital planes of this and the outer SMBH. 
This complex interaction will play a crucial role in the convergence and merging of paired SMBHs and may shorten the merging times significantly. Of course, when examining this mechanism we note that some SMBHBs observed today could potentially be residuals of former triple systems. 
Simulations suggest that such tiered mergers could play an essential role in the evolution of galaxies \citep{Begelman1980, Khan2016, Hopkins2008}.

In this context, such sufficiently close triple SMBH systems, whilst observationally rare, if and when detected would provide an extraordinary opportunity to study a distinct set of astrophysical processes which serve to catalyze SMBHB mergers. 
While SMBH mergers are an essential ingredient in galaxy evolution, close triple SMBHs offer a unique opportunity for studying the dynamics of three-body interactions in GR. 
In a hierarchical triple SMBH system, the orbital eccentricity of the binary system in the inner orbit will exhibit variations over time, allowing it to grow to very large values (e.g., $\mathrm{e}>0.9$). 
The effect of the impact of a third SMBH seems likely to be detectable with low-frequency GW observations, especially at the periphery \citep{Merritt2013}.
Any direct evidence that can be provided from low-frequency GW observations for ngEHT triple candidates will in turn provide us with new information concerning binary SMBH coalescence.

\subsection{Taxonomy of SMBHBs}
\begin{landscape}
\begin{table}[htbp!]
\caption{Selected SMBHB candidates.
We list the estimated masses for both objects in the SMBHB.
In some cases, only the total mass of the potential binary system is known.
We also include below the first confirmed triple system.
References are: (1) \cite{2020MNRAS.495.4061H}, (2) \cite{2019Galax...7...72B}, (3) \cite{2018MNRAS.478.3199B}, (4) \cite{Dey-et-al-2018-b}, (5) \cite{2019MNRAS.487.3990B}, (6) \cite{2021MNRAS.501...50S}, (7) \cite{oneill}, (8) \cite{Rodriguez_2006}, (9) \cite{Bansal:2017izy}, (10) \cite{kharb_2017}, (11) \cite{liu_2019}.
}
    \centering{
    \begin{tabular}{c|c|c|c|c|c}
    \hline
    \hline
    Type of signal & AGN   & $z$ & Angular separation &  Masses ($M_{\odot}$) & References\\
    \hline
    \hline
    \makecell{\textit{Transiting Binaries} \\Self-lensing signal \\}  & 
    \makecell{``Spikey'' \\ KIC 11606854}
    &0.918&& $10^{7.4} $, $10^{6.7}$ &(1)\\
    \hline
    \textit{Astrometric Binaries} & & & &\\
    Jet precession & 3C~84&0.0176 & ? & $\lesssim 8\times 10^8$ & (2)\\
    Jet precession & OJ~287 & 0.306 & 0.001/0.01~pc & 3.96/2.96 $\times 10^{8}$, 4$\times10^{6}/1.04\times10^{8}$ & (3)\\
    Optical flaring & OJ~287 & 0.306 & $\sim$20.3 $\mu$as & $1.5\times 10^{8}$, $1.835\times 10^{10}$ & (4)\\
    QPOs & Mrk~501 & 0.0337 & 5~millipc & 7$\times10^{8}-10^{9}$ & (5)\\
     QPOs & 3C~454.3 & 0.859 & & $7\times 10^{9}-5.7\times 10^{10}$ & (6)\\
   QPOs & PKS 2131-021 & 1.285 & 0.001-0.01 pc & &\\
    \hline
    \textit{Telescopic binaries} & 4C +37.11 (B2 0402+37) & 0.055 & 7.3 pc & $15\times 10^{9}$ & (8), (9)\\
    & NGC 7674 (Mrk 533) & 0.028924 & 0.35 pc & $3.63\times 10^{7}$ & (10)\\
    \hline
    \textit{Apparent Binaries} & 1038+52\{A,B\} & & 33'' & &\\
    \hline
    Triple System & SDSS J0849+111 & 0.078 & within 5 kpc radius & $\sim 10^{7.5}$, $10^{6.4}$, $10^{6.7}$ & (11)\\
    \hline
    \hline
    \end{tabular}}
    \label{candidates}
\end{table}
\end{landscape}
These objects are mostly selected by their observed signatures, which could in principle overlap. 
At 1 mm wavelengths a 25--meter radio telescope has a beam width of $\sim$8 arcseconds.
In this instance the primary systems of interest are pairs of objects that could be observed simultaneously within the same beam-width.
Most of these correspond to classical astronomical binaries, but the relativistic nature of the orbiting objects introduces new effects and additional sets of observable quantities that are not available with main sequence stars. 
We list the best candidate sources in Table \ref{candidates}.

For some of these classes of binary candidates there may be degeneracies with regard to the physical origin of the observed periodic signal. 
Jet precession as well as periodic light-curve flaring can also be caused by the LT effect due to the frame-dragging of the rotating Kerr BH \citep{Thirring, LT}.
\subsubsection{Telescopic binaries} \label{sssec:telescopic}
In telescopic binaries the primary and secondary can both be seen directly and their motion measured. For the ngEHT, this would require both SMBHs to be radio-loud.
\begin{itemize}
\setlength\itemsep{0.5em}
        \item[\textbullet] Image or visibility stacking with source frequency phase referencing (SFPR), e.g., \cite{rioja_2011}, which could improve the detectability of weaker targets. This would help to reveal the secondaries in some sources, especially if there are ngEHT or other independent estimates of the orbital parameters. 
        \item[\textbullet] If observable they would represent the ``gold standard'' in SMBHBs. 
        \item[\textbullet] Large separation binaries are likely to be easily resolvable by the ngEHT out to cosmological distances, but their periods will be $\gg$ 1 per decade and their motions will therefore be harder to detect. 
 \end{itemize}
The ngEHT with a nominal angular resolution in the range of 5--15~$\mu$as and monitoring duration ranging from weeks to $\sim$10 years will be able to study telescopic SMBHBs within a broad range of masses and orbital geometries, at various redshifts inaccessible to other existing and prospective observatories.
\subsubsection{Astrometric binaries}

With an astrometric binary only one source is typically visible, and the binary is detected through a periodic ``wobble'' in the positions of its radio-loud jets, jet components, core, or BH shadow. 
From a detected periodic occurrence one can infer that the perturbation occurs due to the gravitational influence of an unseen companion. 
At least one BH within the binary system is required to be radio-loud, which means it is required to have a radio-loud jet. 
For this reason the expectation is that more ngEHT binaries will fall into this category than telescopic binaries. 
The best candidates are those AGN where the jet precesses (e.g., OJ287, 3C 84) or/and where periodic flux-density changes are observed (e.g., Mrk 501, 3C 454.3), and where the model parameters predict close separations (see Table \ref{candidates}). Astrometric binary candidates have been proposed for, e.g., 3C 279 \citep{AbrahamCarrara}, 3C 273 \citep{AbrahamRomero}, PKS 0735+178 \citep{Britzen2010}, 2200+420 (BL Lac) \citep{Caproni2013}, PG 1553+113 \citep{Caproni2017}, 3C 345 \citep{CaproniAbrahama2004}, 3C 120 \citep{CaproniAbrahamb2004}, 1308+326 \citep{Britzen1308}, TXS 0506+056 \citep{Britzentxs}, PKS 1502+106 \citep{Britzen1502}. Recently, a candidate SMBHB J2102$+$6015 has been identified on the basis of astrometric VLBI monitoring \citep{Titov+2023}. If confirmed as a SMBHB, this source would become an example of a  synergistic multimessenger bridge between ngEHT and prospective GW facilities \citep{Gurvits+IAUS375-2023}. 
All cases of astrometric SMBHB candidates need to demonstrate noticeable motion in order to be detected, and thus will observationally be selected by both the period and the angular separation of the binary. Long-term VLBI monitoring can resolve the orbital motion of a binary \citep{D_Orazio_2018} and this can be achieved both in the image and visibility domains.

We note two practical features of potential applications of the ngEHT for studies of astrometric binaries.
\begin{itemize}
\setlength\itemsep{0.5em}
\item[\textbullet] An extra factor of $(1+z)$ between the proper motion distance and the better determined angular diameter distance should be taken into account. The same factor results in the ``time dilation'' of the observed period of the binary over the actual period in its rest frame.
\item[\textbullet] As opposed to the case of telescopic binaries considered in Sec.~\ref{sssec:telescopic}, detection of an astrometric SMBHB by the ngEHT would require only one component of the binary to be a sufficiently strong radio emitter. 
In this case, the SMBHB observational signature would be a peculiar astrometric behavior or ``wobbles'' on an angular scale smaller than the nominal array resolution. 
An example of such an ``astrometric'' detection of a potential SMBHB is offered by the source J2102$+$6015 \citep{Titov+2023}.
\end{itemize}
Astrometric signatures of precession in AGN jets persist on long time-scales and are comparable to clocks or metronomes. \citet{britzen_23} argue that most of the blazar variability (morphology as well as light-curve) may be due to precession-induced phenomena (except for M87, Sgr A* and TeV blazars). These signals should not be confused with other interesting fluctuations due to plasma instabilites (current-driven or kink) of the flow which develop and disappear on shorter time scales and are lower energy phenomena which do not dominate the observational data.
%
\subsubsection{Spectroscopic binaries}
Spectroscopic binaries are detected as either ``single-peaked'' sources (with spectral emission from only one source) through the detection of periodically varying spectral line frequencies, or as ``double-peaked'' sources with two sets of spectral lines, one from each source.
\begin{itemize}
\setlength\itemsep{0.5em}
    \item[\textbullet] Single-peaked sources can in practice only be detected if the orbital period is short enough to see periodic variations in spectral frequencies.
    \item[\textbullet] Double-peaked spectral source lines would have frequency separations changing with the orbital phase, but could be detected as binaries even for very long orbital periods.
    \citet{Rubiner-2019-a} surveyed 20 double-peaked AGN in [O III] at $\sim$500 nm with the VLA whilst searching for telescopic binaries and found that ``one of them is a dual AGN (DAGN), while the other two could be either DAGN or AGN+ star-forming nuclei pairs.''
    An imaging snapshot ngEHT survey of double-peaked radio sources with indications of compact cores might be an extremely efficient detection strategy to find additional binary candidates for further study.
    \item[\textbullet] The large spectral shifts that would be found with sources approaching inspiral (at $\sim$ 0.3 c) suggests that this method should be utilized as a means of finding candidate multi-messenger binaries.  
    \item[\textbullet] An optical search for broad absorption line (BAL) quasars with periodically changing line frequencies might also be an efficient way to find binaries nearing inspiral.
\end{itemize}
We note that ``discoverability'' of spectroscopic binaries with the ngEHT might be problematic due to the likely long orbital periods (thousands of years). 
However, a potential ngEHT role in studies of spectroscopic binaries discovered by other facilities/techniques is warranted by the unique resolving power of ngEHT imaging in the otherwise unreachable range of angular scales.
\subsubsection{Relativistic transiting binaries}
Transiting binaries would appear superficially as photometric binaries, with quasi-periodic flaring.
However, the changes in brightness would be because of ``self-imaging'', one component imaging the photon orbit region of the other \citep{Davelaar_raytracing,Davelaar:2021eoi, 2021MNRAS.508.2524K}.
\begin{itemize}
\setlength\itemsep{0.5em}
\item[\textbullet] The photon orbit region of one component would be magnified by the other during the transit due to lensing. 
If the transit was exact, this would include a central drop, as the focal point passed over the shadow itself. 
\item[\textbullet] Transiting binaries would make it possible to accurately estimate the mass, orbital phase, and the non-Keplerian variations of the system's orbital parameters, and also permit super-resolving the shadow region of each component. 
A transiting binary with two visible sources (i.e., one that is also an optical binary) would offer a very sensitive laboratory for the study of fundamental gravitational physics.
\end{itemize}
    
\subsubsection{Reverberation binaries}
SMBHs act as omni-directional mirrors, reflecting some part of incoming light from any direction to any other direction, possibly after multiple orbits around the BH. 
These multiple orbits impose delays and thus temporal correlations on radiation received by a remote observer \citep{Chesler:2020gtw,Andrianov-et-al-2022-a}. Such binaries could be detected in the time domain, through the study of time delays due to photon orbits around the two photon spheres in a binary system. This section assumes Schwarzschild BHs, and approximation which does not change the qualitative nature of the BH reverberations. 

No radiation is received from the BH shadow itself, but around the shadow there is a series of increasingly sharp sub-rings, produced by photons that travel around the BH multiple times near the bound photon orbit \citep[see][and Sec.~\ref{sec:Light_ring}]{Chesler:2020gtw}.
Suppose there is a flare in the accretion disk of a solitary SMBH: a distant observer would first see a primary burst from the direct light ($n = 0$), then a delayed lensed burst coming from light partially orbiting the BH, delayed by a time $<$ T ($n = 1$), followed by a series of $n>1$ successively delayed light echoes, each separated by T. 
In the Schwarzschild approximation, for Sgr A* the scale is T$\sim$668 s (11.1 minutes), while for M87* the scale is T$\sim$1.03 $\times$10$^{6}$ s ($\sim$12.0 days).
     
There are several different ways that reverberation binaries may be detected and studied. 
Denoting the two binary components by ``1'' and ``2'', we start with the assumption that only component 1 is directly visible from the Earth.
\begin{itemize}
\setlength\itemsep{0.5em}
\item[\textbullet] A simple reverberation binary could be detected through the presence of two sets of autocorrelations in the source photometry. 
If the components have very different masses the two correlation trains would be well separated in delay space (e.g., 44 days versus 9 hours for the red shifted components of OJ287) and thus could be separated even for a one pixel source.
\item[\textbullet] If both components 1 and 2 are visible from Earth, the physical separation between 1 and 2 will introduce yet another delay due to path length differences in the time correlations. 
For example, for OJ287 the T = 44 day delay for one photon orbit of the primary is comparable to the one-way propagation delay between the components, which varies between $\sim$22 days and 132 days with the orbital phase. 
Observing a reflection of one component at another would therefore provide a direct determination of the distance between the primary and secondary, and thus a novel way of determining the inclination of the orbit. 
\item[\textbullet] Suppose the component 2 jet is not bright enough to be seen directly from Earth, but its jet is pointed directly at component 1 during its orbit (this would happen twice during each orbit, due to the dual sided nature of SMBH jets). 
This could then cause multiple quasi-periodic flares: from where the photon flux from the component 2 jet reaches component 1, the charged particle flux from component 2's jet reaches component 1 and excites its jet, alongside the various delays of the component 1 echos. 
\item[\textbullet] The delay of these secondary echos (the light travel time plus the partial photon orbit echo) will vary with the orbital phase of the two components, both because the orbital distance can be changing, and because the length of the $n = 1$ photon orbit 
will change with orbital phase. 
If the SMBH orbit is circular but not aligned with component 1's equatorial rotation plane, it may be possible to use the variation in this total delay to determine the spin of component 1.  
\end{itemize}

\subsection{Multi-messenger detection of supermassive black hole binaries}\label{sub-sec:Multi-mess_SMBHB}
SMBHB systems must emit GWs accompanied by electromagnetic counterparts, especially when they are in gas-rich environments \citep{Schnittman_2011, Dotti2012AdAst, Burke-Spolaor:2013aba, Kocsis:2007yu} which could be a result of galaxy mergers \citep{Sanders1988ApJ, Mayer:2007vk}.

We anticipate scientific synergies between the ngEHT and several key multi-messenger facilities that are already operational, as well as those that will become operational concurrently with the ngEHT. 
\begin{itemize}
\setlength\itemsep{0.5em}
\item[\textbullet] The EHT \citep{EventHorizonTelescope:2019dse} is already probing the SMBHB candidate OJ287 with highest resolution at 230 GHz. Quasars exhibiting broad emission lines with $\geq$ 1000 km ${\rm s}^{-1}$ velocity offsets with respect to the host galaxy rest frame have been discovered. 
These velocity-offset broad lines could be due to the dynamics of a SMBHB \citep[e.g.,][]{Breiding2021ApJ}.
\item[\textbullet] The ngVLA will be able to probe these system at angular scales as small as $\sim$ 3 mas--80 $\mu$as for the respective frequency range of 2.4--93 GHz \citep{2018ASPC..517.....M}.
\item[\textbullet] While Gaia observations use a novel technique to search for binary quasars at previously unreachable sub-kpc scales \citep[e.g.,][]{shen}, the unprecedented near-IR sensitivity, spatial resolution, and spectral coverage of the James Webb Space Telescope (JWST) will enable detailed study of the gas dynamics in binary quasars at high redshifts \citep{yuzo}. 
\item[\textbullet] The Athena (Advanced Telescope for High ENergy Astrophysics) mission has the broad aim of understanding the hot and energetic universe and will help unravel accretion processes and jet physics in SMBHBs \citep{piro}.
\item[\textbullet] The Cherenkov Telescope Array (CTA), a next generation ground-based very-high-energy gamma-ray observatory, will be a key instrument for multi-messenger astrophysics in the very-high energy (VHE, i.e., $>$ 100 GeV) range \citep{cta}. Due to its unprecedented sensitivity, rapid response, and capability to monitor a large sky area via a scanning mode of operation, SMBHs might be detectable by their specific flaring properties.
\item[\textbullet] On-going PTA projects aim at detecting GWs in the nanohertz band, dominated by the gravitational radiation emitted by SMBH binaries with masses in the range $10^{8}$--$10^{10}~M_{\odot}$ inspiraling at sub-parsec separations ($\sim$ $0.01 \text{pc}$).
\item[\textbullet] Future space-based GW detectors such as LISA \citep{LISA2017arXiv}, Tianqin \citep{Luo_2016}, and Taiji \citep{Wu_2017} are designed to detect the millihertz GWs of SMBH binaries with masses in the range $10^{5}$--$10^{7}~M_{\odot}$, emitting during their late inspiral and merger phases.
\end{itemize}
There are many proposed channels through which SMBHBs could emit EM counterparts, either simultaneously with their GW emission phase or afterward. 
The interaction between a SMBH binary with the gaseous disc in a gas-rich environment is likely to be the primary mechanism for the dissipation of orbital angular momentum by a SMBH binary in its final parsec stage (i.e., shrinking down from several parsec to sub-parsec separations, or, during the GW-emitting stage), effecting the merger within a Hubble time \citep[see, e.g.,][]{khan}. 
If one or both components of the SMBHB continue to accrete gas the coalescing binary may emit elctromagnetic counterparts with a periodic variability that would be detectable both in the PTA band \citep[see e.g., ][for reviews]{NANOGrav:2019tvo} and in the LISA band \citep{Kocsis_2006, armitage2002accretion}. 
Some of the SMBHB candidates detected with electromagnetic counterparts are listed in Table \ref{candidates}, and the proposed emission models are discussed in Sec.~\ref{Emission_models}. 

The joint detection of SMBHBs with GWs and electromagnetic counterparts stands to fundamentally transform our understanding of the important role played by astrophysics, and will offer cosmological probes via a new class of standard sirens while also revealing fundamental aspects of gravity \citep[see][for reviews]{Schnittman_2011, NANOGrav:2019tvo, Mangiagli:2020rwz}.

\subsubsection{Multi-messengers with pulsar timing arrays}
A stochastic isotropic GW background (GWB) is predicted from the GWs emitted by the population of SMBHBs at sub-parsec separations across the whole universe. 
The GWB detected by PTAs will be characterized by a common spectrum and interpulsar spatial correlations: the Hellings \& Downs, or HD correlations \citep{Hellings1983ApJ}, which is the ‘smoking gun’ signature for GWB signals \citep{Tiburzi2015}. 
Recently, evidence for a spatially uncorrelated common-spectrum process was detected in the 12.5 yr NANOGrav data set \citep{NANOGrav:2020bcs}, and later confirmed by the Parkes PTA \citep{Goncharov_2021}, the European PTA \citep{Chen:2021rqp}, and International PTA \citep{Antoniadis:2022pcn} in their corresponding data releases. 
The GWB production process is modelled as an additional time-correlated term with a similar power spectrum with GWB in all of the pulsars. 
While evidence is currently lacking in support of the existence of spatial HD correlations \citep{NANOGrav:2020bcs, Goncharov_2021, Chen:2021rqp, Antoniadis:2022pcn}, we cannot yet declare a detection of GWB. 
The reported signal offers a hint of the existence of a GWB from SMBHBs, but on-going analysis of the 15~yr data is likely to improve matters \citep{Middleton:2020asl,Mingarelli2019Nat}. 
The current non-detection is already providing interesting constraints on both SMBH binary populations \citep{Grahammnras1726, NANOGrav:2019tvo} as well as on individual candidates \citep{Jenet_2004}. 
With monitoring by the SKA, even a small number ($\sim 20$) of high-quality millisecond pulsars will be able to deliver valuable information about the redshift evolution of SMBHBs.
Calculations show that within 30 years of operation, about 60 individual SMBHB detections with $z<0.05$ and more than 104 with $z<$1 can be expected \citep{feng}.

\subsubsection{Multi-messengers with LISA}
Contrary to PTAs detecting GWs of individual SMBHBs in the local Universe ($z<0.5$), LISA, with its high GW detection sensitivity, could reach to the highest redshifts over a large mass range \citep{LISA2017arXiv}. 
Additionally, the detection of the full chirping signal of LISA sources will help to break the degeneracy between the chirp mass and luminosity distance for SMBHBs. 
The ability to measure the luminosity distance and sky-position of SMBHBs with the full inspiral-merger-ringdown waveform means LISA could make a relatively precise prediction of the source locations \citep{Cutler1998,Barack2004, Vecchio2004, Berti_will2005, holz2005using, Kocsis_2008, Mangiagli:2020rwz}.
Parameter estimation from the inspiral waveform could provide an early warning system for upcoming merger events, on the order of a week to up to a month in advance of the merger itself \citep{holz2005using, Lang_2008, Kocsis_2008, Mangiagli:2020rwz}, thereby enabling monitoring and detection of any associated electromagnetic counterparts. 
The predicted merger events of SMBH binaries detectable by LISA is uncertain at present and stands between about 5--100 per year \citep{Klein:2015hvg} as it is closely related to the poorly determined galaxy merger rates and the ill-determined dynamical evolution time scales of SMBHBs in their post-merger environments.
%
\subsubsection{The role of VLBI in multi-messenger studies}

Despite the various multi-band detections of SMBHB candidates at separations close to or larger than a parsec, sub-parsec scale-bound SMBHBs can only be spatially resolved with radio VLBI. 
The array size and observing frequency sets a fundamental limit on the ability of a VLBI network to resolve a telescopic binary. 
For a well distributed network, VLBI image resolutions primarily depend on the size of the network, and are thus expected to be $\sim$ 50, 20 and 15 $\mu$as at the proposed observing frequencies of 86, 230 and 345 GHz, respectively.
For a source at a redshift of $z=1$, these correspond to 0.41, 0.16 and 0.12 pc, respectively. 
These resolution limits are all smaller than the radius where dynamical accretion disk drag is thought to become ineffective (leading to the so-called ``final parsec'' problem), but much larger than the ISCO of $\sim$0.3 milliparsec for a $10^{9}~M_{\odot}$ SMBHB at that redshift. 
A SMBHB of that mass would therefore need to be at $z\lesssim 0.03$ in order for its ISCO to be resolveable by the ngEHT. 

The array size and observing frequency also sets a limit on the astrometric accuracy of even in-beam SMBHB astrometry. The candidate binary source 0402+379 ($z = 0.055$) was observed by the VLBA over 12 years and at frequencies between 5 and 22~GHz, revealing a motion consistent with a binary orbit with an orbital velocity of $0.0054 \, c \, \pm \, 0.0003 \, c$ and an orbital period of 3 $\times$ 10$^{4}$~yr, implying a total binary mass of 1.5 $\times$ 10$^{10}~M_{\odot}$ \citep{Bansal:2017izy}.
This source is likely an astrometric binary with a separation of $\sim$7.5 pc.
Astrometry of this source or other parsec-scale candidate sources with the ngEHT could be significantly more precise, but it would likely be impossible to observe non-linear motions (and thus fit an orbit) with the planned duration of the ngEHT program.

The enhanced resolution of the ngEHT is promising to resolve SMBHBs at sub-parsec separation through relative astrometry \citep{D_Orazio_2018, Fang:2021xab}. 
By fitting the time-varying visibility function of point-like emitters, one can in principle recover the orbital parameters of SMBHBs \citep{Fang:2021xab}. 
The joint detection of SMBHBs with VLBI and PTAs will help to break the degeneracy between characteristic parameters \citep{Fang:2021xab}. 
In the future, multi-messenger and/or multi-band detections of SMBHBs will enable us to confirm or rule out candidate sources. 

\subsection{Emission models of supermassive binary black holes}
\label{Emission_models}
The ngEHT as a VLBI array will measure complex visibilities in the Fourier domain depending on their mutual separation and its orientation with respect to the source (i.e., Earth aperture synthesis). 
Any theoretical model must make predictions for these or derived data products. 
We discuss several distinct modeling routes for the emission from a binary BH source and discuss prospects for detection and characterization of the source parameters.
\subsubsection{Hybrid modeling}

A critical and unavoidable complication for the ngEHT will be that any binary source will produce other emission components that are not directly related to the orbital motion of the BH binary. This will in general cause a mismatch between data and model that must be investigated and where possible mitigated. With a hybrid imaging+modeling technique  \citep[see][]{2020ApJ...898....9B} the ngEHT will be able to harness its exquisite signal-to-noise ratio to search for even faint binary components buried in diffuse emission beyond the nominal resolution.

\subsubsection{Model selection: binary vs Lense-Thirring}
A Bayesian framework like Themis allows for a statistically meaningful way to select models based on fit quality and degrees of freedom via the Bayesian evidence or standard information criteria (BIC, AIC, etc.).
More concretely, given two competing emission models, both can be fit to the ngEHT or simulated data and the statistically preferred model can be inferred.

\subsubsection{Simple emission models}
A simple but effective emission model that is implemented in Themis \citep{BroderickEtAl2020-ThemisCodePaper-b} is a double Gaussian source on a circular orbit. 
Obvious extensions to eccentric and PN orbits are planned. 
For any given value of the free parameters the model can generate the set of complex visibilities as a function of observing time and uv coordinate.
These simple emission models do not rely on the detailed physical properties of the source and instead utilize a simple representation of a physical scenario via a geometric approach.

\subsection{Simplified physical models: approximate, semi-analytic emission models}

In moving towards more realistic modelling approaches one can develop phenomenological models utilising semi-analytic schemes, e.g., by considering a SMBBH system orbiting within  semi-analytic accretion disk model of the surrounding circumbinary disk.

The simplest model of a binary BH is an analytic spacetime metric which is a superposition of two stationary Kerr BHs, with a binary separation large enough to ensure that the spacetime is approximately Minkowskian at the centre of mass of the system and tidal and secular effects induced between the binary pair are negligible.
Whilst such a metric does not formally satisfy the EFEs, we are only concerned with mildly relativistic orbital velocities and the spins of the BHs are anti-aligned such that precession is formally zero and the binary orbit is confined to a two-dimensional plane (adiabatic approximation).
Since ngEHT binary targets will have to contain SMBHs with orbital separations sufficient large that the distinct BHs can be spatially resolved, the above assumptions are appropriate.
Furthermore, this spacetime metric accurately incorporates the effects of gravitational self-lensing of the binary pair, which is a key observable from synchrotron-emitting plasmas in future ngEHT images.
The morphologies of self-lensing flares (SLFs) using a superimposed binary BH metric have been studied by \citet{Davelaar_raytracing}. 
In the case where the binary orbit is observed nearly edge-on,
a distinct feature in the light curve is imprinted by the shadow around the larger of the two BHs.
This method of SLF measurement could make it possible to infer and extract BH shadows that are spatially unresolvable by high-resolution VLBI.

\subsubsection{Physical models: GRMHD simulations of accreting SMBH binaries}

\begin{figure}
    \centering
    \includegraphics[width=\textwidth]{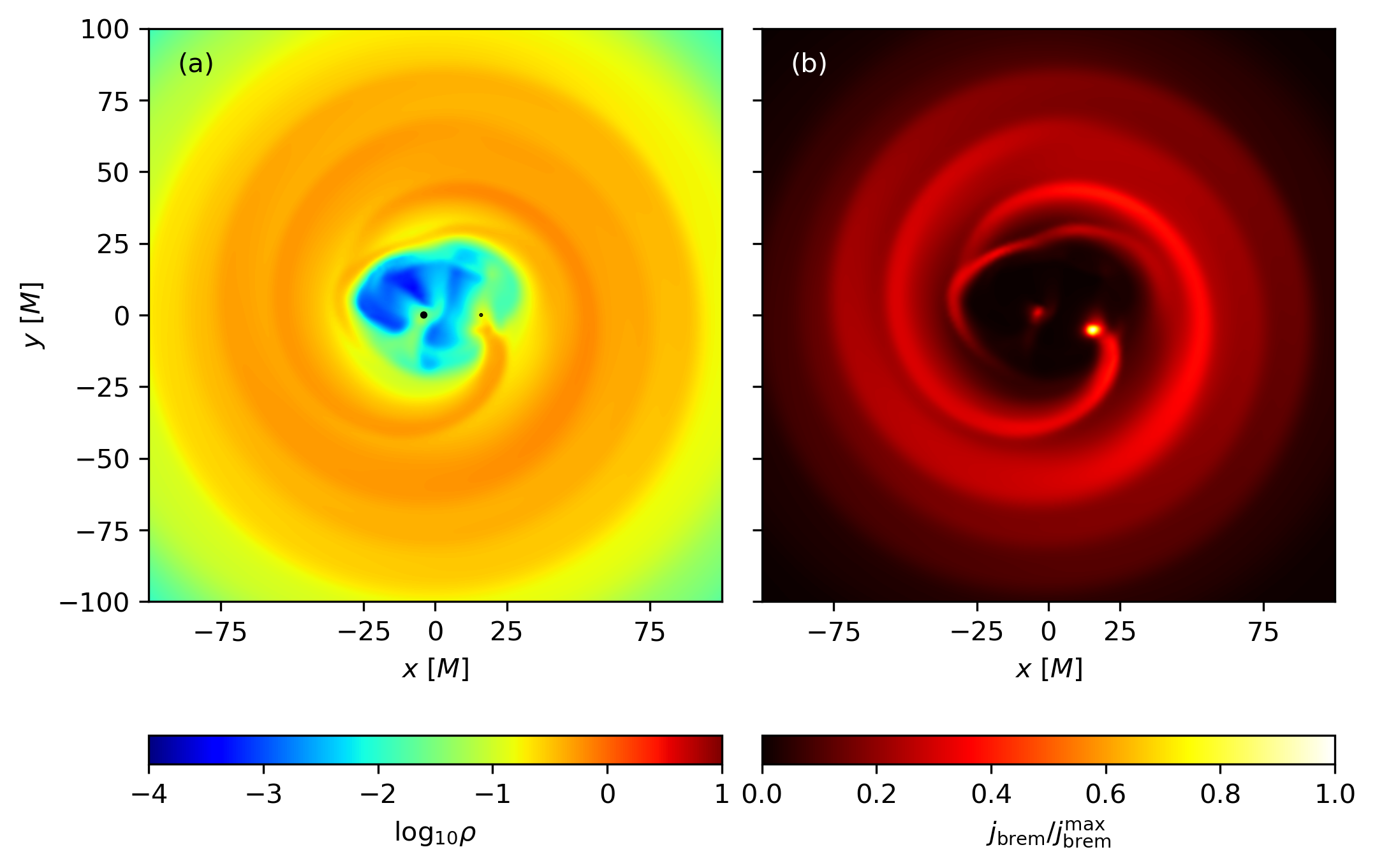}\\
    \includegraphics[width=\textwidth]{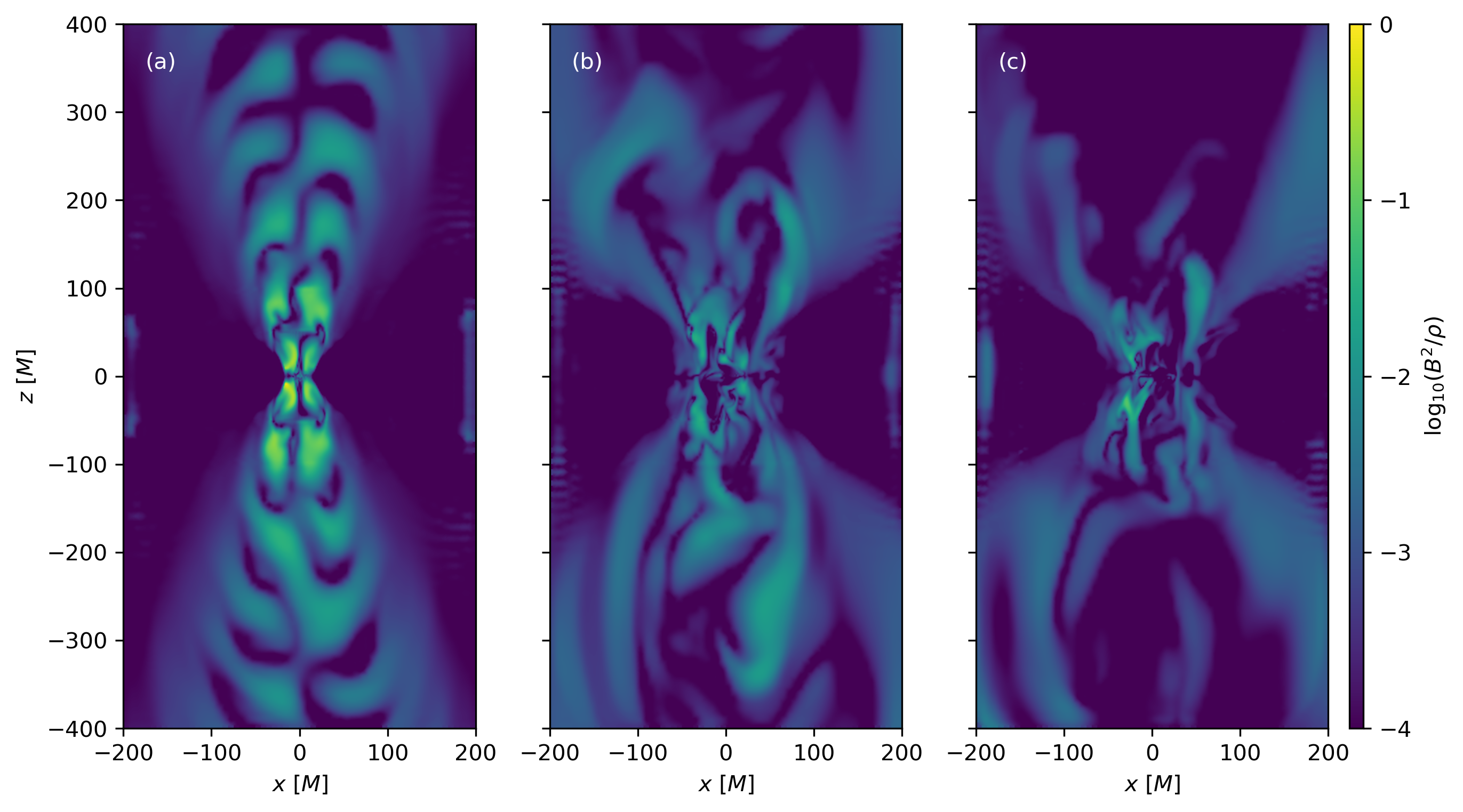}
    \caption{GRMHD simulations of SMBHBs. Top: simulation viewed in in the equatorial plane for a binary with mass ratio 1:4, colored by logarithm (base 10) of plasma rest mass density (left) and normalized bremsstrahlung emissivity ($j_{\rm brem}$, right). 
    A spiral shock is produced by the smaller mass secondary, increasing $j_{\rm brem}$.
    Bottom: the structure of the jet (colored by magnetization of the plasma) for three cases with different mass ratios: (a) 1:1 with zero spin, (b) 1:4 with zero spin, and (c) 1:4 with spin $a_{*}=0.7$ for both BHs. The jets form a ``braid'' structure for the 1:1 case, while the other cases show a more disordered structure
    \citep{olivares_binary_inprep}.}
    \label{fig:SMBBH_sim}
\end{figure}

\begin{figure}
    \centering
    \includegraphics[width=0.5\textwidth]{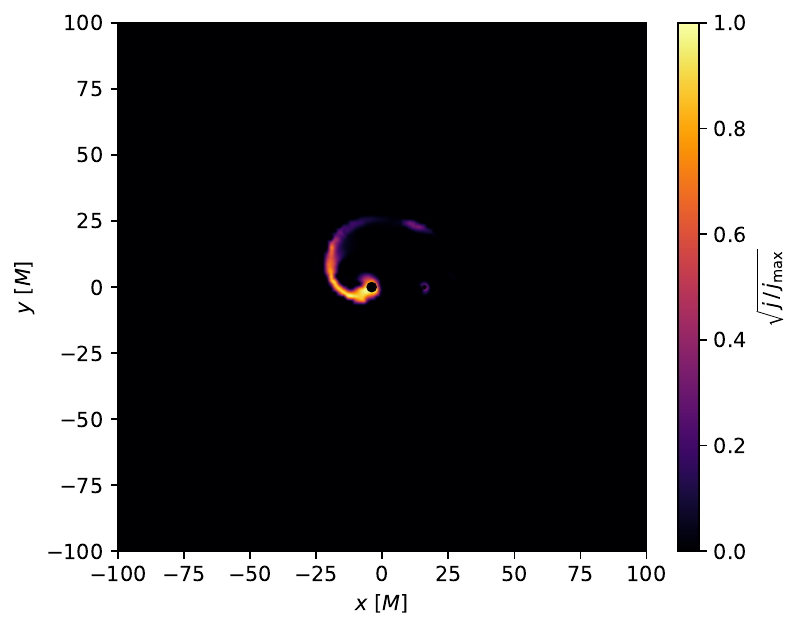}\includegraphics[width=0.5\textwidth]{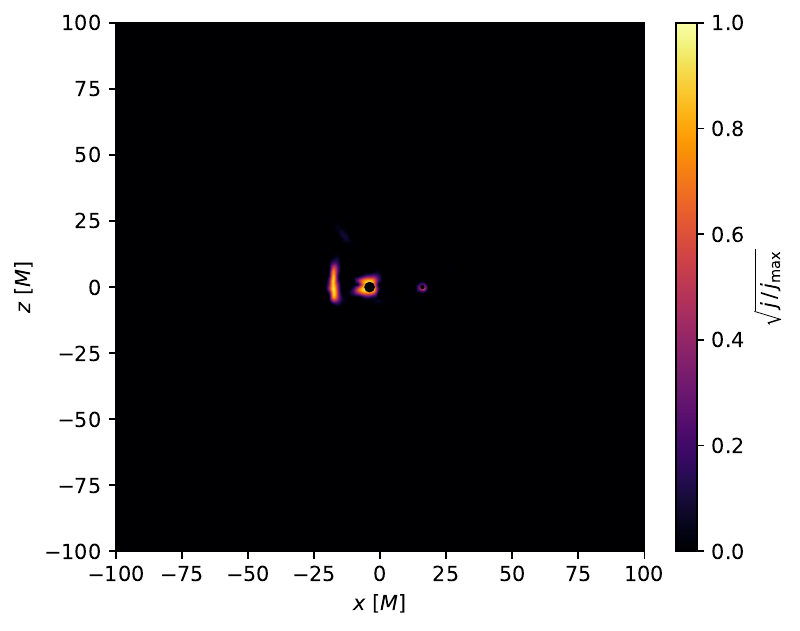}
    \caption{Proxy for synchrotron emissivity at 230 GHz \citep[equation (18) of][]{EventHorizonTelescope:2019pcy} for a GRMHD simulation of accretion onto a binary with mass ratio 1:4 and aligned spins with $a_{*}=0.7$, viewed in the equatorial plane ({\it left}) and meridional plane ({\it right}).
    It is possible to observe emission from the spiral shock, and contrary to the Bremsstrahlung case (c.f., Figure \ref{fig:SMBBH_sim}), the primary appears brighter than the secondary \citep{olivares_binary_inprep}.
    }
    \label{fig:SMBBH_sim_synch}
\end{figure}

 GRMHD simulations in combination with general-relativistic radiative transfer (GRRT) codes \citep[e.g.,][]{Dexter2009,Younsi2012,Chan2013,Younsi2015,Pu2016,Dexter2016,IPOLE_2018,Chan2018,Bronzwaer2018} are another approach for modeling emission from SMBHBs. 
 In contrast to geometric and semi-analytic modeling, they aim to model the observable phenomenology from `first principles' by directly solving the equations that describe the behavior of plasma, magnetic fields, and radiation, in the vicinity of BHs. 
 While this is desirable from the point of view of physical accuracy, their main disadvantage is their computational cost, which makes it impossible to sample the parameter space as efficiently as the aforementioned approaches. 
 This comes in large part from the necessity of simulating small length and time scales, such as those required to resolve the MRI turbulence on horizon-scales, alongside the much larger scales of the system such as those governing the dynamics of the binary.

While GRMHD simulations become prohibitive for binaries with large separations, where semi-analytic modeling can be sufficiently accurate, they can yield more detailed information at smaller separations, where the timescales of the accretion flow and the binary dynamics become comparable. 
This makes the two approaches complementary. 
GRMHD simulations can also provide information on the connection between large scale features observed in SMBHB candidates and the physical processes originating from within them, which is very difficult to establish using semi-analytic models.

In addition to considering the plasma as a test fluid which does not back react on the metric, GRMHD simulations of SMBHBs in the literature use different approximations that are valid at different separations and evolutionary stages of the binary.

At large binary separations, a metric constructed from the superposition of two boosted Kerr BH spacetimes approximates sufficiently well the corresponding solution to the EFEs.
GRMHD simulations on spacetimes constructed in this way have been performed by \citet{Combi2021} and explored in a series of papers \citep{armengol_circumbinary_2021,combi_minidisk_2022,gutierrez_electromagnetic_2022,noble_mass-ratio_2021}. 
The approximation is shown to be accurate up to separations as small as $\sim 10~r_{\rm g}$, when the BHs are moving with trajectories prescribed by PN equations. 
From the observational point of view, among the most interesting predictions is a periodic modulation in lightcurves caused by an asymmetric accretion disk, which in principle could be used to detect SMBHBs and constrain their masses \citep{noble_mass-ratio_2021}.

In order to obtain higher accuracy for the spacetime in a regime where gravitational radiation can still be neglected, another approach is to use a conformal thin sandwich approximation \citep{gold_accretion_2014,olivares_binary_inprep}. 
This enables finding a SMBHB spacetime that fulfills the Einstein constraints and is stationary in a reference frame co-rotating with the binary. 
Consequently, standard techniques used for GRMHD simulations in stationary spacetimes can be used. 
Figure \ref{fig:SMBBH_sim} displays snapshots of simulations of binaries performed this way for different values of the mass ratio and spin, reproduced from \citet{olivares_binary_inprep}. 
They show some of the features that are expected in accretion onto SMBHBs, such as spiral shocks producing bremsstrahlung radiation (top panels), and helical jets that appear more ordered for the equal mass case (bottom panels). 
The spiral shock may also be detectable in synchrotron emission at 230 GHz, as shown in Figure~\ref{fig:SMBBH_sim}, which employs a proxy for synchrotron emission described in \citet{EventHorizonTelescope:2019pcy}.

When gravitational radiation becomes dominant as a mechanism for decreasing the orbital separation, the spacetime needs to be evolved using full numerical relativity.
Besides being more physically accurate, this permits simultaneous retrieval of the GW signal and the electromagnetic counterpart, e.g., as done in \citet{gold_accretion_2014}.

After the merger, the spacetime settles down to that of a Kerr BH. The spacetime is stationary again in a frame where the BH is at rest. 
However, the accretion flow will likely still show signatures of the recent merger. 
It has been shown that after the release of linear momentum in the form of GWs, the resulting BH can have a residual recoil velocity from hundreds to thousands of kilometers per second \citep{baker_modeling_2007}.
This can produce spiral shocks emmitting Bremsstrahlung radiation, indicating a recent merger.
Simulations of these electromagnetic counterparts have been performed by \citet{Zanotti2010} and \citet{Meliani2017}.

The ngEHT will bring the possibility to refine theoretical models of circumbinary accretion flows by allowing comparison of simulations with higher precision observations.
Possible directions for simulation-based modeling of these systems in the near future are to specialize to the parameters of the most promising SMBHB candidates,
and to refine predictions of observational signatures that enable identification at different stages in their evolution.

\subsection{Summary of prospects of probing SMBH binaries with the ngEHT}
Supermassive BH binaries in sub-parsec separation phase evolution located at Gpc distances typically have angular separations of $1$--$10\mu$as.
This close separation is beyond the current resolution of most telescopes.
The ngEHT, with its higher resolution, promises to resolve these supermassive BH binaries through relative astrometry \citep{D_Orazio_2018, Fang:2021xab}. 

To track the orbital path of the components in the binary, both individual SMBHs are required to be bright enough to be detectable independently, or if one component is
bright, then a calibrator nearby is necessary, as it will be required for successful relative astrometry.
The orbital period could be inferred from the regularity of the periodic variations, which is possible when the orbital period is shorter than the detection duration of the ngEHT lifetime. 
The upper limit on the detection duration is the designed lifetime of ngEHT, which is 10 years. 
Furthermore, the validity of the second criterion requires that the mm-wavelength emission region (roughly the light-travel distance within the duration of the shortest mm-variability timescales) should be smaller than the binary separation.
It is suggested that the low-luminosity AGN (LLAGN) with mm-emission regions comparable to the size of the event horizon meet the requirement of the variation timescale in the emission region \citep{D_Orazio_2018}. 
By modeling the fraction of SMBHBs in the distribution function of LLAGN, \citet{D_Orazio_2018} predict that the abundance of SMBHBs resolvable by ngEHT at redshift $z\le 0.5$ is about 100, assuming the orbital period is less than the ten-year lifetime ngEHT. 

The quantity measured by EHT is the visibility function, which is the Fourier transformation of the intensity function of the image, realized by the technology of VLBI. 
In the LLAGN scenario of SMBHB candidates, the emission region of the individual sources is small compared to the binary separation. 
The simplest assumption for the intensity function of the components is to model them with point-like luminosities.
In this point-like luminosity approximation, the amplitude of the visibility function is proportional to $\sqrt{{I_1}^2 + {I_2}^2 + 2 I_1 I_2 \cos{2\pi {\bf u}\cdot {\boldsymbol \theta}} }$, where $I_1$ and $I_2$ are the intensity amplitudes for the two components, and ${\bf u}$ and ${\boldsymbol \theta}$ are the baseline vector and angular separation vector of the binary in the sky plane, respectively.
Due to the orbital motion of the binary, the projected relative position vector ${\boldsymbol \theta}$ changes its direction with respect to the baseline ${\bf u}$, and varies in amplitude if there is eccentricity or inclination in the binary orbit. 
Thus the variation of the visibility function is modulated by the binary orbital motion with the same period. 
By considering the binary intensities as point-like emitters, \citet{Fang:2021xab} make a proof-of-concept estimate that the orbital motion of the SMBHB could be traced and recovered by fitting the time-varying visibility function. 
They report that orbital tomography of the binary orbital motion is possible even if the binary orbital period is larger than the lifetime of ngEHT \citep{Fang:2021xab}. 
This could greatly increase the orbital period or semi-major axis of SMBHBs resolvable by ngEHT, and as a consequence increase the source abundance from $\sim$100 up to several 1000s, as inferred form the LLAGN scenario modeled by \citet{D_Orazio_2018}.

\subsection{Challenges \& future prospects}

To prepare for future detections and monitoring of SMBHBs with the ngEHT, further detailed simulations are required. 
\begin{itemize}
\setlength\itemsep{0.5em}
\item[\textbullet] The preparation of simulations testing the imaging quality of different ngEHT-arrays (e.g., number of telescopes) using different geometric models will be pivotal. In particular, emission blobs with various parameters (e.g., different intensities, angular separation, sizes, field of view) and their reconstructed images will need to be probed and studied in detail.
\item[\textbullet] Orbital and light curve tomographies will need to be ray-traced whilst taking additional physical effects and interactions into account.
\item[\textbullet] Modeling of different types of binary systems, i.e., two radio-loud objects versus a combination of a radio-loud and a radio-quiet object, will be essential. 
The role of jets, circumbinary accretion disk, and other radiating material media for specific candidate sources will need to be prepared. 
In this context, inspecting the specific signatures and time scales of LT-precession and helical magnetic fields will be important.
\item[\textbullet] Simulation of lensing signatures will be of crucial importance, i.e., if the accretion disk has (spiral) structure visible in the radio/synchrotron emission, the ngEHT may be able to see that structure lensed in the BH shadow (i.e., the emission is a gravitationally-lensed image of the accretion disk, which would have a spiral structure here). Further self-lensing effects with the ngEHT need to be studied.
\item[\textbullet] The influence of GW emission on the closest pairs and their appearance in ngEHT imaging also needs to be explored further.
\end{itemize}